\documentclass[11pt]{JHEP3}

\usepackage{amsmath,amssymb, amsfonts}
\usepackage{cite}
\usepackage{graphicx}

\allowdisplaybreaks

\makeatletter
\newcommand{\fmslash}[2][0mu]{%
  \mathchoice
    {\fmsl@sh\displaystyle{#1}{#2}}%
    {\fmsl@sh\textstyle{#1}{#2}}%
    {\fmsl@sh\scriptstyle{#1}{#2}}%
    {\fmsl@sh\scriptscriptstyle{#1}{#2}}}
\newcommand{\fmsl@sh}[3]{%
  \m@th\ooalign{$\hfil#1\mkern#2/\hfil$\crcr$#1#3$}}
\makeatother

\def\bra#1{\mathinner{\langle{#1}|}}
\def\ket#1{\mathinner{|{#1}\rangle}}
\def\braket#1{\mathinner{\langle{#1}\rangle}}
\DeclareMathOperator{\tr}{tr}

\newcommand{\sbraket}[1]{\lbrack #1\rbrack}

\newcommand{\dalpha}{\dot{\alpha}}
\newcommand{\dbeta}{\dot{\beta}}
\newcommand{\dgamma}{\dot{\gamma}}
\newcommand{\ddelta}{\dot{\delta}}

\newcommand{\dk}{\mathrm{d} k}
\newcommand{\CP}{\mathbb{CP}}
\newcommand{\dbar}{\bar\partial}

\allowdisplaybreaks
\numberwithin{equation}{section}
\numberwithin{table}{section}
\numberwithin{figure}{section}

\title{Deriving CSW rules for massive scalar legs\\ and pure Yang-Mills loops}
\author{Rutger Boels \\ Niels Bohr Institute, Niels Bohr International Academy,\\ Blegdamsvej 17, DK-2100 Copenhagen, Denmark\\ E-mail: \email{boels@nbi.dk}}
\author{ Christian Schwinn\\ Institut f\"ur Theoretische Physik E,\\ RWTH Aachen University, D - 52056 Aachen, Germany\\ Email: \email{schwinn@physik.rwth-aachen.de}}

\abstract{
This article provides two complementary detailed derivations of Cachazo-Svr\v{c}ek-Witten-style Feynman rules for Yang-Mills gauge theory coupled to a massive coloured scalar as presented in earlier work. These proceed through a direct canonical transformation method on space-time and through a gauge transformation in an action constructed on twistor space. It is shown explicitly that the field transformations are identical in both cases. Some simple tree-level examples of our rules are given and we comment on the application of them to the calculation of the rational part of one-loop pure glue amplitudes. A possible direct quantum completion of pure glue CSW rules based on dimensional regularisation motivated by these results is sketched. Finally, it is shown how to derive CSW rules for effective Higgs-gluon and Higgs-matter couplings proposed in the literature directly from the action. This derivation yields additional towers of vertices which generate a subset of the contributions to effective multi-Higgs scattering amplitudes.
}
\keywords{QCD, gauge symmetry, Higgs physics, NLO calculations}
\preprint{PITHA 08/10 \\ SFB/CPP-08-25 \\arXiv:0805.1197 [hep-th]  }

\begin{document}

\maketitle


\section{Introduction}

Scattering amplitudes in Yang-Mills theories can take much simpler forms than suggested by individual textbook Feynman diagrams. A prominent example are the maximally helicity violating (MHV) tree-level amplitudes with two negative helicity
gluons and an arbitrary number of positive helicity gluons that are given by a simple closed-form expression~\cite{Parke:1986gb,Berends:1987me}. Although Nair was able to interpret this simple formula in terms of a $\mathcal{N}=4$ supersymmetric two-dimensional sigma-model~\cite{Nair:1988bq} on a degree one curve embedded in twistor space, until recently no systematic way was known to exploit these results for calculations of general scattering amplitudes. This situation improved when Witten generalised Nair's observation and obtained a precise relation between all tree level Yang-Mills scattering amplitudes on space-time and, in general, higher degree curves in twistor space~\cite{Witten:2003nn}. Since twistor space is the space on which the conformal group acts linearly, this reflects the underlying conformal invariance of Yang-Mills theory at tree level. 

Witten's twistor observations and his speculations about a possible underlying twistor string theory led to the development of new Feynman-like rules by Cachazo, Svr\v{c}ek and Witten (CSW)~\cite{Cachazo:2004kj} where the MHV amplitudes serve as vertices.  The tree-level CSW rules were applied also to amplitudes with massless quarks and scalars in supersymmetric theories~\cite{Georgiou:2004wu,Georgiou:2004by,Wu:2004fb}, currents with external $W$ and $Z$ vector bosons or the Higgs boson~\cite{Dixon:2004za,Badger:2004ty,Bern:2004ba}
 and found numerous further applications \cite{Bena:2004ry, Kosower:2004yz,Ozeren:2005mp,Dinsdale:2006sq}.
At the one-loop level, a construction of the cut-constructable
pieces of amplitudes from MHV vertices was possible~\cite{Brandhuber:2004yw,Bedford:2004py,Quigley:2004pw,Bedford:2004nh,Brandhuber:2005kd,Badger:2007si,Glover:2008ff}
while the CSW representation of the rational piece of amplitudes was initially unknown.    In pure Yang-Mills theory, the tree-level CSW formalism was shown to follow from the on-shell recursion relations of Britto, Cachazo, Feng and Witten~(BCFW) \cite{Britto:2004ap,Britto:2005fq} both indirectly~\cite{Britto:2005fq} and directly~\cite{Risager:2005vk}. However, subsequently it was understood how to derive the CSW rules directly from a Lagrangian, independent of the BCFW rules~\cite{Gorsky:2005sf,Mansfield:2005yd,Boels:2007qn}. These action based approaches led to several proposals for the construction of the rational parts of one loop amplitudes~\cite{Brandhuber:2006bf, Boels:2007gv, Brandhuber:2007vm,Ettle:2007qc} and were furthermore applied to supersymmetric theories~\cite{Boels:2006ir,Feng:2006yy} and to Einstein gravity~\cite{Ananth:2007zy} or self-dual $\mathcal{N}=8$ supergravity~\cite{Mason:2007ct}.

In a previous paper~\cite{Boels:2007pj} we have presented CSW-style rules for amplitudes with colored massive
scalars, incorporating for the first time propagating massive particles in
this approach. The rules have been obtained using the two different Lagrangian
approaches in the literature~\cite{Mansfield:2005yd,Boels:2007qn}, demonstrating that these
methods are useful to derive new results. Apart from providing a simple testing ground for ideas, amplitudes with massive scalars are intrinsically interesting since they are related by supersymmetry to phenomenologically relevant tree amplitudes with massive quarks~\cite{Schwinn:2006ca} and to the rational part of pure Yang-Mills amplitudes at one loop~\cite{Dixon:1996wi}.  The scalar mass term in the Lagrangian was shown to give rise to a new tower of vertices with an arbitrary number of positive helicity gluons, in addition to the MHV vertices present already for massless scalars.
Like the MHV vertices in massless theories, the new vertices resulting from the mass term only contain ``holomorphic'' spinor products and therefore localise on lines in twistor space, whereas the structure of
the simplest on-shell scattering amplitudes of massive scalars is more
involved~\cite{Badger:2005zh, Forde:2005ue, Ferrario:2006np}. Therefore, in
contrast to the massless case, it appears difficult to derive the rules given in~\cite{Boels:2007pj} from the BCFW relations along the lines of~\cite{Risager:2005vk}. On the contrary, the holomorphic representation allows to give a far more direct proof of the BCFW recursion for amplitudes with massive scalars~\cite{Boels:2007pj} compared to earlier treatments~\cite{Badger:2005zh, Schwinn:2007ee}. 

In this paper we provide full details of the derivation of the massive CSW rules through the two different action based methods available in the literature. In one approach~\cite{Mansfield:2005yd} a canonical transformation is used to bring the Yang-Mills Lagrangian in light-cone gauge to a form which appears to involve only MHV vertices, as has been explicitly verified for the first five vertices~\cite{Ettle:2006bw}. The second approach~\cite{Boels:2007qn} starts from an action written directly on the twistor space~\cite{Mason:2005zm} that reduces to the space-time Yang-Mills action in a particular gauge~\cite{Boels:2006ir,Boels:2007qn} while a different gauge choice results in the CSW rules. The two 
approaches have been known to be closely related~\cite{Boels:2007gv} and in this article we will show equivalence explicitly by deriving the canonical transformation coefficients of~\cite{Ettle:2006bw} from the twistor
lifting formulae.

We also study the applicability of our rules to the calculation of the rational terms of one-loop amplitudes in pure Yang-Mills theory by demonstrating explicitly that the four-point amplitude with positive helicity gluons is correctly reproduced, without the need to take ``equivalence theorem violations''~\cite{Ettle:2007qc} into account or to abandon working with dimensional regularisation~\cite{Brandhuber:2007vm}. We argue that this feature persists for general amplitudes. As a further example of the derivation of CSW-rules we derive the rules of~\cite{Dixon:2004za,Badger:2004ty} resulting from an effective Higgs-gluon coupling and obtain additional multi-Higgs-gluon CSW vertices not noted in the previous literature.

The rest of this paper is organised as follows: We begin in section \ref{sec:glue} by setting up our conventions and notation, proceeding to review the canonical transformation method and the construction of a twistor action to derive the CSW rules. This will culminate in a direct comparison between the field transformations which are shown to coincide exactly. The section is closed by a simple observation about violations of the equivalence theorem. In the next section we derive the CSW rules for a massive colored scalar~\cite{Boels:2007pj} in the same two ways as demonstrated earlier for glue. Results obtained here are then applied in section \ref{sec-loops} to calculate one-loop scattering amplitudes in Yang-Mills theory. A final section on effective Higgs-gluon couplings, section \ref{sec:higgs}, leaves nothing but the conclusions and some technical appendices.

\section{Lagrangian based derivations of CSW rules}
\label{sec:glue}

\subsection{Spinor, light cone and twistor conventions}
\label{sec:notation}

We begin by setting up our general conventions for two-component spinors,  the light-cone decomposition used for the space-time derivation of the CSW Lagrangian and the conventions used for the twistor Yang-Mills approach.

To every four-momentum we can associate a matrix $p^{\alpha\dot\beta}=p_\mu \bar\sigma^{\mu \alpha \dot\beta}$.  For a
light-like momentum the matrix $p^{\alpha\dot\alpha}$ factorises into
a product of two spinors $\pi_p^\alpha$ and $\pi_p^{\dot\alpha}$ that
are determined up to a rescaling $\pi^{\dot\alpha}\to\lambda \pi^{\dot
\alpha}$ and $\pi^{\alpha}\to\lambda^{-1} \pi^{\alpha}$ with
$\lambda\in\mathbb{C}$ that leaves the momentum $p^{\dot\alpha \beta}$
invariant.  Spinors with lower indices are defined by
$\pi_{\dot{\beta}} = \pi^{\dot{\alpha}}
\varepsilon_{\dot{\alpha}\dot{\beta}}$ and $\pi_{\beta} = \pi^{\alpha}
\varepsilon_{\alpha\beta}$.  We also use the bra-ket notation
$\pi_{p\dot\alpha}=\ket{p+}$, $\pi_p^{\alpha}=\ket{p-}$,
$\pi_{p\alpha}=\bra{p+}$ and $\pi_p^{\dot\alpha}=\bra{p-}$ familiar
from the QCD literature.  We will refer to the dotted
spinors as holomorphic spinors and the un-dotted as anti-holomorphic
ones. These conventions
follow~\cite{Boels:2007qn} where the role of dotted and un-dotted
indices is reversed compared to~\cite{Witten:2003nn}.   Spinor products are denoted as
\begin{equation}
\label{eq:braket}
 \braket{ p q } = \braket{ p - | q +} = 
 \pi_p^{\dot \alpha} \pi_{q\,\dot\alpha},\qquad
 [ q p ] = \braket{ q + | p - } = \pi_{q\, \alpha} \pi_p^\alpha.
\end{equation}

Given a basis $(\eta^{\dot \alpha},\kappa^{\dot \alpha})$ 
of the holomorphic spinors and an anti-holomorphic basis
 $(\eta^{\alpha},\kappa^{\alpha})$ normalised according to $\braket{\eta\kappa}=\sbraket{\kappa\eta}=\sqrt 2$  one can expand
any momentum in terms of light cone components $(p_+,p_-,p_z,p_{\bar z})$ according to
\begin{equation}
\label{eq:eta-light-cone}
  p^{\alpha\dot\alpha}=p_-\,\eta^{\alpha}\eta^{\dot\alpha}
  +p_+\,\kappa^{\alpha}\kappa^{\dot\alpha}
  +p_z\,\kappa^{\alpha}\eta^{\dot\alpha}
  +p_{\bar z} \,\eta^{\alpha}\kappa^{\dot\alpha}
\end{equation}
The components of the momenta can be projected out  by
\begin{equation}
\label{eq:lc-components}
\begin{aligned}
 2 p_-=  \kappa_{\alpha} p^{\alpha\dot\alpha}\kappa_{\dot\alpha}&,&
 2 p_+= \eta_{\alpha} p^{\alpha\dot\alpha}\eta_{\dot\alpha}&,&
 2 p_z=- \eta_{\alpha} p^{\alpha\dot\alpha}\kappa_{\dot\alpha}&,&
 2 p_{\bar z}= -\kappa_{\alpha} p^{\alpha\dot\alpha}\eta_{\dot\alpha}
\end{aligned}
\end{equation}
With this conventions the Minkowski product is expressed in light-cone
components as
\begin{equation} 
p_\mu k^\mu= 
\frac{1}{2}p^{\alpha \dot\alpha}k_{\dot\alpha\alpha}
=p_+k_-+p_+k_-- p_zk_{\bar z}-p_{\bar z}k_{z}
\equiv p_+k^+ + p_- k^-+p_z k^z +p_{\bar z}k^{\bar z}
\end{equation}
The two-component spinors can be expanded in the $(\eta,\kappa)$ bases
as
\begin{equation}
\label{eq:define-spinors}
  \pi_p^\alpha=p_+^{-1/2} 
(p_{\bar z}\,\eta^{\alpha}+p_+\,\kappa^{\alpha})
\;,\quad
\pi_p^{\dalpha}=p_+^{-1/2} 
(p_{z}\,\eta^{\dalpha}+p_+\,\kappa^{\dalpha})
\end{equation}
up to an arbitrary phase. For negative or complex $p_+$ the square
root should be interpreted as $p_+^{1/2} =|p_+|^{1/2}e^{i\phi_p/2}$
with the phase defined by $p_+=|p_+|e^{i\phi_+}$.
 The expressions of spinor-products in
terms of the light-cone components are given by
\begin{equation}
\label{eq:lc-brakets}
  \begin{aligned}
    \braket{pk}&=
\sqrt 2\, p_+^{-1/2} k_+^{-1/2} (p_z k_+-k_zp_+)
    \equiv \sqrt 2\, p_+^{-1/2} k_+^{-1/2} (p,k)\\
    \sbraket{kp}&= 
    \sqrt 2\, p_+^{-1/2} k_+^{-1/2} (k_{\bar z}p_+-p_{\bar z} k_+)
    \equiv  \sqrt 2\, p_+^{-1/2} k_+^{-1/2}\{k,p\} 
  \end{aligned}
\end{equation}
Here we also have introduced  momentum brackets $(p,k)$ and $\{p,k\}$ as used
in~\cite{Ettle:2006bw}.
We also have the relations
\begin{equation}
\label{eq:braket-eta}
  \braket{ \eta k}=\sqrt {2 k_+}
=\sbraket{k \eta }
\end{equation}
for Minkowski momenta. 

In the light-cone formalism one can also implement an off-shell
continuation of spinor products~\cite{Mansfield:2005yd}.  Since
the momentum component $p_-=(p^2 +2 p_zp_{\bar z})/(2p_+)$ 
does not appear in the explicit representation of
the spinors given in~\eqref{eq:define-spinors}, these expressions
are also defined for off-shell momenta. 
This corresponds to an off-shell continuation of the spinors
defined by a light-cone projection of
the momentum $p$ according to
\begin{equation}\label{eq:spinormomentumtrick}
\pi_p^{\dot \alpha} \pi_p^{\alpha}=p^{\dot\alpha \alpha}
-\frac{p^2}{2p_+}\eta^{\dot\alpha}\eta^{\alpha}
\end{equation}
This prescription is equivalent~\cite{Kosower:2004yz} to the original 
off-shell continuation in the CSW rules~\cite{Cachazo:2004kj}.

For polarisation vectors satisfying the gauge condition $\eta\cdot
\epsilon^\pm=0$ we use the usual expressions of the spinor-helicity
formalism
\begin{equation}
\label{eq:polarization}
  \epsilon^{+,\alpha\dot\alpha}(k)=\sqrt 2  \,
 \frac{ \pi_k^{\alpha} \eta^{\dot\alpha}}{\braket{\eta k}} \quad;\quad
 \epsilon^{-,\dot\alpha\alpha}(k)=\sqrt 2  \,
 \frac{ \eta^{\alpha} \pi_k^{\dot\alpha}}{\sbraket{k\eta}} 
\end{equation}
The reference spinors $\eta^\alpha$ and $\eta^{\dalpha}$ 
are chosen to be identical for all gluons and are taken to be the same as in the light-cone decomposition~\eqref{eq:eta-light-cone}. This has been called the 'space-cone' gauge in~\cite{Chalmers:1998jb}.
The  relevant light-cone components of the polarisation vectors 
 are trivial in our spinor conventions 
\begin{equation}
\label{eq:lc-eps}
  (\epsilon^+(k))_{z}=-\frac{\sbraket{\eta  k}}{\braket{\eta k}}=1\qquad
  (\epsilon^-(k))_{\bar z}=-\frac{\braket{k\eta}}{\sbraket{k\eta}}=1
\end{equation}
Therefore the component $A_z$ of a gauge field $A$ can be identified with 
the positive helicity mode, the component $A_{\bar z}$ with the negative
helicity. 
The further non-vanishing components $(\epsilon^\pm)_-$ will play no role in the remaining discussion.

\subsubsection*{Twistor space}
Ordinary space-time is the space on which the Poincare group acts linearly. In a Lagrangian it is therefore natural to integrate over this space in order to display the Poincare symmetry of the theory explicitly. Twistor space is the space on which the conformal group acts linearly. It is therefore a natural question whether one can formulate an action on this space directly, since it is known that Yang-Mills theory at the tree level enjoys a conformal invariance. Concretely, the question is whether there is a way to write a Lagrangian which involves an integration over twistor space instead of ordinary space-time. Since the Poincare group is a subgroup of the conformal group, it can be expected that this can be done. 

For the purpose of the construction of actions on twistor space it is useful for technical reasons (see below) to work in Euclidean space instead of Minkowski space\footnote{A Minkowski space version of the action follows by simple Wick rotation. More formally, one can work on the total space of the spin-bundle.}. In this article we will be concerned with an off-shell version of the Penrose transform. For a review of on-shell twistor methods see~\cite{Woodhouse:1985id}. The twistor space for flat four dimensional space-time is $\CP^3$, with homogeneous coordinates $Z$,
\begin{equation}
Z = \left\{ (\omega^{\alpha},\pi_{\dalpha}) \in \mathbb{C}^4 | Z \sim \lambda Z \, \forall \lambda \in \mathbb{C}^*\right\}
\end{equation}
Here $\omega$ and $\pi$ are two two-component spinors which transform naturally under two $SL(2,\mathbb{C})$ subgroups. Space-time arises as the space of holomorphic lines in this space by the so-called incidence relation
\begin{equation}\label{eq:incidence}
\omega^{\alpha} = X^{\alpha \dot\alpha} \pi_{\dot \alpha} 
\end{equation}
It is easy to verify using this equation that the natural action of the generator of the linear symmetry group of $\CP^3$,
\begin{equation}
J^{\tilde{\alpha}}_{\tilde{\beta}} \sim Z^{\tilde{\alpha}}\frac{\delta}{\delta  Z^{\tilde{\beta}}},
\end{equation}
on the coordinates corresponds to the conformal group in four dimensional complexified space-time. This is of course a simple consequence of the complex algebra isomorphism $\textrm{SU}(4) \sim \textrm{SO}(6)$.  

Solving the incidence relation for $X$ requires a reality condition on this coordinate. In this article we will take Euclidean reality conditions, which are enforced by requiring reality under Euclidean conjugation denoted by a hat,
\begin{equation}
\widehat{\left(\begin{array}{c} \pi_1 \\ \pi_2 \end{array} \right)} = \left(\begin{array}{c} - \bar{\pi}_2 \\ \bar{\pi}_1 \end{array} \right).
\end{equation}
What is special about Euclidean signature is that the solution to \eqref{eq:incidence} is actually unique in this case: the resulting equation for $X$ is a projection map and $\CP^3$ can therefore be interpreted as the total space of a $\CP^1$ fibre bundle over $R^4$. Therefore, it is useful in Euclidean signature to think of $\CP^{3}$ as $R^{4} \times \CP^1$ with coordinates $(x,\pi)$ as in figure \ref{fig:fibrebundle}.
\FIGURE[t]{
  \includegraphics[scale=0.6]{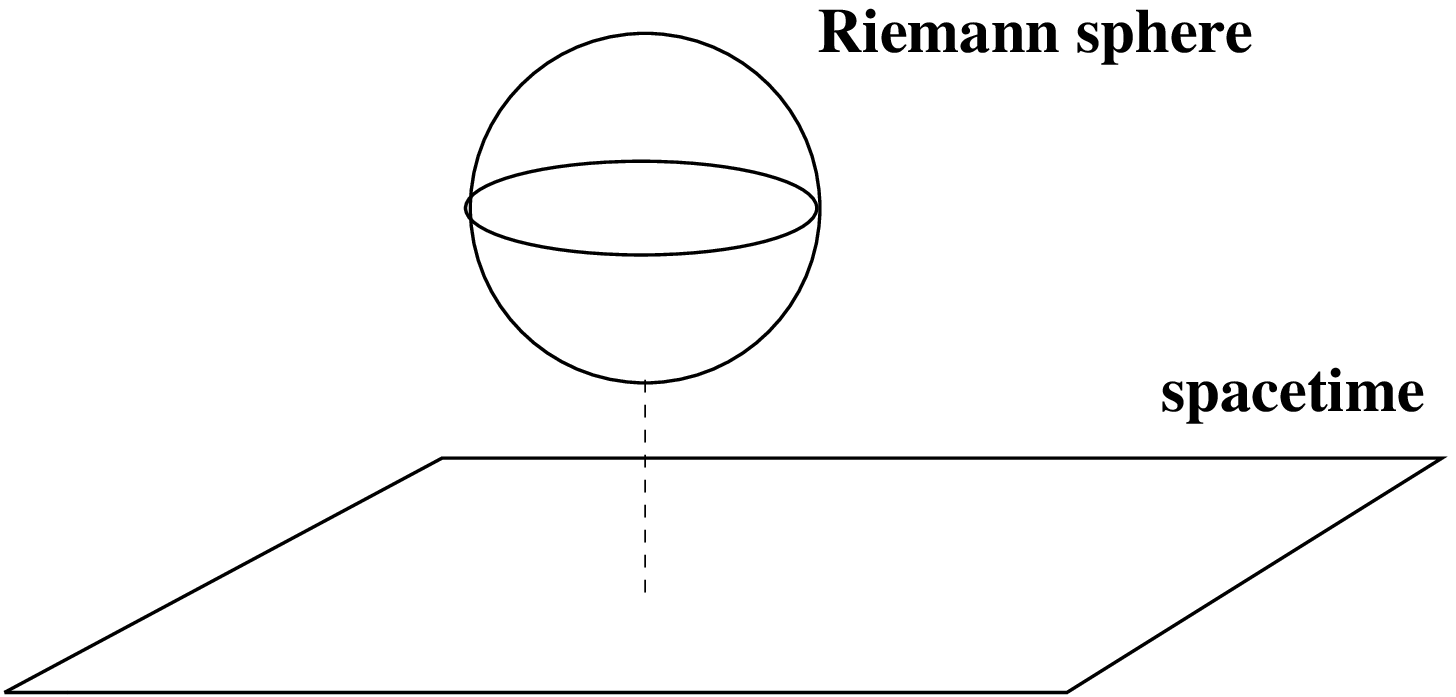}
  \caption{Twistor space in Euclidean signature is the total space of a $\CP^1$ fibre bundle}
  \label{fig:fibrebundle}
}
With the Euclidean conjugation we have the symmetric spinor products $\braket{\pi\hat\pi}=\braket{\hat\pi\pi}=\pi_1\bar\pi_1+\pi_2\bar\pi_2$. We can then take $\kappa=\hat\eta$ in the decomposition~\eqref{eq:eta-light-cone} for the euclidean signature, resulting in the replacements $2\to \braket{\eta\hat\eta}\sbraket{\hat\eta\eta}$ and a plus sign
in the expressions for $p_z$ and $p_{\bar z}$ in~\eqref{eq:lc-components}.

The next step\footnote{We apologise to the mathematically inclined reader for the following slightly ham-fisted explanation of Dolbeault cohomology classes on twistor space.} in the twistor construction is the definition of the natural fields on the twistor space. This is most natural in terms of the homogeneous coordinates $Z$ in $\mathbb{C}^4$. In order for these fields to be well-defined on the projective space, they need to have definite weight under the scaling $Z \rightarrow \lambda Z$. Furthermore, a notion of 'holomorphicity' is very important on the twistor space. $\CP^1$ is one complex dimensional, so has a holomorphic and a anti-holomorphic direction. The $R^4$ is roughly $\mathbb{C}^2$, so has two holomorphic and two anti-holomorphic directions. An 'anti-holomorphic vector' therefore has three components: two along space-time and one up the fibre direction. In more mathematical terms these 'anti-holomorphic vectors' are (dual to) $(0,1)$ forms on the twistor space. It is useful from a space-time point of view to write the needed $(0,1)$ forms in a basis with elements
\begin{equation}
\bar{e}_0 = \frac{\hat{\pi}^{\dalpha} d \hat{\pi}_{\dalpha}}{(\pi \hat{\pi})^2} \quad  \bar{e}^\alpha = \frac{ dx^{\alpha \dalpha} \hat{\pi}_{\dalpha}}{(\pi \hat{\pi})}
\end{equation}
Note that $\bar{e}_0$ points along the fibre, while the other $2$ directions point along space-time. With the above basis of one-forms, anti-holomorphic one-form fields $B$ can be expanded as
\begin{equation}
B =  \bar{e}^\alpha B_\alpha+  \bar{e}^0 B_0
\end{equation}
The $B_{\alpha}$ and $B_0$ therefore make up the components of what could be called a anti-holo\-mor\-phic vector. Note that $B_{\alpha}$ and $B_0$ have holomorphic weight $+1$ and $+2$ respectively compared to the original weight of $B$ with respect to scaling of the fibre coordinates. In addition, from a Lorentz point of view this $(0,1)$ form consists of a scalar and a spinor representation. 

We will furthermore need a notion of anti-holomorphic vector derivation, denoted $\dbar$. As can be verified by contracting into the above choice of basis of $(0,1)$ forms its components are,
\begin{equation}
\dbar_0 = (\pi\cdot\hat\pi)\pi_{\dalpha}{\partial\phantom{\pi}\over\partial\hat\pi_{\dalpha}}
\quad\quad\quad\quad
\dbar_\alpha = \pi^{\dalpha}{\partial\phantom{\pi}\over\partial x^{\alpha \dalpha}}
\end{equation}

The most basic twistor theory result now equates Dolbeault cohomology classes on twistor space with solutions to the wave equation (see e.g.~\cite{Woodhouse:1985id}). More concretely, if a $(0,1)$ form $B$ of weight $n$ satisfies
\begin{equation}\label{eq:twistoreq}
\dbar B = 0
\end{equation} 
than it corresponds to a solution of the wave equation on space-time with helicity $\frac{n+2}{2}$. Note that this equation can be solved up to invariance $B \rightarrow B + \dbar f$ for some scalar field $f$, which we will interpret as a gauge symmetry. Identifying solutions up to this invariance is the definition of the cohomology class and yields a one-to-one map between this class and solutions to the wave-equation. Note that equation \eqref{eq:twistoreq} in the basis introduced earlier splits into three equations
\begin{align}
\dbar_0 B_{\alpha} - \dbar_{\alpha} B_0 & = 0 \nonumber \\
\dbar_{\alpha} B^{\alpha}& =0 \label{eq:fieldeqtwi}
\end{align}
In the appendix \ref{app:noetherproc} it is shown explicitly in the case of a massless scalar how the solution to these equations up to gauge equivalence (i.e. the cohomology class) corresponds to a solution of the wave equation.

\subsection{CSW rules from a canonical transformation}
\label{sec:mansfield}

In this subsection we review the canonical
transformation approach  to the derivation of the  CSW rules
\cite{Mansfield:2005yd,Ettle:2006bw} in order to prepare for the 
extension of this method to massive scalars
given  in section~\ref{sec:scalars}.
 The conventions used for
the Yang-Mills fields are collected in appendix~\ref{app:yang-mills}.
Expressing the gauge field in terms of light-cone components~\eqref{eq:eta-light-cone} 
and imposing the light cone gauge $A_+=A\cdot\eta=0$, the Yang-Mills
Lagrangian depends only quadratically on the component $A_-$.
 The equation of motion
reads
\begin{equation}
\label{eq:eom-a-}
  A_-=\frac{1}{\partial_+^2}
\left([D_z,\partial_+A_{\bar z}]+[D_{\bar z},\partial_+A_z]\right)
\end{equation}
where the covariant derivatives are $D_i=\partial_i-ig A_i$ for
$i=z,\bar z$. Using~\eqref{eq:eom-a-}, the component $A_-$ can be integrated out, resulting in a  Lagrangian that contains
only the two physical components of the gauge fields $A_z$ (positive helicity) 
and $A_{\bar z}$ (negative helicity).
The Lagrangian for the physical modes has the form
\begin{equation}
\label{eq:light-cone-lag}
\mathcal{L}_A=
\mathcal{L}^{(2)}_{A_zA_{\bar z}}+\mathcal{L}^{(3)}_{++-}
+\mathcal{L}^{(3)}_{+--}
+\mathcal{L}^{(4)}_{++--}
\end{equation}
with the kinetic term
\begin{equation}
\mathcal{L}^{(2)}_{A_zA_{\bar z}}=
-2\tr A_z\square A_{\bar z}
=-4\tr A_z(\partial_+\partial_-
-\partial_z\partial_{\bar z})A_{\bar z}
\end{equation}
and the interaction terms 
 \begin{align}
\mathcal{L}^{(3)}_{--+}&= 4 ig \tr\left[
 [\partial_+A_{z}, A_{\bar z}] (\partial_+)^{-1}
(\partial_{z}A_{\bar z})\right]\\
\mathcal{L}^{(3)}_{++-}&= 4 ig \tr\left[
 [\partial_+A_{\bar z},A_z](\partial_+)^{-1}(\partial_{\bar z}A_{z})\right]\\
\mathcal{L}^{(4)}_{++--}&=4(ig)^2
\tr\left[[\partial_+A_{\bar z},A_z](\partial_+)^{-2}
[\partial_+A_z,A_{\bar z}]\right]
\end{align}

Note that for 'mostly +' CSW rules one would as a first step like to eliminate the non MHV-type coupling $\mathcal{L}^{(3)}_{++-}$. This vertex together with the kinetic term correspond \cite{Chalmers:1996rq} to the self-dual version of Yang-Mills theory. Eliminating this vertex therefore corresponds to solving self-dual Yang-Mills by mapping it to a free theory. This will be important when comparing to the twistor Yang-Mills formulation. 

Concretely, the non-MHV vertex can be eliminated by a transformation to new variables
$B$ and $\bar B$~\cite{Mansfield:2005yd} that satisfies the condition
\begin{equation}
\label{eq:transform-L}
\int d^3 x \left[\mathcal{L}^{(2)}_{A_{\bar z}A_z}+\mathcal{L}^{(3)}_{++-}
\right]
=\int d^3 x\mathcal{L}^{(2)}_{\bar B B}
\end{equation}
In order to have a unit Jacobian, the transformation
of the fields and conjugate momenta $(A_{z},\Pi_{A_{z}}\sim
\partial_+ A_{\bar z})$ to new fields and momenta $(B,\Pi_B\sim
\partial_+\bar B)$ has to be canonical. This requirement can be
satisfied by a transformation with the generating functional of the
form~\cite{Mansfield:2005yd,Ananth:2007zy}
\begin{equation}
  G[A_z,\Pi_B]=\int d^3 y B[A_z(\vec y)]\Pi_B(\vec y)
\end{equation}
Here $x^-$ is treated as time variable common to all fields and
the remaining coordinates are collected in the vector
$\vec x =(x^+,x^z,x^{\bar z})$.
The generating functional induces the  transformations of the fields,
 $ B=\delta G/\delta \Pi_B$,  and of the momenta
$\Pi_{A_z}= \delta G/\delta A_z$
resulting in the transformation 
\begin{equation}
\label{eq:momentum-trafo}
A_z\to B[A_z] \;,\quad \partial_+ A_{\bar z}(x)
=\int d^3 y \frac{\delta B(x^-,\vec y)}{\delta A_z(x^-,\vec x)} 
\partial_+ \bar B(x^-,\vec y)
\end{equation}
 This transformation of the momentum guarantees
that the structure of the kinetic term $A_{\bar z} \partial_+\partial_-A_z$
remains intact, provided $ B[A_z]$ depends on $x^-$
 only implicitly through $A_z$.
Using the form of the transformation~\eqref{eq:momentum-trafo}, the condition for the elimination of the non-MHV vertex~\eqref{eq:transform-L} can be written as~\cite{Mansfield:2005yd}
\begin{equation}
\label{eq:mansfield-def}
  \int d^3y \Bigl(\frac{\partial_z\partial_{\bar z}}{\partial_{+}}
  A_z(\vec y) -ig \left[A_z(\vec y),
    \frac{\partial_{\bar z}}{\partial_+}A_z(\vec y)\right]
       \Bigr)\frac{\delta B(\vec x)}{\delta A_z(\vec y)}
    = \frac{\partial_z\partial_{\bar z}}{\partial_{+}} B(\vec x)
\end{equation}
The dependence on the common light-cone time $x_-$ has been left
implicit.  In momentum space, the equations for $A_z$ and
$\partial_+A_{\bar z}$ can be solved using the Ans\"atze
\begin{align}
  A_{p,z}&=\sum_{n=1}^\infty\int \prod_{i=1}^n
   \widetilde{d k_i}\;
\mathcal{Y}(p,k_1,\dots,k_n)B_{-k_1}\dots B_{-k_n}
\label{eq:b-trafo}\\
p_+ A_{p,\bar z}&=\sum_{n=1}^\infty\sum_{s=1}^n \int \prod_{i=1}^n 
\widetilde {dk_i}\;
\mathcal{X}^s(p,k_1,\dots,k_n)B_{-k_1}\dots (k_{s+}) \bar B_{-k_s}\dots B_{-k_n}
\label{eq:bbar-trafo}
\end{align}
with the integration measure $\widetilde dp = dp_+ dp_z dp_{\bar z}/(2\pi)^3$ 
and keeping a momentum conserving delta-function
 $(2\pi)^3\delta^3(p+\sum_ik_i)$ implicit.

The condition~\eqref{eq:mansfield-def}
 can be used to extract a recursion relation for the
coefficients $\mathcal{Y}$\cite{Ettle:2006bw}:
\begin{equation}
\label{eq:y-recursion}
\omega_{1,n} \mathcal{Y}(k_1,\dots,k_n)=
 \sum_{j=2}^{n-1} g (\zeta_{k_{j+1,n}}-\zeta_{k_{2,j}})
 \mathcal{Y}(-k_{2,j},k_2,\dots,k_j)  
\mathcal{Y}(-k_{j+1,n},k_{j+1},\dots,k_n) 
\end{equation}
where a number of abbreviations have been defined as follows:
\begin{equation}
k_{i,j}=\sum_{\ell=i}^j k_\ell\quad,\quad
\omega_p=\frac{p_{\bar z} p_{z}}{p_+}\quad,\quad
\omega_{i,j}=\sum_{k=i}^j\omega_k\quad,\quad
  \zeta_p=\frac{p_{\bar z}}{p_+}
\end{equation}
Note that the prefactor
$ (\zeta_{k_{j+1,n}}-\zeta_{k_{2,j}})$ appearing in~\eqref{eq:y-recursion}
is proportional to the 
momentum-space vertex arising from the interaction $\mathcal{L}^{(3)}_{++-}$
that the canonical transformation is designed to eliminate~\cite{Ettle:2007qc}.
This recursion relation is similar to the
Berends-Giele relation for the one-particle off-shell current with positive helicity gluons~\cite{Berends:1987me,Dixon:1996wi} and can be solved with
similar methods.
The difference is that in the present case all gluons can be off-shell, 
in contrast to the case of the Berends-Giele relation.

The solution 
of the conditions~\eqref{eq:y-recursion} and~\eqref{eq:momentum-trafo}
has been found by Ettle and Morris (EM)~\cite{Ettle:2006bw} and reads
in our conventions  
\begin{align}
\label{eq:em-coeff}
 \mathcal{Y}(p,k_1,\dots,k_n)&=  (-ig)^{n-1} 
\frac{i^{n+1} p_{+}k_{2+}\dots k_{(n-1)+}}{(1,2)\dots((n-1), n)}
= \frac{ (g \sqrt 2)^{n-1}\braket{\eta p}^2}{\braket{\eta 1}
  \braket{12}\dots\braket{(n-1) n}\braket{n \eta}}\\
\mathcal{X}^s(p,k_1,\dots,k_n)&=-\frac{k^s_+}{p_+} 
\mathcal{Y}(p,k_1,\dots,k_n)
 =-\frac{ (g \sqrt 2)^{n-1}\braket{\eta s}^2}{\braket{\eta 1}
  \braket{12}\dots\braket{(n-1) n}\braket{n \eta}}
\label{eq:em-coeff-bbar}
\end{align}

Inserting the expressions~\eqref{eq:b-trafo} and~\eqref{eq:bbar-trafo} 
into the light-cone Lagrangian~\eqref{eq:light-cone-lag} leads
to a tower of  vertices that contain two $\bar B$ fields and
an arbitrary number of $B$ fields:
\begin{equation} 
\mathcal{L} =\mathcal{L}^{(2)}_{B\bar B}+
\sum_{n=3}^\infty \mathcal{L}^{(n)}_{+\dots +--}
\end{equation}
The interaction vertices in momentum space take the form
\begin{equation}
\mathcal{L}^{(n)}_{+\dots +--}=
\frac{1}{2} 
\sum_{j=2}^n
\int \prod_{i=1}^n 
\widetilde{d k_i}
\mathcal{V}_{\bar B_1, B_2,\dots \bar B_j,\dots B_n}
\tr \left( \bar B_{k_1}\dots\bar B_{k_{j}}\dots B_{k_n}\right)
\end{equation}
It has been argued~\cite{Mansfield:2005yd} and explicitly checked up
 to $n=5$~\cite{Ettle:2006bw} that the coefficients $
 \mathcal{V}_{\bar B_1,\dots \bar B_i,\dots \dots B_n}$ are just the
 MHV amplitudes continued off-shell according to the CSW prescription.

 Since the field-redefinitions~\eqref{eq:b-trafo}
 and~\eqref{eq:bbar-trafo} begin with a linear term, i.e.
 $A_z=B+\mathcal{O}(B^2)$ and $A_{\bar z}=\bar B+\mathcal{O}(\bar B
 B)$, standard reasoning (see e.g. \cite{'tHooft:1973pz}) suggests
 that on-shell matrix elements of the new and old fields coincide up
 to a possible wave-function renormalisation (this has been shown to
 be absent at one loop~\cite{Ettle:2006bw}) since the non-linear terms
 naively do not contribute to single particle poles. This result is sometimes known as the 'equivalence theorem'. However, due to the non-local character of the field-redefinition which violates one of the assumptions in the theorem the situation is more subtle. This will be discussed in more detail in~\ref{sec:violate}. 

\subsection{CSW rules from twistor Yang-Mills theory}
\label{sec:twistor}
As stated before, the Penrose correspondence relates Dolbeault cohomology classes for $(0,1)$ forms of specified weight with solutions to the wave equation on space-time with specified helicity. This means that the most natural action on twistor space is first order. Based on this information and gauge-invariance we will review the construction of an action on twistor space from the ordinary space-time one by means of an off-shell version of the Penrose transform. 

The basic building block is a $(0,1)$ form $B$. Interpreting this as a gauge field, three natural independent $(0,2)$ form curvatures can be defined by commuting covariant derivatives, $F_{\alpha \beta} = \epsilon_{\alpha \beta} F$ and $ F_{0 \alpha}$. Since we are only interested in curvature along space-time, we will impose $F_{0 \alpha}=0$. Note the close connection of this constraint to the constraints used in harmonic superspace methods\cite{harmonicsuperspace}. From this one can derive~\cite{Boels:2007gv} the 'lifting' formula 
\begin{equation}\label{eq:liftforA}
-i g A_{\alpha \dalpha}(x) \pi^{\dalpha}  = H^{-1}\left( \dbar_\alpha - \sqrt{2} i g B_{\alpha}  \right)H 
\end{equation}
Here $H$ are holomorphic frames such that $(\dbar_0 -i \sqrt{2} g B_0)H=0$. These frames are defined at each space-time point $x$ and always exists perturbatively since Yang-Mills theory on the 'extra' two dimensional sphere $\CP^1=S^2$ has no local degrees of freedom. Working out the remaining curvature, one arrives at the intriguing formula
\begin{align}\label{eq:liftforselfdualcurv}
\frac{1}{2} H^{-1} \left[\dbar_\alpha -i \sqrt{2} g B_\alpha, \dbar^{\alpha} -i \sqrt{2} g B^{\alpha} \right] H  &=  -i \sqrt{2} g H^{-1}\left( \dbar_\alpha B^{\alpha} -i \sqrt{2} g B_\alpha B^\alpha \right)H \nonumber \\ & = -i g F_{\dalpha \dbeta}(x) \pi^{\dalpha} \pi^{\dbeta} 
\end{align}
where $F_{\dalpha \dbeta}$ is the self-dual part of the space-time curvature. As stated before, the most natural action on twistor space is first order. With this fact and given equation (\ref{eq:liftforselfdualcurv}), it is natural to study the Chalmers and Siegel action~\cite{Chalmers:1997ui}, 
\begin{equation}\label{eq:chalsiegaction}
S_{\textrm{Chalmers and Siegel}} =  \tr \int dx^4 C_{\dalpha \dbeta} F^{\dalpha \dbeta}_+ - \frac{1}{2} C_{\dalpha \dbeta} C^{\dalpha \dbeta}.
\end{equation}
Here $C$ is an auxiliary self-dual two-form. Demanding that the first term is local on twistor space then yields the lifting formula for the self-dual field $C$, 
\begin{equation}\label{eq:liftforB}
C_{\dalpha \dbeta} = \sqrt{2} \int_{\CP^1}\dk\, H^{-1} \bar{B}_0 H \pi_{\dalpha} \pi_{\dbeta} 
\end{equation}
Here $\dk$ is the natural volume form on $\CP^1$,
\begin{equation}
\dk = \frac{\braket{\pi d\pi} \braket{\hat{\pi} d\hat{\pi}}}{\braket{\pi \hat{\pi}}}
\end{equation}
Note that the above transform for $C$ can also be arrived at by more carefully considering~\cite{Mason:2005zm} the $C^2$ term: since this will be non-local on twistor space, to keep gauge invariance a gauge link operator between the two operators has to be inserted. This Wilson line operator is naturally formed by the frames and because it is a link in $2$ dimensions it is independent of the path taken.   

Putting in the transforms and the constraint $F_{0\alpha}=0$ by Lagrange multiplier then yields the full twistor action for Yang-Mills theory
\begin{align}\label{eq:twym}
S =  &  2 \tr \int d^4\!x \dk  \bar{B}_0 \left(\dbar^{\alpha} B_{\alpha}-i \sqrt{2} g B^{\alpha} B_{\alpha} \right) + \bar{B}^{\alpha} \left(\dbar_{\alpha} B_0 - \dbar_0 B_\alpha -i \sqrt{2} g [B_\alpha,B_0]\right) \nonumber\\
& - \tr \int d^4\!x \dk_1 \dk_2 H^{-1}_1 \bar{B}^0(\pi_1) H_1 H^{-1}_2 \bar{B}^0(\pi_2) H_2 \braket{\pi_1 \pi_2}^2
\end{align}
This action is invariant under $B \rightarrow B+\dbar_B f^0, \bar{B} \rightarrow \bar{B} - i \sqrt{2} g [\bar{B},f^0]$, as well as $\bar{B} \rightarrow \bar{B}+\dbar_B f^{-4}$ for two independent gauge parameters of the indicated weight and the covariant derivative $\dbar_B=\dbar -ig \sqrt 2 B$. Note that the second symmetry arises from the usual 'quantising with constraints' procedure. It is therefore natural to interpret $\bar{B}$ as a second $(0,1)$ form, and the action can be written as
\begin{align}\nonumber S = \tr & \int d^4x \dk \,  \bar{B} \wedge \left(\dbar_B B \right)  \\ & -  \tr \int d^4x \dk_1 \dk_2 \wedge \, \braket{\pi_1 \pi_2}^2 \left(H^{-1} \bar{B} H\right)_{(1)} \wedge \left(H^{-1} \bar{B} H\right)_{(2)} 
\end{align}
Since both gauge symmetry parameters live on the (six real dimensional) twistor space, the gauge symmetry is \emph{larger} than on space-time. In order to calculate correlation functions for scattering amplitudes from this action, it needs to be fixed and two gauge choices will be discussed below. 

The action derived here can also be derived from the original twistor string proposal~\cite{Witten:2003nn} by truncating to single trace terms, as is in principle clear already from~\cite{Cachazo:2004kj}. As a historical aside, note that the same action follows from simply adding the result on MHV amplitudes in~\cite{Nair:1988bq} to the local twistor space action of~\cite{Sokatchev:1995nj}. 

\subsubsection*{Space-time gauge}
Since the part of the gauge freedom which is not present on space-time resides on the $\CP^1$ fibre, it is natural to try to fix this extra symmetry with a Lorenz-like gauge
\begin{equation}
\dbar^{\dagger} B_0 = \dbar^{\dagger} \bar{B}_0 =0
\end{equation}
Working out this gauge choice turns out to reduce the action in \eqref{eq:twym} back to ordinary Yang-Mills\cite{Boels:2006ir}. In particular the infinite tower of interactions in the second term reduces to a quadratic term. The ghosts associated to this gauge condition can be shown to decouple. The residual gauge symmetry left from the above gauge-fixing is exactly ordinary space-time gauge symmetry. Fixing this will, of course, lead directly to ordinary Yang-Mills perturbation theory. 

\subsubsection*{CSW gauge}
More general gauges are possible. Picking an arbitrary spinor $\eta_{\alpha}$, the CSW gauge is the axial gauge choice
\begin{equation}
\eta_{\alpha} B^{\alpha} = \eta_{\alpha} \bar{B}^{\alpha} =0
\end{equation}
An important consequence of this choice is that it eliminates the vertex in the 'CF' part of the action. Since this term came from the interaction in $F^{+}$, this gauge choice eliminates the original 'self-dual' vertex. The propagators and on-shell fields can be calculated by an ordinary Fourier transform on the space-time coordinate. This changes coordinates,
\begin{equation}
\{x_{\alpha \dalpha},\pi_{\dbeta}\} \rightarrow \{i p_{\alpha \dalpha}, \pi_{\dbeta} \}.
\end{equation}
The propagator relevant here\footnote{There are more propagators, but these can be shown to be irrelevant for calculating amplitudes because the only vertex in the game contains only $B_0$ and $\bar{B}_0$ fields.} is 
\begin{equation}\label{eq:propab}
:B_0 \bar{B}_0: = \frac{i \delta(\eta \pi_1 p) \delta(\eta \pi_2 p)}{p^2}.
\end{equation}
The only vertices left in the theory are now the infinite tower in the part which originated from $C^2$. These already have the right shape to be the MHV scattering amplitudes, as an expansion in terms of the fields gives,
\begin{equation}
C^2 = - \tr \sum_{n=2}^\infty (- i\sqrt 2 g)^{n-2} \int \sum_{p=2}^n \frac{\braket{\pi_{1} \pi_{p}}^4}{\braket{\pi_1 \pi_2} \braket{\pi_2 \pi_3} \ldots \braket{\pi_n \pi_1}} {\rm tr}\left( \bar{B}_0^1 B_0^2 \ldots B_0^{p-1} \bar{B}_0^p B_0^{p+1}\ldots B_0^n \right) 
\end{equation}
To calculate the scattering amplitudes correctly, one needs the relation between correlators and scattering amplitudes given by LSZ. This can be worked out~\cite{Boels:2007qn} for the twistor action starting from  ordinary Yang-Mills using \ref{eq:liftforA}. There is a subtlety here related to the application of the 'equivalence theorem' which will be addressed below in section \ref{sec:violate}. With this taken into account, the remaining vertices generate exactly MHV amplitudes, and from there it is a small step to see the complete CSW rules. Note how the CSW spinor arises here from the gauge fixing spinor, and the independence of this spinor needed for Lorentz invariance translates therefore to gauge invariance. 

One should note the close similarity of the above gauge choices with harmonic superspace methods. Harmonic superspace has a similar extended gauge symmetry and basically the same gauge choices are used. The main difference is that where the 'CSW' gauge here breaks one of the $SL(2,\mathbb{R})$ Lorentz subgroups, in the harmonic superspace case the corresponding gauge only breaks the $SU(2)$ $R$-symmetry. 

\subsection{Comparison of the two approaches}
\label{sec:compare}

Above we reviewed two seemingly very different action-based derivations of the CSW rules. A natural question is how
these two are related.  On the level of perturbation theory it is clear that the two approaches will be equivalent since both lead to the same expressions for Feynman diagrams. They are also based on the same principle: both methods
trivialise the self-dual Yang-Mills equations by a canonical field transformation. However, this does not show that the lightcone approach to MHV diagrams is the same as the twistor one on the level of the amplitudes. The remaining gap is that the canonical transformation method has only been checked to give off-shell MHV vertices up to 5 gluons, although there is no reason to really doubt the general result. In the twistor-action, on the other hand, the all-multiplicity vertices can be easily derived but the action is formulated in terms of fields on twistor space instead of space time. One of the ways to complete the argument on the level of the action is to start with the twistor action, impose CSW gauge and integrate out all the components of the gauge fields which do not appear in the actual rules. That this can be done in principle
is clear. However, one encounters some obstacles as, for instance, the conjugate component to $\eta\eta A$ is a non-linear combination of twistor fields. On the other hand, the field equation for the component of $\bar{B}$ which is gauged to zero will impose some constraint which need to be taken into account. A second approach would be to take the Chalmers and Siegel action and apply the canonical transformation method directly to this.

Instead of attempting to show equivalence on the level of the action, here we relate the field transformations used in the two approaches and show that the canonical transformation is the space-time interpretation of the twistor gauge transformation. This implies that the twistor space gauge symmetry is the linear version of the non-linear symmetry probed by the canonical transformation. Some evidence that this is true was uncovered in \cite{Boels:2007gv}; below we complete the argument.

The space-time interpretation of the twistor gauge transformation utilised above to derive the original CSW rules follows directly from the lifting formula for the gauge field~\eqref{eq:liftforA}. In CSW gauge we have,
\begin{equation}
-i g \eta^{\alpha} A_{\alpha \dalpha}(x) \pi^{\dalpha}  = \eta^{\alpha}  H^{-1}(\pi)\left( \dbar_\alpha\right)H (\pi)
\end{equation}
which is almost an axial gauge on space-time. This observation can be made precise by considering the frames $H$: these are the solutions to $(\dbar_0 - i \sqrt{2} g B_0) H =0$. For a definite solution of this differential equation, a boundary condition must be chosen. If one chooses $H(\eta) = 1$ for an arbitrary point $\eta^{\dalpha}$, 
\begin{equation}
\eta^{\alpha} \eta^{\dalpha} A_{\alpha \dalpha} = 0
\end{equation}
holds. Hence the space-time interpretation of the twistor CSW gauge is light-cone gauge, with $\eta^{\alpha} \eta^{\dalpha}$ the gauge-fixing vector. The main advantage of light-cone gauge on space-time is that the physical polarisations of the gluon are manifest and are given
in terms of twistor fields by:
\begin{equation}\label{eq:firsttrans}
A_z \equiv \frac{1}{2} \eta^{\alpha} \hat{\eta}^{\dalpha} A_{\alpha \dalpha}(x)=  \frac{i}{2 g} H^{-1}(\hat{\eta}) \left(\eta^{\alpha} \hat{\eta}^{\dalpha} \partial_{\alpha \dalpha} \right)H(\hat{\eta}) 
\end{equation}
and
\begin{equation}\label{eq:secondtrans}
A_{\bar{z}} =  \frac{1}{2} \hat{\eta}^{\alpha} \eta^{\dalpha}A_{\alpha \dalpha}(x) = \frac{\sqrt{2}}{2} \hat{\eta}^{\alpha} B_\alpha(\eta)
\end{equation}
The $\bar{B}_0$ field equation from the action \eqref{eq:twym} can then be used to give for the second equality
\begin{equation}
i(\eta_\alpha \eta_{\dalpha} p^{\alpha\dalpha})
 A_{\bar{z}}(p) =  \frac{1}{\sqrt{2}} (\hat{\eta}^{\alpha} \eta_\alpha) \int_{\CP^1}\dk\, H(\eta)H^{-1} \bar{B}_0 H H^{-1}(\eta) \braket{\pi \eta}^2
\end{equation}
Let us focus on equation (\ref{eq:firsttrans}) first. The frames can be expanded into $B_0$ fields using
\begin{equation}
\label{eq:expand-h}
H(\hat{\eta}) = H(\hat{\eta}) H^{-1}(\eta) = \frac{H(\hat{\eta}) H^{-1}(\eta)}{\braket{\hat{\eta} \eta}} \braket{\hat{\eta} \eta} = \left(\frac{1}{\dbar_0 - i \sqrt{2} g B_0} \right)_{\hat{\eta} \eta}\braket{\hat{\eta} \eta}
\end{equation}
Note that this formula is local on space-time. Now 
\begin{equation}
\delta \left(\frac{1}{\dbar_0 - i \sqrt{2} g B_0} \right)_{\hat{\eta} \eta} =  i\sqrt{2} g \int_{\pi_k} \frac{ H(\hat{\eta})  H^{-1}(\pi_k)}{\braket{\hat{\eta} \pi_k}} \delta B_0 \frac{H(\pi_k) H^{-1}(\eta) }{\braket{\pi_k \eta}}
\end{equation}
can be used to find
\begin{equation}\label{eq:emforA}
A_z(x) =  -\frac{1}{\sqrt{2}} \braket{\hat{\eta} \eta} \int_{\pi_k} \frac{H(\eta) H^{-1}(\pi_k)}{\braket{\eta \pi_k}}(\frac{\braket{\eta \pi_k}}{\braket{\hat{\eta} \pi_k}} \eta^{\alpha} \hat{\eta}^{\dalpha} \partial_{\alpha \dalpha} B_0) \frac{H(\pi_k) H^{-1}(\eta) }{\braket{\pi_k \eta}}
\end{equation}
This expression can be expanded into components. Observe that when calculating scattering amplitudes through Feynman diagrams in the twistor action approach, all the $B_0$ fields in the above expression will only be contracted with $\bar{B}_0$ fields, leading to delta-functions of $(\eta_{\alpha} \pi_{\dalpha} p^{\alpha\dalpha})\equiv\braket{\pi(\eta p)}$ on the fibre. To see what the above expression corresponds to on space-time we can therefore insert the delta functions everywhere and perform the integrals. Expanding to third order for instance yields
\begin{align}
\label{eq:em-twistor}
A_z(p) = \frac{-i}{\sqrt{2}} \braket{\hat{\eta} \eta} \left( \frac{B_0(q_1)}{\braket{\eta (\eta q_1)}}
 -i g \int_{q_1,q_2}\braket{\eta (\eta p)}
\frac{B_0(q_1) B_0(q_2)}{\braket{\eta (\eta q_1)}\braket{(\eta q_1) (\eta q_2)}
\braket{\eta (\eta q_2)}}\right. \nonumber\\
- \left. \sqrt{2} g^2\int_{q_1,q_2,q_3} \braket{\eta (\eta p)}
\frac{B_0(q_1) B_0(q_2) B_0(q_3)}{\braket{\eta (\eta q_1)}
\braket{(\eta q_1) (\eta q_2)}
\braket{(\eta q_2) (\eta q_3)} \braket{\eta (\eta q_3)}} +
\ldots \right)
\end{align}
Here momentum conservation was used to sum the numerator. 
The integral shorthand is defined by $\int_{q_1,\dots q_n}=\int d^4 q_1\dots d^4 q_n \delta^4 (p+q_1+\dots +q_n)$.
 Proceeding further shows that the full sum will be exactly
the solution to the recursion relations in equation
\eqref{eq:em-coeff-bbar} originally found by Ettle and
Morris\cite{Ettle:2006bw}. To see this more explicitly, note first
that the linear term in this expansion shows that one has to apply a
normalisation factor $\braket{\eta(\eta q_i)}$ for each external $B_0$
leg in order to obtain the correctly normalised $A_{z}$ correlation
functions.  Furthermore the expression for the EM-coefficients
\eqref{eq:em-coeff-bbar} assumes that the external wave-function
normalisation is trivial.  Therefore through inserting the correct pre-factor
for the polarisation factors \eqref{eq:polarization} the field $B_0$
in the twistor approach is related to the field $B$ in the canonical
approach by a normalisation factor
\begin{equation}
\label{eq:b0tob}
  B(q_i) \Rightarrow  \frac{ i}{\sqrt{2}}
\frac{B_0(q_i)}{\braket{\eta(\eta q_i)}(\epsilon^+(q_i))_z } =
  - i \frac{B_0(q_i)}{\sbraket{\eta i}^2 } 
\end{equation}
where the spinor products in the last term are 
continued  off-shell using~\eqref{eq:spinormomentumtrick}.
Indeed,  inserting the external polarisation factor of $A_z$
into~\eqref{eq:em-twistor} precisely reproduces the
expression \eqref{eq:em-coeff-bbar} for the EM-coefficients for
the correctly normalised fields:
\begin{align}
(\epsilon^+(p))_z^{-1}  A_z(p) = &
\frac{-i B_0(q_1)}{\sbraket{\eta 1}^2} +
\int_{q_1,q_2} (\sqrt{2} g) \frac{\braket{\eta p}^2 }{\braket{\eta 1}\braket{1 2}\braket{2 \eta}} \frac{(-i B_0(q_1))}{\sbraket{\eta 1}^2} \frac{ (-iB_0(q_2))}{\sbraket{\eta 2}^2} \nonumber \\
&  +\int_{q_1,q_2,q_3} (\sqrt{2} g)^2 \frac{\braket{\eta p}^2 }{\braket{\eta 1} \braket{1 2} \braket{2 3} \braket{3 \eta}} \frac{(-iB_0(q_1))}{\sbraket{\eta 1}^2} \frac{(-iB_0(q_2))}{\sbraket{\eta 2}^2} \frac{(-iB_0(q_3))}{\sbraket{\eta 3}^2} + \ldots
\end{align}
Note that the denominator of that expression simply follows here from the general structure of the lifting formula~\eqref{eq:liftforA} in terms of link operators. Furthermore, note that all the momenta in the above expression do not depend on the momentum-component $\hat{\eta}^{\alpha} \hat{\eta}^{\dalpha} p_{\alpha \dalpha}$, so the momentum integrals along this component of the momenta can all be performed trivially. This explains the difference between the $4$ dimensional momentum integrals utilised  above and the $3$ dimensional momentum integrations natural in lightcone Yang-Mills. 

The other component can be treated using similar techniques, as we can write 
\begin{equation}\label{eq:twistorforbarB}
i (\eta \eta p) A_{\bar{z}}(p) = - \frac{1}{\sqrt{2}} (\hat{\eta}^{\alpha} \eta_\alpha) \int_{\pi_k}\dk\,  \frac{H(\eta) H^{-1}(\pi_k)}{\braket{\eta \pi_k}} \bar{B}_0 (\pi_k) (\braket{\pi_k \eta})^4  \frac{H(\pi_k) H^{-1}(\eta)}{\braket{\pi_k \eta}} 
\end{equation}
Expanding and evaluating will yield the same coefficients as found in the canonical transformation case. The field $\bar{B}_0$ will appear in the combination
\begin{equation}
\sim i \sbraket{\eta 1}^2 \bar{B}_0 (q_1)
\end{equation}
which again is a consequence of the polarisation vector for $\bar{B}_0$. In more detail, expanding the above equation yields,
\begin{multline}
(\epsilon^-(p))_{\bar z}^{-1} (\eta \eta p) A_{\bar{z}}(p)  = i \frac{\sbraket{\eta p}}{\braket{\eta p}} \left( \bar{B}_1 \sbraket{\eta 1}^2 \braket{\eta 1}^2 \right.\\
\left. - i \sqrt{2 g}\int_{q_1,q_2} \left(
\frac{\braket{\eta 1}^3}{ \braket{1 2} \braket{2 \eta} }\frac{\sbraket{\eta 1}^2 \bar{B}_1 B_2}{\sbraket{\eta 2}^2} +
 \frac{\braket{\eta 2}^3}{ \braket{\eta 1} \braket{1 2}  }
\frac{B_1 \bar{B}_2 \sbraket{\eta 2}^2 }{\sbraket{\eta 1}^2}
\right)
 + \ldots \right)
\end{multline}
The difference to ~\eqref{eq:bbar-trafo} is fully contained in the convention of \eqref{eq:braket-eta}. In particular, this makes the prefactor in the above equation disappear. The other change needed to see the equivalence is 
\begin{equation}
\braket{\eta p_{\bar{B}}}^2 \rightarrow (k_s)_+  
\end{equation}

Note that in order to derive equation
\ref{eq:twistorforbarB}, the field equation for $\bar{B}_0$ was used, which will be important later.
We like to stress that above we have used the fact that $B_0$ and
$\bar{B}_0$ are always accompanied by a $\delta$ function on the
fibres in this gauge and the above identification should be understood
in that way. By the above calculation, in this sense it is shown that the canonical
transformation coefficients can be derived from the twistor lifting
formulae.

In our view the above calculation clearly identifies the non-local and
non-linear canonical transformation employed in the 'canonical'
approach to MHV diagrams with a local and linear gauge transformation
on twistor space. In the following we might sometimes illustrate a
point in one formalism, but by the above argument the same point could
at least in principle be made in the other. Both have advantages and
disadvantages. The canonical approach is calculational more intense
and does not easily yield the full MHV Lagrangian, but is conceptually
transparent. The twistor action approach requires some mathematical
insight and requires one to choose the 'right' first order space-time
action to lift\footnote{Applying lifting formulae to $F^2$ directly
  for instance does not immediately lead to nice results!}, but makes
the underlying symmetry and geometry transparent and as such
immediately yields complete Lagrangians. These advantages and
disadvantages manifest themselves for instance in the range of
applicability of both methods: the canonical method has been applied
to Einstein gravity with some success for the four point interaction~\cite{Ananth:2007zy},
where the same has not been done yet through a twistor action. In
contrast, the twistor action has already yielded the complete Nair
super-vertex~\cite{Boels:2006ir}, where this has only been derived up to four point
vertices~\cite{Feng:2006yy} in the canonical approach.

\subsection{Equivalence theorem violations}
\label{sec:violate}
The results discussed in the section so far leave two questions
open. 
The first is the three-point googly-MHV vertex
\begin{equation}
\label{eq:googly-three}
  V(A_{k_1}(+),A_{k_2}(+),A_{k_3}(-))=
i\sqrt 2 \frac{\sbraket{21}^3}{\sbraket{32}\sbraket{13}}
 = i\sqrt 2 \sbraket{21}
 \frac{\braket{3\eta}^2}{\braket{1\eta}\braket{2\eta}}
\end{equation}
that has been transformed away through the canonical transformation 
or the choice of CSW gauge in the twistor Yang-Mills action. 
This vertex controls a vanishing amplitude on-shell for real momenta in Minkowski signature but this amplitude is
non-vanishing for complex momenta or for Euclidean
signature~\cite{Witten:2003nn}.
The other missing pieces in pure Yang-Mills are the rational parts
of one-loop amplitudes, including the purely rational amplitudes
with less than two negative helicity gluons.
One resolution noted in the literature requires to relate carefully the correlation functions of the original fields $(A_z,A_{\bar z})$ to that of the new fields $(B ,\bar B)$~\cite{Ettle:2007qc}.
Here we will review the argument for the recovery of the 
 googly three particle MHV amplitude.

Using the original field variables, the vertex~\eqref{eq:googly-three} is obtained by the LSZ formula from a correlation function of $A$ fields as
\begin{equation}
\mathcal{A}_3(A_{k_1}(+),A_{k_2}(+),A_{k_3}(-))
=\prod_{i=1}^3\lim_{k_i^2\to 0}(-ik_i^2)
\braket{0|(\epsilon^+_{k_1}\cdot A_{k_1})
  (\epsilon^+_{k_2}\cdot A_{k_2})(\epsilon^-_{k_3}\cdot A_{k_3})|0}
\end{equation}
Evaluating the right-hand side in the light-cone gauge using
$(\epsilon^\pm\cdot A)=-(\epsilon^\pm)_{z/\bar z}A_{\bar z/z}$  leads to
consider the Green's function $
\braket{0|A_{k_1,\bar
      z}A_{k_2 \bar z}A_{k_3 z}|0}$.
Note that this still contains propagators connecting $A_z$ and $A_{\bar z}$ fields.
To derive the same amplitude from the MHV Lagrangian one
considers the Green's function as a function of the new fields
by inserting
 the field-redefinitions~\eqref{eq:b-trafo} and~\eqref{eq:bbar-trafo} 
or equivalently the relations to the twistor-fields~\eqref{eq:firsttrans} 
and~\eqref{eq:secondtrans}:
\begin{multline}
\label{eq:gf-trafo}
\braket{ 0|A_{p_1, \bar
      z}A_{p_2,\bar z}A_{p_3,z}|0}
=\Bigl\langle 0|\bar B_{p_1} \bar B_{p_2} B_{p_3}
+\int \widetilde{d k_1}\widetilde d{k_2}
\left[
 \left(\tfrac{k_{1+}}{p_{1+}}
\mathcal{X}^1(p_1,k_1,k_2) \bar B_{-k_1 } B_{-k_2}\right) 
\bar B_{p_2} B_{p_3}
 \right. \\
\left.+\bar B_{p_1}
\left(\tfrac{k_{2+}}{p_{2+}}
\mathcal{X}^2(p_2,k_1,k_2) B_{-k_1 }\bar B_{-k_2}\right) B_{p_3}
+\bar B_{p_1} \bar B_{p_2}
\left(\mathcal{Y}(p_3,k_1,k_2) B_{-k_1 }B_{-k_2}\right)
\right]+\dots|0\Bigr\rangle
\end{multline}
where two more contributions in the expansion of the $A_{\bar z}$ fields
have not been shown.
 Since there is no vertex for a single $\bar B$
field in the MHV Lagrangian, the first term  vanishes.
 Usually the non-linear terms in the field redefinitions are argued not to contribute to the on-shell scattering amplitudes after the LSZ reduction since higher order expansions at a given site will generically lead to multi-particle poles and not to the single particle poles required for non-trivial LSZ contributions. This is the content of the so-called 'equivalence theorem' (see e.g. the discussion in \cite{'tHooft:1973pz}). 
In~\cite{Ettle:2007qc} it was shown that even at tree level one needs to be careful in applying the theorem: the non-local nature of the canonical transformation violates one of the assumptions, and summing seemingly vanishing contributions can lead to non-vanishing amplitudes. 

 A quick (admittedly slightly non-rigorous) way of finding the missing $3$ point amplitude is to take asymmetric on-shell limits in the LSZ procedure.
Take for instance the LSZ reduction (and the on-shell limit) on site $1$ and $2$, leaving site 3 off-shell. Then only the last contribution will survive.
Now note that for $p_1^2=p_2^2=0$
the three-point EM-coefficients are simply proportional to 
the  three-point googly-MHV vertex:
\begin{equation}
\label{eq:y-lsz}
  \mathcal{Y}(p_3,p_1,p_2)= g \sqrt 2 \frac{\braket{\eta 3}^2}{
\braket{\eta 1}\braket{12} \braket{2\eta}}
= g \frac{i}{\braket{12}\sbraket{21}}
   V(A_{p_3,\bar z},A_{p_1,z},A_{p_2,z})
\end{equation}
Since legs $1$ and $2$ are on-shell, the denominator becomes
 $\braket{12}\sbraket{21}=2(p_1\cdot p_2)=p_3^2$ so despite appearances the last term in~\eqref{eq:gf-trafo}
\emph{has} a pole in $p_3^2$. Performing now the LSZ reduction
on this leg gives a non-vanishing result and leads to the correct three-point 
amplitude. Once can check that the same result is reached by a different
order of these asymmetrical LSZ-limits.

To see why these asymmetric LSZ-limits work, consider the case that
all particles are off-shell. Using~\eqref{eq:spinormomentumtrick} 
one can express the denominator in~\eqref{eq:y-lsz} as
\begin{equation}
\label{eq:simplify-ij}
\begin{aligned}
\braket{12}\sbraket{21}
&=\frac{\braket{\eta+|\fmslash p_1\fmslash p_2 \fmslash p_1|\eta+}}{
2(p_1\cdot\eta)}
-2 p_2^2 \frac{(p_1\cdot \eta)}{(p_2\cdot \eta)}\\
&=-\frac{1}{(p_1\cdot\eta)(\eta\cdot p_2)}\left[p_1^2 (\eta\cdot p_2)^2
+ p_2^2 (p_1\cdot \eta)^2
-2 (p_1\cdot p_2)(\eta \cdot p_1)(\eta\cdot p_2)\right]
\end{aligned}
\end{equation}
Applying~\eqref{eq:simplify-ij} e.g. to the denominator in~\eqref{eq:y-lsz} one sees that there is no pole in $p_3^2$ as long as $p_1$ and $p_2$ are not \emph{both} on-shell so this term doesn't contribute if the LSZ reduction is performed for, say, legs $2$ and $3$ first. However, as demonstrated in~\cite{Ettle:2007qc}, if all legs are initially kept off-shell the sum of the three terms in~\eqref{eq:gf-trafo} survives the LSZ reduction although the separate terms don't contain three single-particle poles explicitly.

Note that only for the three-point case the right hand side of \eqref{eq:simplify-ij} can be expressed entirely in terms of the squares of the external momenta and  a pole in $p_3^2$ can develop for $p_1^2=p_2^2\to 0$. This indicates that, at tree level, 'equivalence theorem evasion' is only expected to play a role in the three particle case. In this case the two legs connected to the 'vertex' must be on-shell (and null) to give a contributing pole.  This is analogous to the non-vanishing of the three-point non-MHV vertex for complex kinematics~\cite{Witten:2003nn,Britto:2004ap} while the one-particle off-shell currents with one negative helicity gluon and more than two positive helicity ones~\cite{Berends:1987me,Dixon:1996wi} vanish upon LSZ reduction, also for complex kinematics.

\section{CSW rules for massive scalars}
\label{sec:scalars}
\subsection{Summary of the rules}
\label{sec:scalar-rules}

In this section we summarise the CSW rules for scalar matter coupled to glue~\cite{Boels:2007pj} and present further examples of their applications. These rules hold for a scalar  $\phi$ in the fundamental representation of the gauge group described by the space-time Lagrangian
\begin{equation}
\label{eq:phi-lag}
\mathcal{L}=
\mathcal{L}(A)+
\mathcal{L}_\phi
 =\mathcal{L}(A)+ (D_\mu\phi)^\dagger D^\mu\phi-m^2\phi^\dagger \phi
\end{equation}
where $D_\mu=\partial_\mu-ig A$. Some further explicit examples of the application of the CSW rules derived from this Lagrangian at tree level are given in~\ref{sec:tree}. In sections ~\ref{sec:scalar-mansfield} and~\ref{sec:scalar-twistor} we offer two detailed derivations of CSW-like rules starting from the
Lagrangian~\eqref{eq:phi-lag}, based on the canonical transformation in
the light-cone gauge and on the twistor Yang-Mills approach,
respectively.

In both approaches used to derive the CSW rules a field-redefinition $\phi\to \xi$ of the
scalars is performed that "dresses" them with an infinite number of positive
helicity gluons and eliminates the non-MHV coupling of two scalars and a
positive  helicity gluon. The key idea in both derivations is to
perform the \emph{same} transformation of the massive scalars as in
the massless case. In this way we arrive at a formalism that includes roughly the same vertices as the CSW formalism for massless scalars~\cite{Georgiou:2004wu,Georgiou:2004by} derived by supersymmetry arguments:
\begin{align}
\label{eq:csw-glue}
  V_{\text{CSW}}(\bar B_1,B_2,\dots \bar B_{i},\dots, B_n)&=
i2^{n/2-1}  \frac{ \braket{1i}^4}{
  \braket{12}\dots\braket{(n-1)n}\braket{n1}}\\
\label{eq:csw-2phi}
  V_{\text{CSW}}(\bar\xi_1,B_2,\dots \bar B_{i},\dots \xi_n)&=
-i2^{n/2-1}  \frac{ \braket{in}^2\braket{1i}^2}{
  \braket{12}\dots\braket{(n-1)n}\braket{n1}} \\
  V_{\text{CSW}}( \bar\xi_1,B_2,\dots \xi_{i},\bar\xi_{i+1}\dots, \xi_n)&=
-i2^{n/2-2}  \frac{
\braket{1i}^2\braket{(i+1)n}^2}{
  \braket{12}\dots\braket{n1}}
 \left(1+\frac{ \braket{1(i+1)}\braket{in}}{\braket{1i}\braket{(i+1)n}}\right)
 \label{eq:csw-4phi}
\end{align}
and an additional tower of vertices with a pair of scalars and an arbitrary
number of positive helicity gluons that is
generated from the transformation of the mass term:
 \begin{equation}
\label{eq:csw-mass}
V_{\text{CSW}}(\bar \xi_1,B_2,\dots, \xi_n)= i 2^{n/2-1}
\frac{- m^2 \braket{1n}}{
\braket{12}\dots\braket{(n-1)n }} 
\end{equation}
These vertices are connected by the usual massive scalar propagator
\begin{equation}
D_{\bar\xi\xi}(p^2)= \frac{i}{p^2-m^2}
\end{equation}
and scalar propagators $i/p^2$ connecting positive and negative-helicity
gluons. Spinors corresponding to off-shell gluons and both on-shell and off-shell
scalars are understood as usual in the CSW rules~\cite{Cachazo:2004kj} and are obtained from the momentum by contraction with an arbitrary but fixed anti-holomorphic spinor $\eta^{a}$:
\begin{equation}
\label{eq:csw-continue}
k_{\dot \alpha}=k_{\dot\alpha\alpha}\eta^{\alpha}
\end{equation}
When these quantities appear in spinor products we use the equivalent notations
$\braket{pk}=\braket{p-|\fmslash k|\eta-}=\braket{p(k\eta)}$.  

The difference of the above rules with those in~\cite{Georgiou:2004wu,Georgiou:2004by} is that their formalism applies to a scalar space-time action which include a $\phi^4$ space-time vertex. This coupling is required by the supersymmetry used there to derive the formalism. In contrast, our rules are those for a minimally coupled scalar.
For a scalar in the fundamental representation there are also subleading color
structures for amplitudes with four or more external scalars
that will not be considered explicitly in this paper.

The twistor-space structure of massive amplitudes resulting from the 
above rules was discussed in~\cite{Boels:2007pj} where it was also
verified that the amplitudes with a pair of massive scalars
and only positive helicity gluons derived from the CSW rules satisfy the required on-shell recursion
relation~\cite{Forde:2005ue, Ferrario:2006np}.
The rules were also used to simplify the proof of the BCFW recursion relations for amplitudes with massive
scalars~\cite{Badger:2005zh}. Since the extra vertices are proportional to the mass, the holomorphic
representation also provide a simple method to obtain the leading part of massive amplitudes in the small-mass limit~\cite{Boels:2007pj}.

 One could also contemplate a
transformation that transforms the mass term into a quadratic term of
the $\xi$ fields instead of the tower of vertices~\eqref{eq:csw-mass}
 and therefore removes the non-MHV type couplings completely.
However, this apparently also removes the amplitudes with only
positive helicity gluons. Since they are known to be non-vanishing, for such 
a transformation they are
likely to be generated by the ``equivalence theorem violation''
mechanism~\cite{Ettle:2007qc} reviewed in section~\ref{sec:violate}. Here we do not attempt to
find such a transformation.
\subsection{Tree level examples}
\label{sec:tree}

As input for higher-point amplitudes, we note that inserting the prescription~\eqref{eq:csw-continue} and using momentum conservation, the cubic mass-vertex~\eqref{eq:csw-mass} can be written as
\begin{equation}
\label{eq:csw-cubic}
V(\bar \xi_1,B_2,\xi_3)= -\sqrt 2im^2
 \frac{  \braket{13}}{\braket{12}\braket{23}}
= \frac{\sqrt 2 im^2  \sbraket{2\eta}}{ \braket{2-|\fmslash k_3|\eta-}}
\end{equation}
where momentum conservation was used.  This expression can be shown to
be equivalent to the corresponding vertex resulting from the original
action~\eqref{eq:phi-lag} provided all particles are on-shell~\cite{Boels:2007pj}.  
In order to simplify  calculations, it is useful to note that this vertex vanishes if the spinor $\ket{\eta
  -}$ is chosen in terms of the momentum of the gluon entering the
vertex (this is not possible for the on-shell  three particle amplitude~\cite{Boels:2007pj}).

In~\cite{Boels:2007pj} the CSW rules for massive scalars have been verified
for three- and four-point scattering amplitudes. As a more stringent check, 
in this paper we consider some five point-functions.  As shown in figure~\ref{fig:5point}, there are four diagrams contributing to the five point function with only positive helicity gluons.
\FIGURE[t]{
  \includegraphics[width=0.9\textwidth]{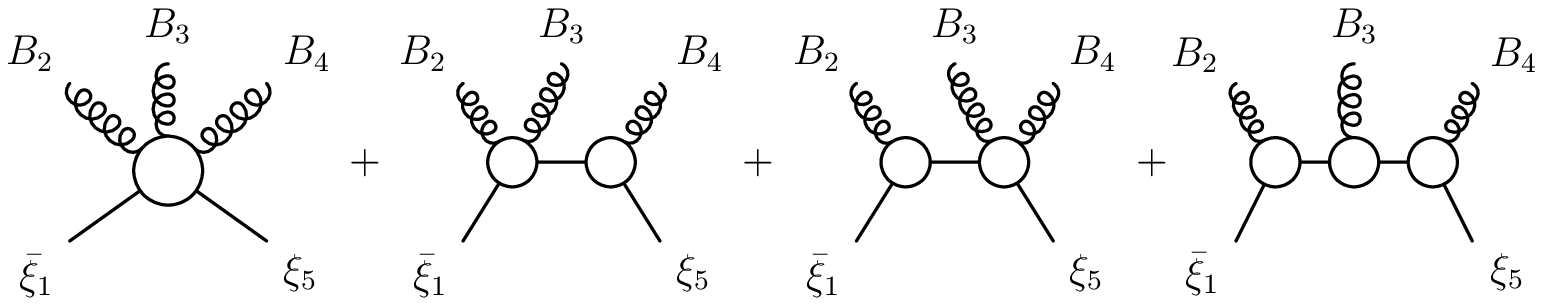}
  \caption{CSW-diagrams for the five-point amplitude with
   positive helicity gluons}
  \label{fig:5point}
}
 Setting 
$\ket{\eta-}=\ket{2-}$ eliminates the last two diagrams that contain
$g_{2}^+$ entering a three point vertex.

Using the prescription~\eqref{eq:csw-continue} 
 for both the internal off-shell momenta and  the external scalars, 
 the remaining two diagrams
give (using the notation $y_{i,j}=k_{i,j}^2-m^2$)
\begin{equation}
\begin{aligned}
  A_5( \bar \xi_1^+, B_2, B_{3}, B_4,\xi_{5})&=
\frac{-2^{3/2}im^2\braket{15}}{\braket{12}\braket{23}\braket{34}\braket{45}}
+\frac{-2im^2\braket{1k_{1,3}}}{\braket{12}\braket{23}\braket{3k_{1,3}}}
  \frac{i}{k_{1,3}^2-m^2} 
\frac{-\sqrt 2 im^2\braket{k_{1,3}5}}{
\braket{k_{1,3}4}\braket{45}}\\
&=\frac{-2^{3/2}im^2
(y_{1,3}\braket{2+|\fmslash k_1\fmslash k_5|2-}
+m^2\braket{2+|\fmslash k_3\fmslash k_4|2-})}
{y_{1,2}y_{1,3}
\braket{23}\braket{34}\braket{4-|\fmslash k_5|2-}}
\end{aligned}
\end{equation}
This can be shown to be equivalent to the known result~\cite{Bern:1996ja,Badger:2005zh,Forde:2005ue}.

The structure of the amplitudes with one negative helicity gluon is more
involved. Consider the amplitude $ A_5( \bar \xi_1, \bar B_2, B_{3}, B_4, \xi_{5})$. One class of diagrams contributing to the five point amplitude is similar to figure~\ref{fig:5point}  but with the vertex coupled to $\bar B_2$ replaced by a massless MHV vertex. Setting  $\ket{\eta+}=\ket{4+}$ kills the second and the last diagram. In addition there are two diagrams with an internal gluon line connecting the scalar mass-vertex with a gluonic MHV vertex (see figure~\ref{fig:mhv5})
One gets
\begin{multline}
  A_5( \bar \xi_1, \bar B_2, B_{3}, B_4, \xi_{5})
=\frac{2^{3/2}i\braket{12}^2\braket{25}^2}{
  \braket{12}\braket{23}\braket{34}\braket{45}\braket{51}}
+\frac{ 2^{3/2} im^2 \braket{12}\braket{2k_{1,2}}\braket{k_{1,2}5}}{\braket{1k_{1,2}}
(k_{1,2}^2-m^2)\braket{k_{1,2}3}\braket{34}\braket{45}}\\
+\frac{2^{3/2}im^2 \braket{15} \braket{(-k_{2,4})2}^3}{
\braket{1 k_{2,4}}\braket{k_{2,4}5}k_{2,4}^2
\braket{23}\braket{34}\braket{4(-k_{2,4})}}
+\frac{2^{3/2} im^2 \braket{15}\braket{(-k_{2,3})2}^3 }{\braket{1k_{2,3}}\braket{k_{2,3}4}\braket{45}k_{2,3}^2
\braket{23}\braket{3(-k_{2,3})}}
\end{multline}
Making the choice $\ket{\eta+}=\ket{4+}$ explicit this becomes
\begin{multline}
 A_5( \bar \xi_1, \bar B_2, B_{3}, B_4, \xi_{5})
=\frac{2^{3/2}i\braket{4+|\fmslash k_1|2+}\braket{2-|\fmslash k_5|4-}^2}{
 y_{4,5}\braket{23}\braket{34}
\braket{4+|\fmslash k_5\fmslash k_1|4-}}+
\frac{ 2^{3/2} im^2
\braket{4+|\fmslash k_1|2+}\sbraket{43}}
{y_{1,2}y_{4,5}\sbraket{24}\braket{34}}\\
-\frac{2^{3/2}im^2\sbraket{34}^3\braket{23}^2 }{
\braket{4+|\fmslash k_{1}\fmslash k_5|4-}k_{2,4}^2
\braket{34}\braket{4-|\fmslash k_{2,3}|4-}}
+\frac{2^{3/2}im^2 \sbraket{34}^3}{\braket{4+|\fmslash k_{2,3}|4+}
y_{4,5}\sbraket{32}\sbraket{42}}
\end{multline}
This expression agrees numerically with the known result~\cite{Badger:2005zh,Forde:2005ue}.
\FIGURE[t]{
  \includegraphics[width=0.9\textwidth]{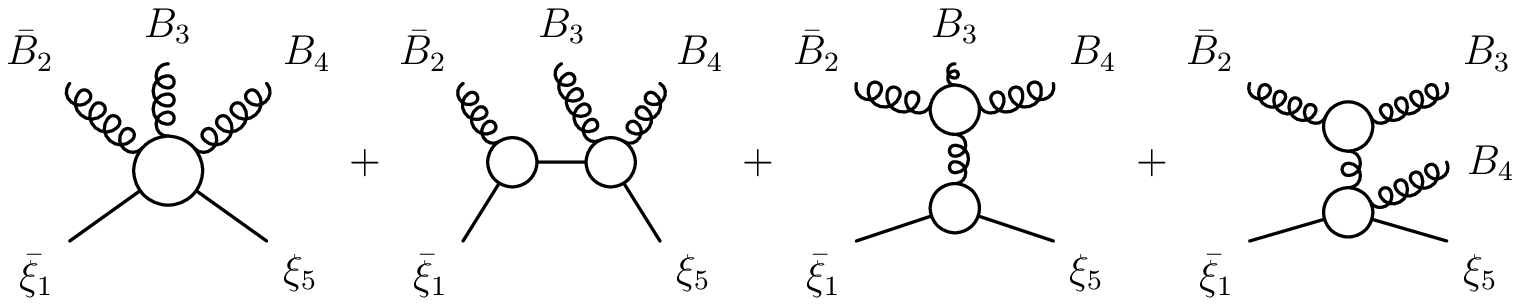}
  \caption{Non-vanishing CSW-diagrams for the five-point amplitude with
   one negative helicity gluon for the off-shell continuation $\ket{\eta-}=\ket{4-}$}
  \label{fig:mhv5}
}
Although yielding less compact results than BCFW recursion relations,
the calculation is still much simpler than the one using conventional
Feynman rules, since no more combinatorics associated to the gluon-vertices has to be performed.

\subsection{Light-cone derivation of CSW rules for massive scalars}
\label{sec:scalar-mansfield}

In this subsection we derive the CSW rules for massive scalars in the
framework of canonical transformations in the light-cone gauge.  We
obtain the light-cone gauge Lagrangian for a scalar and derive the
field redefinitions.  Since, as emphasised before, the same field
redefinition as for massless scalars is used, the structure of the
transformation is similar to that for massless quarks proposed
in~\cite{Mansfield:2005yd}\footnote{{\textbf{Note added in proof:}} The transformation for quarks has been worked out explicitly very recently in~\cite{Ettle:2008ey} which appeared while this article was in the final stages of preparation.}. 
 The explicit form of the expansion
coefficients is related in a simple way to that for
gluons~\cite{Ettle:2006bw} as suggested by supersymmetry~\cite{Feng:2006yy}.  These results are then used to
derive the vertex~\eqref{eq:csw-mass} arising from the transformation
of the mass term. The origin of the difference of the four-scalar
vertex~\eqref{eq:csw-4phi} to the results in the literature obtained
from supersymmetry~\cite{Georgiou:2004wu,Georgiou:2004by} is also
discussed.

\subsubsection*{Light-cone Yang-Mills with a massive scalar}

In the first step of the derivation of the canonical transformation,
the sum of the Yang-Mills Lagrangian and the Lagrangian of the massive
scalars~\eqref{eq:phi-lag} is expressed entirely in terms of the
physical gluon components ($A_{z}$/$A_{\bar z}$) and the scalars.  An
analogous discussion for the light-cone gauge Lagrangian of a scalar
in the adjoint representation can be found
in~\cite{Brandhuber:2006bf}.  Imposing the light-cone gauge condition
$A_+=0$ on the gluon, the scalar Lagrangian~\eqref{eq:phi-lag} can be
expressed in terms of the surviving light-cone components $A_-$ and
$A_{z/\bar z}$.  Making the group indices of the vector potential and
the scalars explicit, the result is
\begin{equation}
\label{eq:scalar-light-cone}
\mathcal{L}_\phi =
-\phi^\dagger_i(\square+m^2) \phi_i
+ig \left[ A_{-,ij}(\phi^\dagger_i\overleftrightarrow{\partial_+}\phi_j)
-A_{\perp,ij}(\phi^\dagger_i\overleftrightarrow {\partial_\perp} \phi_j)\right]
+
(ig)^2(\phi^\dagger A_\perp)_i (A_{\perp}\phi)_i
\end{equation}
Here the abbreviation $\phi\overleftrightarrow\partial\psi=
\phi(\partial\psi)-(\partial\phi)\psi$ and the scalar product of the
transverse components $A_\perp B_\perp\equiv A_zB_{\bar z}+A_{\bar
  z}B_z$ have been introduced.  As in the purely gluonic case, the
component $A_-$ appears only quadratically in the Lagrangian so it can
be eliminated using the equation of motion. However, it now receives
an additional contribution from the scalars:
\begin{equation}
  A_{-,ij}=\frac{1}{\partial_+^2}
\left([D_\perp,\partial_+A_{\perp}]_{ij}
+i\frac{g}{2}\left[
(\phi_j^\dagger \overleftrightarrow {\partial_+}\phi_i)
-\frac{1}{N}\delta_{ij}(\phi_k^\dagger \overleftrightarrow {\partial_+}\phi_k)
\right]\right)
\end{equation}
To derive this relation the identity
\begin{equation}
  \frac{\delta A_{ij}(x)}{\delta A_{kl}(y)}=\frac{1}{2}
\left(\delta_{il}\delta_{jk}-\frac{1}{N}\delta_{ij}\delta_{kl}\right)
\delta^4(x-y)
\end{equation}
has been used.
Inserting the solution of $A_-$ back
 into the scalar Lagrangian~\eqref{eq:scalar-light-cone}
and the gluon Lagrangian, 
one obtains the interactions of the scalars with 
the  physical modes of the gluons and a quartic self-coupling induced
by the scalar contribution to $A_-$~\cite{Brandhuber:2006bf}:
\begin{equation}
 \mathcal{L}^{(2)}_{\phi\phi}+ 
\mathcal{L}^{(3)}_{\phi A_z\phi}
+\mathcal{L}^{(3)}_{\phi A_{\bar z}\phi}
+\mathcal{L}^{(4)}_{\phi A_zA_{\bar z}\phi}
+\mathcal{L}^{(4)}_{\phi\phi\phi\phi}
\end{equation}
The cubic interactions can be written as
\begin{align}
  \mathcal{L}^{(3)}_{\phi A_z\phi}&=
2ig \left[(\partial_{\bar z}\phi^\dagger) A_z \phi 
- (\partial_+\phi^\dagger)\left(\frac{\partial_{\bar z}}{\partial_{+}} 
  A_{z}\right)\phi\right]\\
\mathcal{L}^{(3)}_{\phi A_{\bar z}\phi}&=
2ig \left[(\partial_{ z}\phi^\dagger) A_{\bar z} \phi 
- (\partial_+\phi^\dagger)\left(\frac{\partial_{z}}{\partial_{+}} 
  A_{\bar z}\right)\phi\right]
\end{align}
Going to momentum space and translating to spinor notation 
 one obtains the familiar
 vertices in the helicity formalism~\cite{Badger:2005zh}, for instance
the coupling
to a positive helicity gluon:
\begin{equation}
\label{eq:bgks}
V(\phi^\dagger_1, A_{z,2},\phi_3)
=(2 i)(\epsilon(k_2)^+)_z
 \frac{p_{1+}p_{2{\bar z}}-p_{1{\bar z}}p_{2+}}{p_{2+}}
=\sqrt 2 i\frac{\braket{2+|\fmslash k_1|\eta+}}{\braket{\eta 2}}
\end{equation}
For completeness we also give the quartic interactions
for a scalar in the fundamental representation
\begin{align}
\mathcal{L}^{(4)}_{\phi A_zA_{\bar z}\phi}=&(ig)^2\left[
(\phi^\dagger A_\perp)_i (A_{\perp}\phi)_i +
\left([\partial_+A_{\perp},A_\perp]_{ij}\right)(\partial_+)^{-2}
(\phi^\dagger_j\overleftrightarrow{\partial_+}\phi_i)\right]\\
\mathcal{L}^{(4)}_{\phi \phi\phi\phi}=&\left(\frac{ig}{2}\right)^2
 (\phi^\dagger_i\overleftrightarrow{\partial_+}\phi_j)(\partial_+)^{-2}
 (\phi^\dagger_k\overleftrightarrow{\partial_+}\phi_l)
\left(\delta_{il}\delta_{jk}-\frac{1}{N}\delta_{ij}\delta_{kl}\right)
\end{align}

\subsubsection*{Field redefinition for scalars}

In the second step, the non-MHV vertices contained in the
gluon and the scalar Lagrangian are eliminated by an appropriate
canonical transformation.
As in the case of the pure Yang-Mills
Lagrangian reviewed in section~\ref{sec:mansfield},  a canonical  transformation to new gluon variables
 $B$ and momenta
$\partial_+\bar B$ can be used in order to eliminate the cubic 
gluonic non-MHV coupling.
In addition, there is now the non-MHV type 
scalar interaction $\mathcal{L}^{(3)}_{\phi A_z\phi}$
that can be eliminated by transforming also 
to new scalar field variables and momenta 
$\xi$ and $\Pi_\xi\sim\partial_+\bar\xi$.

Following closely the
proposal for the incorporation of massless quarks
in~\cite{Mansfield:2005yd}, 
 the transformation of the positive helicity gluon is taken independent
of the scalars
so that  the generating functional
  of the transformation is of the form
\begin{equation}
  G[A_z,\phi,\Pi_B,\Pi_\xi]=\int d^3 y\Bigl( B[A_z(\vec y)]\Pi_B(\vec y)
    +\xi[\phi(\vec y),A_z(\vec y)] \Pi_\xi(\vec y)\Bigr)
\end{equation}
A common dependence on the fixed 'time' $x^-$ has been suppressed.
The transformation of the scalars are induced via the relations
$\xi = \delta G/\delta \Pi_\xi$ and $\Pi_\phi=\delta G/\delta \phi$:
\begin{equation}
\label{eq:phi-canonical}
  \phi\to \xi[\phi,A_z]\qquad 
\Pi_{\phi}\sim \partial_+\phi^\dagger(x^-,\vec x)
=\int d^3 y\; \partial_+ \bar \xi(x^-,\vec y)\frac{\delta \xi (x^-,\vec y)}{\delta \phi(x^-,\vec x)} 
\end{equation}
Since $\xi$ depends on the $A_z$ there is  also a contribution from the scalars
to the canonical momentum $\partial_+ A_{\bar z}\sim \delta G/\delta A_z$, in addition to~\eqref{eq:momentum-trafo}:
\begin{equation}
\label{eq:d-bara}
 \partial_+ A_{\bar z}(x^-,\vec x)
=\int d^3 y \left(\frac{\delta B(x^-,\vec y)}{\delta A_z(x^-,\vec x)} 
\partial_+ \bar B(x^-,\vec y)
+\frac{1}{2}\partial_+ \bar \xi(x^-,\vec y)
\frac{\delta \xi(x^-,\vec y)}{\delta A_z(x^-,\vec x)} 
\right)
\end{equation}
The second term ensures that the sum
of the kinetic terms is invariant, i.e. 
\begin{equation}
\int d^3 x (2\partial_+\bar
B\partial_-B + \partial_+\bar\xi\partial_-\xi)=\int
d^3x(2\partial_+A_{\bar
  z}\partial_-A_{z}+\partial_+\phi^\dagger\partial_-\phi)
\end{equation}
 since $\xi$
depends on $x^-$  implicitly through both $\phi$ and $A_z$ so that 
\begin{equation}
\partial_-
\xi(x^-,\vec x)=\int d^3 y\left[ 
\frac{\delta\xi(\vec x)}{\delta\phi (\vec y)} \partial_-\phi(x^-\vec y)
+\frac{\delta\xi(\vec x)}{\delta A_z(\vec y)} \partial_-A_z(x^-,\vec y)\right]
\end{equation}  
 The
$\xi$-dependent contribution to the conjugate gluon-momentum
$\partial_+A_{\bar z}$  will result in  an additional term
in the condition of the gluonic transformation~\eqref{eq:mansfield-def} 
so the transformations of
the scalars and the gluons cannot be treated separately.  Therefore
the condition to eliminate the non-MHV type vertices is
\begin{equation}
\label{eq:transform-Lphi}
\int d^ 3 x \; \left[4A_{\bar z} 
\partial_z\partial_{\bar z}  A_z
+2\phi^\dagger \partial_z\partial_{\bar z} \phi
+\mathcal{L}^{(3)}_{++-}
+\mathcal{L}^{(3)}_{\phi A_z\phi}\right]
=\int d^3 x\;\left[ 4\bar B\partial_z\partial_{\bar z}B
+2 \bar\xi \partial_z\partial_{\bar z} \xi\right]
\end{equation}
Since the mass term of the scalar has not been included in this
definition, the transformation has
the same form for massive and massless scalars.
Consequently the mass term $m^2\phi^\dagger\phi$ will not be invariant
but instead be transformed
into a tower of vertices $m^2\bar \xi B\dots B\xi$ resulting in the
CSW vertices~\eqref{eq:csw-mass}. 

The conditions~\eqref{eq:transform-Lphi} and~\eqref{eq:phi-canonical}
can be solved by methods similar to the ones used in pure Yang-Mills theory
in~\cite{Ettle:2006bw}. 
The transformation of the positive helicity gluons
$A_z[B]$ will be taken identical to  the pure Yang-Mills case defined
by~\eqref{eq:mansfield-def}.  Inserting the expressions for
the canonical momenta~\eqref{eq:phi-canonical} and~\eqref{eq:d-bara}
into~\eqref{eq:transform-Lphi} 
 the vanishing of the coefficient
of  $\partial_+\bar\xi(\vec x)$ leads to the condition 
\begin{multline}
\label{eq:xi-cond-1}
\frac{\partial_z\partial_{\bar z}}{\partial_+}\xi(\vec x)
+\int d^3 y d^3 w
\tr\left[ 
 \left(\frac{\partial_{\bar z}\partial_z}{\partial_+}B(\vec w)\right)
\frac{\delta \xi(\vec x)}{\delta \phi(\vec y)}
\frac{\delta \phi(\vec y)}{\delta B(\vec w)}
\right]\\
=
\int d^3 y 
\left[
\frac{\partial_z\partial_{\bar z}}{\partial_+}
\phi(\vec y) 
+ig
\left(\frac{\partial_{\bar z}}{\partial_+} A_z(\vec y)\right)\phi(\vec y)
-ig\frac{\partial_z}{\partial_+}
\Bigl( A_z(\vec y)\phi(\vec y)\Bigr)
\right]\; 
\frac{\delta \xi (\vec x)}{\delta \phi(\vec y)} 
\end{multline}
The second term on the left-hand side  
arises from inserting the scalar 
contribution to the canonical momentum~\eqref{eq:d-bara}
 into the cubic gluon vertex $\mathcal{L}^{(3)}_{++-}$
and the gluon kinetic term. To arrive at this form
it has been used that $A_z[B]$ satisfies the same
 relation~\eqref{eq:mansfield-def}
 as in the purely gluonic case. This implies the relation
\begin{multline}
\int d^3 y
 \left(\frac{\partial_z\partial_{\bar z}}{\partial_+} A_z(\vec y)
      -ig\left[A_z(\vec y), \frac{\partial_z}{\partial_+} A_z(\vec y)\right]
       \right)\frac{\delta \xi(\vec x)}{\delta A_z(\vec y)}\\
=- \int d^3y d^3 z
 \left(\frac{\partial_{\bar z}\partial_z}{\partial_+}B(\vec z)\right)
\frac{\delta \xi(\vec x)}{\delta \phi(\vec y)}
\frac{\delta \phi(y)}{\delta B(\vec z)}
\end{multline}
where it was  used that the new coordinates $B[A_z]$ and $\xi[\phi,A_z]$ are
independent variables so that
$0=\delta \xi_x/\delta B_z = \int d^3 y
( \delta\xi_x/\delta \phi_y) (\delta \phi_y/\delta B_z)
+(\delta \xi_x /\delta A_y)(\delta A_y/\delta B_z)$.

Following~\cite{Ettle:2006bw} a recursion relation can be derived from
the condition~\eqref{eq:xi-cond-1}
by multiplying
with $\delta \phi(\vec z)/\delta\xi(\vec x)$, integrating over $\vec x$
and transforming to momentum space:
\begin{equation}
\label{eq:xi-condition}
\int \widetilde{d p} \left(\frac{\delta \phi_{k_1}}{\delta \xi_p}
  \omega_{p}\xi_p +\frac{\delta \phi_{k_1}}{\delta B_p}
  \omega_{p}B_p\right)
 = \omega_{k_1}\phi_{k_1}+g \int \widetilde{d k_2}
  \left(\zeta_{k_1}-\zeta_{k_2}\right)A_{k_2,z}\phi_{k_1-k_2}
\end{equation}
Note that here the relations $\xi[\phi,A_z]$ and $B[A_z]$ 
have been inverted resulting in 
the old  fields as function of the new ones, $A_z[B]$ and $\phi[\xi,B]$.

To solve the condition~\eqref{eq:xi-condition} we make an Ansatz for
$\phi[\xi,B]$ that is linear in $\xi$:
\begin{equation}
\label{eq:phi-trafo}
 \phi(p)=\sum_{n=1}^\infty\int \prod_{i=1}^n \widetilde{d k_i}
\mathcal{Z}(p,k_1,\dots,k_n)B_{-k_1}\dots B_{-k_{n-1}}\xi_{-k_n}
\end{equation}
Inserting this together with  the solution for
$A_z$~\eqref{eq:b-trafo} into the
condition~\eqref{eq:xi-condition} gives rise to  a recursion relation
for the coefficients $\mathcal{Z}$:
\begin{equation}
\label{eq:z-recursion}
\omega_{k_{1,n}}\mathcal{Z}(k_1,\dots k_n)=g
\sum_{i=2}^{n-1}  \left(\zeta_{k_{2,i}}-\zeta_{k_1}\right)
\mathcal{Y}(-k_{2,i},k_2\dots,k_i)\mathcal{Z}(-k_{i+1,n},k_{i+1}\dots k_n)
\end{equation}
Analogously to the gluon case, the factor multiplying the expansion 
coefficients is proportional to the vertex~\eqref{eq:bgks} that we are trying
to eliminate.
In fact, using momentum conservation it is easily seen that
the factor is proportional to that
in the gluonic case~\eqref{eq:y-recursion}:
\begin{equation}
\zeta_{k_{2,i}}-\zeta_{k_1}
 =-\frac{k_{(2,i)\bar z}k_{(i+1,n)+}-k_{(2,i)+}k_{(i+1,n)\bar z}}{
   k_{(2,i)+}k_{1+}}
=\frac{k_{(i+1,n)+}}{k_{1+}}\left(\zeta_{k_{i+1,n}}-\zeta_{k_{2,i}}\right)
\end{equation}
Therefore the solution to the recursion relation~\eqref{eq:z-recursion}
is simply proportional to the gluonic expansion coefficients $\mathcal{Y}$.
The solution satisfying the normalisation condition
$\mathcal{Z}(p,-p)=1$ is given by
\begin{equation}\label{eq:lightconescalarstransf}
  \mathcal{Z}(p,k_1,\dots, k_n)=-\frac{k_{n+}}{p_+}
\mathcal{Y}(p,k_1,\dots, k_n)
= (g\sqrt 2)^{n-1}\frac{\braket{\eta n}}{\braket{\eta 1}
\braket{12}\dots\braket{(n-1) n}}
\end{equation}
This expansion can also be inferred from the proposed transformation
of the light-cone $\mathcal{N}=4$ SUSY superfield~\cite{Feng:2006yy}.
The fact that all six real scalars of the $\mathcal{N}=4$
SUSY-multiplet have the same expansion~\cite{Feng:2006yy} suggests
that the expansion coefficients for the conjugate momentum
$\phi^\dagger_p$ are the same as those
for $\phi$:
\begin{equation}
\label{eq:phibar-trafo}
 \phi^\dagger_p=\sum_{n=1}^\infty\int \prod_{i=1}^n \widetilde{d k_i}
\;(g\sqrt 2)^{n-1}
 \frac{\braket{\eta 1}
 \bar\xi_{-k_1}B_{-k_2}\dots B_{-k_n}}{
\braket{\eta n}\braket{12}\dots\braket{(n-1)n}}
\end{equation}
In order to verify that this is the correct expansion one would have to insert
the inverse of the transformation $\phi\to \xi[\phi, A_{z}]$ into the defining
equation of the canonical momentum of the scalars~\eqref{eq:phi-canonical}.
In the same way one would also obtain the coefficients that enter the
expansion of the additional scalar contribution to the momentum of the gluons
in~\eqref{eq:d-bara} in terms of the $B$-fields.
Instead of attempting to invert the transformation, here we follow
the path used in~\cite{Ettle:2006bw} where the transformation
for the conjugate momentum $\partial_+ A_{\bar z}$ was constructed by
demanding that the kinetic term for the gluons remains invariant.
For the case at hand one
can only demand  that the \emph{sum} of the
kinetic terms of scalars and gluons remains invariant:
\begin{equation}
\label{eq:transform-kinetic}
   (p_+\phi^\dagger_p)( p_-\phi_{-p}) +2\tr[( p_+ A_{p,\bar z})(p_-A_{-p,z})]
=  (p_+\bar\xi_{p})(p_-\xi_{-p}) +2\tr[( p_+ \bar B_p) (p_-B_{-p})]
\end{equation}
As demonstrated in appendix~\ref{app:derive-a-bar}, this requirement together with 
the expansions~\eqref{eq:phi-trafo} and~\eqref{eq:phibar-trafo} fixes the
 expansion of the gluon momentum to be
\begin{multline}
\label{eq:bbar-trafo-scalar}
  p_+ A_{p,\bar z}=\sum_{n=1}^\infty\int\sum_{s=1}^n  \prod_{i=1}^n 
\widetilde {dk_i}\;
\frac{- (g \sqrt 2)^{n-1}\braket{\eta s}^2}{\braket{\eta 1}
  \braket{12}\dots\braket{(n-1) n}\braket{n\eta}}\\
B_{-k_1}\dots B_{-k_{s-1}}\Bigl((k_{s})_+\bar B_{-k_s} B_{-k_{s+1}}
  +\xi_{-k_s}(k_{s+1})_+\bar\xi_{-k_{s+1}}
\Bigr)B_{-k_{s+1}}\dots B_{-k_n}
\end{multline}
For the explicit color structure of this expression
see~\eqref{eq:bbar-em-scalar}. This completes the construction of the
canonical transformation for a scalar in the fundamental
representation. The results obtained so far apply equally well to
massless and massive scalars.

\subsubsection*{Derivation of the vertices}

The expansion of the scalars~\eqref{eq:phi-trafo} and~\eqref{eq:phibar-trafo}
can now be used to determine the additional tower
of vertices generated by the transformation of the mass term 
$m^2\phi^\dagger\phi$:
\begin{equation}
-m^2\phi^\dagger_p\phi_{-p}= \sum_{n=2}^\infty\int \prod_{i=1}^n 
\widetilde{d k_i}
\mathcal{V}_{\bar \xi_1,B_{2},\dots B_{n-1},\xi_n}
 \left(\bar\xi_{k_1}B_{k_2}\dots B_{k_{n-1}}\xi_{k_n}\right)
\end{equation}
where the vertex-function can be simplified using
the eikonal identity~\eqref{eq:eikonal}:
\begin{equation}
\begin{aligned}
\mathcal{V}_{\bar \xi_1,B_{2},\dots B_{n-1},\xi_n}&=
\sum_{j=1}^{n-1} 
 \frac{-m^2(g\sqrt 2)^{n-2}\braket{\eta 1}\braket{\eta n}}{
\braket{12}\dots\braket{\eta j}
\braket{\eta(j+1)}\dots\braket{(n-1)n }}
=\frac{ -m^2(g\sqrt 2)^{n-2} \braket{1n}}{
\braket{12}\dots\braket{(n-1)n }} 
\end{aligned}
\end{equation}
For the
case $n=2$ we just obtain a mass-term $-m^2\bar\xi\xi$ that can be
re-summed  into the scalar propagator.  Since in
the convention for colour-ordering~\eqref{eq:color-ordered} a factor
$g^{n-2}$ is stripped off, the colour ordered vertex generated by the
mass term is given by~\eqref{eq:csw-mass}. 

This establishes the main new result of the rules given in
section~\ref{sec:scalar-rules}. The other vertices are not specific
to massive scalars but we briefly comment on their origin in the
light-cone formalism.
The four-scalar vertex~\eqref{eq:csw-4phi} differs from the
result for $\mathcal{N}=4$ supersymmetric theories
\begin{equation}
   \mathcal{V}^{\mathcal{N}=4}_{\bar\xi_1,B_2,\dots \xi_{j},\bar\xi_{j+1}\dots \xi_n}=
-(g\sqrt 2)^{n-2}  \frac{
\braket{1j}^2\braket{(j+1)n}^2}{
  \braket{12}\dots\braket{n1}}
 \end{equation}
 considered previously in the
 literature~\cite{Georgiou:2004wu,Georgiou:2004by,Wu:2004fb}
since the $\mathcal{N}=4$ SUSY action contains a quartic
 scalar coupling, in contrast to the action~\eqref{eq:phi-lag}.
  The transformation of a $\phi^4$ interaction is
 obtained analogously to that of the mass term:
\begin{equation}
\label{eq:csw-4phi-susy}
\lambda(\phi^\dagger\phi)^2= \sum_{n=4}^\infty\sum_{j=2}^{n-1} \int \prod_{i=1}^n 
\widetilde{d k_i}
\mathcal{V}^\lambda_{\bar \xi_1,B_{2},\dots ,\xi_j,\bar\xi_{j+1},\dots \xi_n}
\left(\bar\xi_{k_1}B_{k_2}\dots \xi_{k_j}\right)
\left(\bar\xi_{k_{j+1}}\dots B_{k_{n-1}}\xi_{k_n}\right)
\end{equation}
with the vertex function
\begin{equation}
  \mathcal{V}^{\phi^4}_{\bar \xi_1,B_{2},\dots ,\xi_j,\bar\xi_{j+1},\dots \xi_n}=
\lambda (g\sqrt 2)^{n-4} \frac{\braket{1j}\braket{j(j+1)\braket{(j+1)n}}}{
\braket{12}\dots\braket{(n-1)n }} 
\end{equation}
Subtracting this expression with $\lambda=-g^2$ from the SUSY
vertex~\eqref{eq:csw-4phi-susy} leads precisely to the
vertex~\eqref{eq:csw-4phi} appropriate for the Lagrangian~\eqref{eq:phi-lag}. 
Of course it is also straightforward to obtain the CSW-rules for a scalar
with a genuine $\phi^4$ coupling from these result.

To complete the proof of the CSW rules for scalars in the light-cone
formalism would require to show explicitly that the remaining interaction
vertices are transformed into towers of the massless MHV
vertices~\eqref{eq:csw-2phi} and~\eqref{eq:csw-4phi}. 
Since we use the same transformation for massive scalars as for massless 
ones, this result is expected.
However, the explicit derivation 
is  more complicated 
than in the case of pure Yang-Mills---where it was done
only for the four and five-point vertices~\cite{Ettle:2006bw}---
because of the
the additional scalar contributions in the transformation of
$\partial_+A_{\bar z}$ in~\eqref{eq:d-bara}. Therefore the scalar MHV
vertex~\eqref{eq:csw-2phi} receives contributions from four sources:
the transformed scalar-gluon vertices $\mathcal{L}^{(3)}_{\phi^\dagger
  A_{\bar z}\phi}$ and $\mathcal{L}^{(4)}_{\phi^\dagger A_zA_{\bar
    z}\phi}$ but also from the cubic and quartic gluon vertices
$\mathcal{L}^{(3)}_{++-}$ and $\mathcal{L}^{(4)}_{++--}$. 
Similarly, the vertices $\mathcal{L}^{(3)}_{\phi^\dagger
  A_{\bar z}\phi}$,  $\mathcal{L}^{(4)}_{\phi^\dagger A_zA_{\bar
    z}\phi}$ , $\mathcal{L}^{(4)}_{\phi^\dagger\phi\phi^\dagger\phi}$ and $\mathcal{L}^{(4)}_{++--}$ contribute to the 
four-scalar MHV vertex~\eqref{eq:csw-4phi}.  
 Given the
fact that the form of the vertices~\eqref{eq:csw-2phi}
and~\eqref{eq:csw-4phi} is not specific to massive scalars, which are
our main concern in this paper, and since there is no reason to suspect
that the proposal of~\cite{Mansfield:2005yd} should fail for massless
scalars we do not attempt to derive the MHV vertices here, but the
transformations~\eqref{eq:phi-trafo}
and~\eqref{eq:bbar-trafo-scalar} provide all the necessary ingredients
to do so.

\subsection{Twistor-Yang Mills with massive scalars} 
\label{sec:scalar-twistor}

The twistor action for a scalar coupled to Yang-Mills theory can be derived through the Noether procedure. This is done in full detail in appendix \ref{app:noetherproc}. It is natural to lift a scalar field and its complex conjugate separately to twistor space. This is because a first-order action is preferred\footnote{As stated before basic twistor theory relates weighted Dolbeault cohomology classes with solutions to the wave equation. The closedness condition is a first order differential equation.}. A simpler way to see this is to study the scalar couplings in the $\mathcal{N}=4$ twistor action. This fixes the twistor lift of the covariant kinetic terms in the space-time action. For the mass term we need the explicit lifting formula for the scalar (charged under the fundamental representation of the gauge group),
\begin{align}\label{eq:liftforphi}
\phi(x)= \int_{\CP^1} H^{-1} \xi_0(x,\pi)\\
\phi^{\dagger}(x)= \int_{\CP^1} \bar{\xi}_0(x,\pi) H
\end{align}
so the mass term becomes
\begin{equation}
\mathcal{L}_{\textrm{mass}} = - m^2 \tr \int_{\CP^1 \times \CP^1} \left( \bar{\xi}_0(x,\pi_1) H\right)_1\left( H^{-1}  \xi_0(x,\pi_2)\right)_2
\end{equation}
Putting everything together the action \ref{eq:phi-lag} lifts to 
\begin{equation}\label{eq:massivescalartw}
S[\xi, \bar{\xi}, B, \bar{B}]= S_{YM} + 
 \int \Omega\wedge \left(\, \bar{\xi}\wedge(\dbar\xi -i \sqrt{2} g[B,\xi]) 
 \right)   + S_{\bar{B}, \xi,  \bar{\xi} } + S_{\xi^2 ( \bar{\xi})^2 } + S_{\textrm{mass}}
\end{equation}
with
\begin{eqnarray}
S_{\bar{B}, \xi,  \bar{\xi} }&=  &
\sqrt{2} g\int_{\mathbb{R}^4}\hspace{-0.1cm} d^4x \, 
\int_{(\CP^1)^3} \frac{\braket{\pi_1\pi_2}\braket{\pi_2\pi_3}}{\braket{\pi_1\pi_3}} \left(\bar{\xi}^0_1 H_1 H^{-1}_2 \bar{B}^0_2 H_2 H^{-1}_3 \xi_3^0\right)
\\
S_{\xi^2 ( \bar{\xi})^2 }
&= &
\frac{g^2}{2}\int_{\mathbb{R}^4}\hspace{-0.1cm} d^4x \, 
 \int_{(\CP^1)^4} \left(2 \frac{\braket{\pi_1\pi_2}\braket{\pi_3\pi_4}}{\braket{\pi_2\pi_3}\braket{\pi_4\pi_1}}-1 \right) \, 
\left(\bar{\xi}^0_1 H_1 H^{-1}_2 \xi^0_2  \bar{\xi}^0_3 H_3 H^{-1}_4 \xi^0_4\right )
\end{eqnarray}
as an action on twistor space. In addition to the gauge symmetries for the Yang-Mills part, the full action is invariant under
\begin{equation}
\xi \rightarrow \xi + \dbar_B f^{-2} \quad \bar{B} \rightarrow \bar{B} - i \sqrt{2} g [\bar{\xi}, f^{-2}]
\end{equation}
and
\begin{equation}
\bar{\xi} \rightarrow \bar{\xi} + \dbar_B \bar{f}^{-2} \quad  \bar{B} \rightarrow \bar{B} - i \sqrt{2} g [\xi, \bar{f}^{-2}]
\end{equation}
for twistor gauge parameters of the indicated weight. As before, this symmetry must be fixed in order to derive scattering rules (or more precisely to invert the kinetic operator). 

\subsubsection*{Gauge choices}
A space-time gauge can be imposed on all the twistor fields, including the scalars. The action \ref{eq:massivescalartw} reduces in that case directly to \ref{eq:phi-lag}. This can be derived straightforwardly from~\cite{Boels:2006ir}, up to the subtlety mentioned below. There is in this case no remaining residual gauge symmetry for the scalar fields.

This gauge choice explains the '$-1$' in the vertex for the four point interaction as this factor is necessary to avoid a $\phi^4$ term in the action. In space-time gauge, without the '$-1$', a $\phi^4$ interaction vertex would arise, which the added factor cancels. To see this, note that in space-time gauge 
\begin{equation}
\dbar_0^{\dagger} \xi_0 = 0 \quad \rightarrow \quad \xi_0 = \phi(x) 
\end{equation}
holds. Therefore only the remaining integral over the $\pi$'s must be performed. Now it is convenient to recall the standard integrals
\begin{align}
\int_{\CP^1} \frac{\pi^{\dalpha} \hat{\pi}^{\dbeta}}{\braket{\pi \hat{\pi}}} & = \frac{1}{2} \epsilon^{\dalpha \dbeta}\\
\int_{\CP^1} \frac{\braket{\pi_1 \xi}}{\braket{\pi_1 \nu}} & = \frac{\braket{\xi \hat{\nu}}}{\braket{\nu \hat{\nu}}}
\end{align}
for arbitrary spinors $\xi$ and $\nu$. Therefore
\begin{equation}
\int_{(\CP^1)^4} \frac{\braket{\pi_1\pi_2}\braket{\pi_3\pi_4}}{\braket{\pi_2\pi_3}\braket{\pi_4\pi_1}} = \frac{1}{2}
\end{equation}
and the extra '$-1$' is required to obtain the space-time action we are interested in. The underlying reason this arises is that the action lifted to twistor space is not the minimally coupled one. Interestingly, the $\phi^4$ interaction is required by supersymmetry and, of course, is part of the Higgs potential and arises naturally on twistor space.

The second obvious gauge choice is the CSW-gauge,
\begin{equation}
\eta^{\alpha} \left(B_\alpha, \xi_\alpha, \bar{\xi}_\alpha, \bar{B}_\alpha \right) = 0
\end{equation}
which eliminates the interaction vertices in part of the Lagrangian. In this gauge the scalar fields have, without the mass term, the same interesting propagator as the gluon fields,
\begin{equation}
:\xi_0 \bar{\xi}_0: = i\frac{\delta(\eta \pi_1 p) \delta(\eta \pi_2 p)}{p^2}
\end{equation}
Although this can be verified by direct computation, it is an expression of an effective supersymmetry since the $\mathcal{N}=4$ action has the same kinetic term for all component fields in the local part of the action \footnote{The non-local part is different but can be shown not to contribute to this particular Green's function}. The mass term changes things slightly: for the free propagator the quadratic part has to be split from the other terms in this vertex. It can be verified that the correct propagator now simply is
\begin{equation}
:\xi_0 \bar{\xi}_0: = i\frac{\delta(\eta \pi_1 p) \delta(\eta \pi_2 p)}{p^2-m^2}
\end{equation}
which can be derived by the usual geometric series argument or direct computation. All this does not change much for the EM coefficients, which simply follow directly from the lifting formula as
\begin{align}\label{eq:emcoefscalar}
\phi(x)= \int_{\CP^1} \frac{H^{-1}(\eta) H(\pi)}{\braket{\eta \pi}} \xi_0(x,\pi) \braket{\eta \pi} 
\end{align}
Expanding the two dimensional propagator, plugging in the delta functions and Fourier transforming gives
\begin{equation}\label{eq:emforphi}
\phi(p) = \sum_i \sum_{j=1}^i \int \left(\prod_{k=i}^j dp_k^4 \right) B_0(-p_1) \ldots \xi_0(-p_j)  \mathcal{Y}(p, p_1, \ldots, p_j)
\end{equation}
with
\begin{equation}
\mathcal{Y}(p, p_1, \ldots, p_j) = (i \sqrt{2} g)^{j} \frac{\braket{\eta \eta p_j}}{\braket{\eta (\eta p_1)}\braket{(\eta p_1) (\eta p_2)} \ldots \braket{(\eta p_{j-1}) (\eta p_j) }}
\end{equation}
Using the spinor momentum trick~\eqref{eq:spinormomentumtrick} and absorbing the factors $- i \sbraket{\eta i}^2$ factors into the $B_0$ fields as in~\eqref{eq:b0tob}
this turns into
\begin{equation}
\mathcal{Y}(p, p_1, \ldots, p_j) = (\sqrt{2} g)^{j} \frac{\braket{\eta j}}{\braket{\eta 1}\braket{1 2} \ldots \braket{(j-1)j }}
\end{equation}
It is obvious that an identical result follows for $\phi^{\dagger}$. This is of course nothing but~\eqref{eq:lightconescalarstransf} which was obtained through the canonical transformation.  

Actually, it is interesting to note that the space-time interpretation of the transformation rule for one of the space-time helicities now gets an extra contribution compared to~\eqref{eq:twistorforbarB}. This is in line with the extra contributions in~\eqref{eq:d-bara} in the canonical approach. The reason here is that the $\bar{B}_0$ field equation used in deriving that equation gets an extra contribution from the $\bar{B}_0 \xi_0 \bar{\xi}_0$ vertex. Therefore we obtain the modified equation 
\begin{align}\label{eq:twistorforbarBII}
(\eta \eta p) A_{\bar{z}}& = - \frac{1}{\sqrt{2}} (\hat{\eta}^{\alpha} \eta_\alpha) \int_{\pi_k}\dk\,  \frac{H(\eta) H^{-1}(\pi_k)}{\braket{\eta \pi_k}} \bar{B}_0(\pi_k) (\braket{\pi_k \eta})^4  \frac{H(\pi_k) H^{-1}(\eta)}{\braket{\pi_k \eta}} +\nonumber\\ 
& \int_{\pi_1,\pi_2} \frac{\braket{\eta \pi_1}\braket{\eta \pi_2}}{\braket{\pi_1 \pi_2}} \left(H^{-1}(\eta)H \xi\right)_1 \left(\bar{\xi} H^{-1} H(\eta) \right)_2
\end{align}

Performing the by now familiar (we hope!) operations of inserting the polarisation vector and using the spinor momentum trick, we obtain 
\begin{equation}
\sbraket{\eta p}^2 \bar{A}(p) = \sum_{i,n} \frac{\braket{\eta p_i}^2  (\sqrt{2} g)^{n-1} B_0(p_1) \ldots B_0(p_{i-1}) X_i B_0(p_{i+2}) \ldots B_0(p_n) }{\braket{\eta 1} \braket{2 3} \ldots \braket{(n-1) \eta}}  
\end{equation}
with
\begin{equation}
X_i =  \braket{\eta p_i}^2 \bar{B}(p_i) B(p_{i+1}) +  \braket{\eta p_{i+1}}^2   \xi(p_i) \bar{\xi}(p_{i+1})
\end{equation}
in terms of the spinor products. Again, using \eqref{eq:braket-eta} this is easily seen to correspond to~\eqref{eq:d-bara}. In these equations we have for clarity suppressed the field normalisation constants ($\sim i \sbraket{\eta 1}^2$) and the color factors which are easily reinstated if one bears in mind that the scalars transform in the fundamental.


\subsection{Equivalence theorem for massive scalars}
In massless theories the non MHV-type cubic vertices that appear to be
missing in the CSW Lagrangian are generated by violations of the naive
'equivalence theorem' as discussed in~\ref{sec:violate}. 
For massive scalars the situation is different since
the transformation of the mass term generates a
non-MHV three-point vertex $V(\bar\xi, B,\xi)$
 that is equivalent to the conventional
vertex provided all particles are on-shell~\cite{Boels:2007pj}.  Here
we discuss in more detail why 'equivalence theorem violating'
contributions to this vertex are absent for massive scalars.

In the light-cone gauge the scalar non-MHV vertex arises from the
Green's function $\braket{0|\bar \phi A_{\bar z}\phi|0}$.
  Similar to the discussion of the cubic non-MHV gluon
  vertex~\eqref{eq:gf-trafo} this Green's function is expressed as a
  function of the new fields $(B,\bar B)$ and $(\xi,\bar\xi)$ by
  inserting the transformations~\eqref{eq:phi-trafo} and~\eqref{eq:bbar-trafo-scalar} or equivalently the relations to the twistor-fields~\eqref{eq:emcoefscalar} and~\eqref{eq:twistorforbarBII}:
\begin{multline}
\label{eq:gf-phi-trafo}
\braket{ 0|\bar\phi_{p_1}A_{p_2,\bar z}\phi_{p_3}|0}
=\Bigl\langle 0|\bar \xi_{p_1} \bar B_{p_2} \xi_{p_3}
+\int \widetilde{d k_1}\widetilde d{k_2}
\left[
 \left(\mathcal{Z}(p_1,k_2,k_1) \bar \xi_{-k_1 } B_{-k_2}\right) 
\bar B_{p_2} \xi_{p_3}
 \right. \\
\left.+\bar \xi_{p_1}
\left(\tfrac{k_{2+}}{p_{2+}}
\mathcal{W}^1(p_2,k_2,k_1) \xi_{-k_1 }\bar \xi_{-k_2}\right) \xi_{p_3}
+\bar \xi_{p_1} \bar B_{p_2}
\left(\mathcal{Z}(p_3,k_1,k_2) B_{-k_1 }\xi_{-k_2}\right)
\right]+\dots|0\Bigr\rangle
\end{multline}
For the definition of the $\mathcal{W}$ coefficients see~\eqref{eq:bbar-em-scalar}.
The on-shell scattering amplitude is obtained as usual by multiplying by 
$(p_1^2-m^2)p_2^2 (p_3^2-m^2)$ and taking the on-shell limit.

For a massive scalar the first term is non-vanishing due to the vertex~\eqref{eq:csw-mass}.
 Performing again the LSZ reduction on legs $1$ and $2$
first, the only additional contribution potentially
comes from the last term:
 \begin{equation}
  \mathcal{Z}(p_3,p_2,p_1)=
 g \sqrt 2 \frac{\braket{\eta 1}}{
\braket{\eta 2}\braket{21}} 
=g \sqrt 2\frac{1}{\braket{12}\sbraket{21}}
 \frac{\braket{\eta-|\fmslash k_1|2-}}{\braket{\eta 2}}
\end{equation}
For a massless scalar with $p_1^2=0$ this gives the correct vertex
$V(\bar \phi,A_z,\phi)$ times a propagator $1/(p_1+p_2)^2$.  For a
massive scalar with $p_1^2=m^2$ it can be seen using the identity~\eqref{eq:simplify-ij} that the denominator contains instead
\begin{equation}
  \braket{12}\sbraket{21}
= p_3^2+m^2\frac{(p_3\cdot\eta) }{(p_1\cdot \eta)} 
\end{equation}
so there is no pole for $p_3^2\to m^2$ and the term vanishes upon LSZ reduction.

It is also instructive to see that there is no contribution from the on-shell limit of the gluon leg.
Performing the reduction first on legs $1$ and $3$,
the contribution from leg 2 gives
\begin{equation}
\frac{(p_3)_+}{(p_2)_+}
 \mathcal{W}^1(p_2,p_3,p_1)= g \sqrt 2\frac{(p_3)_+}{(p_2)_+}
 \frac{\braket{\eta 1}}{
\braket{\eta 3}\braket{13}}
= g \sqrt 2 \frac{1 }{\sbraket{31}\braket{13}}
 \frac{\braket{\eta 1}\sbraket{12} }{\braket{\eta 2}}
\end{equation}
Applying~\eqref{eq:simplify-ij} the denominator
can be written as
\begin{equation}
\braket{13}\sbraket{31}
=p_2^2
-m^2 \frac{(\eta\cdot p_2)^2}{ (p_1\cdot \eta)(p_3\cdot \eta)}
\end{equation}
so there is no pole for $p_2^2=0$ in the massive case and again there is no 'equivalence theorem violating' contribution. Hence we have demonstrated explicitly that only for mass-less particles equivalence theorem evasion is expected
to play a role for the field transformation employed in this article.

\section{One-loop amplitudes in pure Yang-Mills }
\label{sec-loops}
It was shown in\cite{Brandhuber:2004yw,Bedford:2004py,Quigley:2004pw,Bedford:2004nh,Brandhuber:2005kd} that the so called cut constructable pieces of
one-loop amplitudes in pure Yang-Mills can be computed using the CSW rules.
For some while the so called rational terms, however, remained elusive from
the CSW point of view although several alternative methods are available to
compute these (see e.g.~\cite{Bern:2008ef} and references therein).
By know several quantum completions of the CSW formalism in pure Yang-Mills
have been proposed in the literature that also reproduce at least some of the rational parts of
one-loop amplitudes. These employ
\begin{itemize}
\item ordinary lightcone loops \cite{Boels:2007gv}
\item a Lorentz-violating regularisation \cite{Brandhuber:2007vm}
\item equivalence theorem violations \cite{Ettle:2007qc}
\end{itemize}
all of which are slightly unsatisfactory for various reasons. The first method
does not offer much advantages over ordinary lightcone methods. The second
method gives a clear argument that the all-plus loop can be interpreted as an anomaly, but the
regularisation used makes it hard in the general case to compare results
obtained in it to ordinary dimensional regularisation for
instance. Equivalence theorem violations show where both tree and loop
amplitudes are hiding, but the formalism in D dimensions employed breaks
supersymmetry explicitly. This obfuscates the use of supersymmetric
decomposition for instance. 

In this section we show that the CSW rules for massive scalars can be useful
to address the issue of one-loop amplitudes in pure Yang-Mills within the
context of conventional dimensional regularisation, in particular in the
four-dimensional helicity scheme variant thereof.
This is suggested by the decomposition
of one-loop amplitudes in pure Yang-Mills theory
 according to the particles which run in the loop~\cite{Bern:1996je}
\begin{equation}\label{eq:susydecomp}
A_{\textrm{pure YM}} = A_{\mathcal{N}=4} - 4 A_{\mathcal{N}=1} + A_{\textrm{scalar}}
\end{equation}
The supersymmetric contributions are cut-constructable, so can be calculated
using four dimensional unitarity or the one-loop MHV rules.
The scalar part is much harder since it involves, from the four dimensional point of view, massive scalars. This is
can be seen by decomposing the $D=4-2\epsilon$ dimensional momentum as
$\ell_D=(\ell,\ell_{-2\epsilon})$ and taking the four-dimensional
part and the $-2\epsilon$ dimensional part as orthogonal so that
\begin{equation}
\ell_{D}^2 = \ell^2 + \ell^2_{-2\epsilon} \equiv \ell^2 -\mu^2
\end{equation}
and decomposing the dimensionally regulated loop integral measure
as\cite{Bern:1996je}
\begin{equation}
\label{eq:measure-dim}
  \frac{d^D\ell}{(2\pi)^D}= \frac{d^4\ell}{(2\pi)^4}
  \frac{d^{-2\epsilon}\mu}{(2\pi)^{-2\epsilon}}
\end{equation}
This suggests that our CSW rules  for massive scalars  given in section~\ref{sec:scalar-rules} should
be suitable for the calculation of the rational parts of loop amplitudes.
We now give several arguments that this is indeed the case and verify in section~\ref{sec:four-point} that the one-loop four-point amplitude with positive  helicity gluons is correctly reproduced. In section~\ref{sec:loop-remarks} we briefly discuss the structure of some other amplitudes in this formalism. In section~\ref{sec:pure-ym} we investigate a possible direct regularisation of the MHV Lagrangian.

Treating the polarisation vectors and momenta of the external gluons
as four dimensional and lifting the space-time Lagrangian of a
massless scalar to $D$-dimensions, the only place where the
$-2\epsilon$ dimensional components of the momentum have to be kept is
the kinetic term where we replace $\square_D^2\to \square+\mu^2$. In
one-loop diagrams with a scalar loop gluons only appear in trees glued
to the loop.  Their momenta will be treated as four-dimensional since
they are given by sums of external momenta. Therefore the interaction
vertex $\phi^\dagger \overleftrightarrow{\partial_\mu}\phi$ is always
contracted with a four-dimensional polarisation vector or momentum so
only the four dimensional scalar momentum has to be kept. This leads
to a Lagrangian of the form~\eqref{eq:phi-lag} with $m^2= \mu^2$,
where loop-integrals are computed using the
measure~\eqref{eq:measure-dim}.  The only difference is that the
scalars in~\eqref{eq:phi-lag} are in the fundamental while the above
argument applies to scalars in the adjoint.  However, the
colour-leading part of amplitudes with adjacent scalars is very easy
to lift from fundamental to adjoint representation, the only
difference being a factor of $N_c$. It would also be straightforward
to extend the derivation to scalars in the adjoint.  Within the
twistor approach, the vertex which is needed for scalars in the
adjoint is
\begin{equation}
\mathcal{L}_{\textrm{mass}} = - m^2 \tr \int_{\CP^1 \times \CP^1} \left(H \phi^{\dagger}_0(x,\pi_1) H^{-1}\right)_1\left( H \phi^{\phantom{\dagger}}_0(x,\pi_2) H^{-1}\right)_2
\end{equation}
This is obtained by considering the twistor lifting formulae for adjoint scalars. Using the techniques of this article it is easy to see how to obtain the expansion coefficients for adjoint scalars from these if desired. Expanding the frames in the above expression using~\eqref{eq:expand-h}  the mass-vertex
 for scalars in the adjoint becomes
 \begin{equation}
\label{eq:csw-mass-adjoint}
V_{\text{CSW}}(B_1,\dots \xi_i,B_{i+1},\dots \xi_j,B_{j+1},\dots B_n)=
 i  2^{n/2-1}
\frac{ m^2 \braket{ij}^2}{
\braket{12}\dots\braket{(n-1)n }\braket{n1}} 
\end{equation}
which reduces to~\eqref{eq:csw-mass} for scalars at position $1$ and
$n$. 

There are two remaining issues to show that the CSW rules for massive scalars
also work at the quantum level. The first  are potential equivalence theorem violations.
 Their absence for massive scalar amplitudes at tree level and the fact that the four-point all-plus
amplitude is reproduced correctly without such contributions (as shown below) suggests that they are also absent at the one-loop level.

The second issue is the absence of anomalies in the Jacobian of the canonical transformation used to derive the CSW-rules.
Such anomalies were suspected to be the origin of the missing rational pieces in pure Yang-Mills amplitudes, but the recovery of the all-plus amplitude from equivalence theorem evasion in~\cite{Ettle:2007qc} suggests that there is no such anomaly in the canonical transformation in the gluon case. The close similarity of the transformation used for the scalars to that of gluons together with our computation of the four-point all-plus amplitude below suggests that there are no such problems in our case as well. Anomalies in the Jacobian would show up as Faddeev-Popov determinants in the twistor action approach. Since an axial gauge choice is employed to derive the rules, ghost contributions are easily seen to be absent.

The reasoning given above uses the fact that the massive scalar Lagrangian~\eqref{eq:phi-lag} is appropriate to compute the scalar piece in the SUSY decomposition~\eqref{eq:susydecomp} in dimensional regularisation
and some evidence that the transformation used to obtain the CSW rules for massive scalars also holds at
the quantum level. We would like to point out that a different line of argument can be based on the fact alone  that the CSW-rules for a massive scalar give the right tree level amplitudes.
The Feynman tree theorem \cite{Feynman:1972mt} reduces the computation of one-loop amplitudes 
to dispersion integrals over tree level amplitudes. This has been used to show the $\eta$-independence of the
one-loop MHV-rules for cut-constructable piece of amplitudes (or amplitudes in supersymmetric theories) from the $\eta$-independence of tree-amplitudes~\cite{Brandhuber:2005kd}. We expect that a similar argument can be given for the massive scalar-loops computed using our CSW rules.
Having established $\eta$-independence and the resulting absence of spurious poles,  one can argue that any unitarity cut of the sum of diagrams will automatically give (dispersion integrals over) on-shell scalar amplitudes.
Altogether this discussion strongly suggests that the rules of section~\ref{sec:scalar-rules} can be used to compute the scalar-piece in the SUSY decomposition~\eqref{eq:susydecomp}.

\subsection{The four-point all-plus amplitude in pure Yang-Mills}
\label{sec:four-point}
An obvious target for the application of our rules to one-loop diagrams is 
the all-plus one-loop amplitude which vanishes in supersymmetric gauge
theories. As a first example, in this article we will discuss the four-point all-plus amplitude~\cite{Bern:1996je}
\begin{equation}
\label{eq:4-plus}
 A_4^{\text{scalar}}(g_1^+,g_2^+,g_3^+,g_4^+)=\frac{8i}{(4\pi)^{2-\epsilon}}
\frac{ \sbraket{12}\sbraket{34}}{\braket{12}\braket{34}} \;K_4
\end{equation}
with 
\begin{equation}
K_4=(-i)(4\pi)^{2-\epsilon}
   \int \frac{d^4\ell}{(2\pi)^4}\frac{d^{-2\epsilon}\mu}{(2\pi)^{-2\epsilon}} 
\frac{\mu^4}{d(\ell)d(\ell_1)d(\ell_2)d(\ell_3)}
= -\frac{1}{6}+\mathcal{O}(\epsilon)
\end{equation}
  We denote the loop
momentum flowing between leg $4$ and leg $1$ by $\ell$ and use the
notation
\begin{equation}
  \ell_i=\ell-k_{1}-\dots- k_i\quad;\quad
  d(\ell_i)= \ell_i^2-\mu^2
\end{equation}
The result~\eqref{eq:4-plus} contains a factor $2$ taking the two degrees
of freedom of a complex scalar into account and differs from the form in~\cite{Bern:1996je} by a factor of $4$ because of the different colour ordering conventions.

We now show how  the result~\eqref{eq:4-plus}
 for the one-loop
four-point amplitude with only positive helicity gluons
 can be obtained using the diagrammatic rules from section~\ref{sec:scalar-rules}.
The contributing diagrams have the form of
 box, triangle (T), bubble (B) and tadpole (Ta) topologies, as shown in figure~\ref{fig:all-plus}.  In the diagrams 
all vertices are given by~\eqref{eq:csw-mass}.  The diagrams are
denoted by the momenta flowing through the propagators. The
tadpole diagram vanishes because the vertex involves the spinor product
$\braket{\ell\ell}=0$ in the numerator. 
\FIGURE[t]{
  \includegraphics[width=0.8\textwidth]{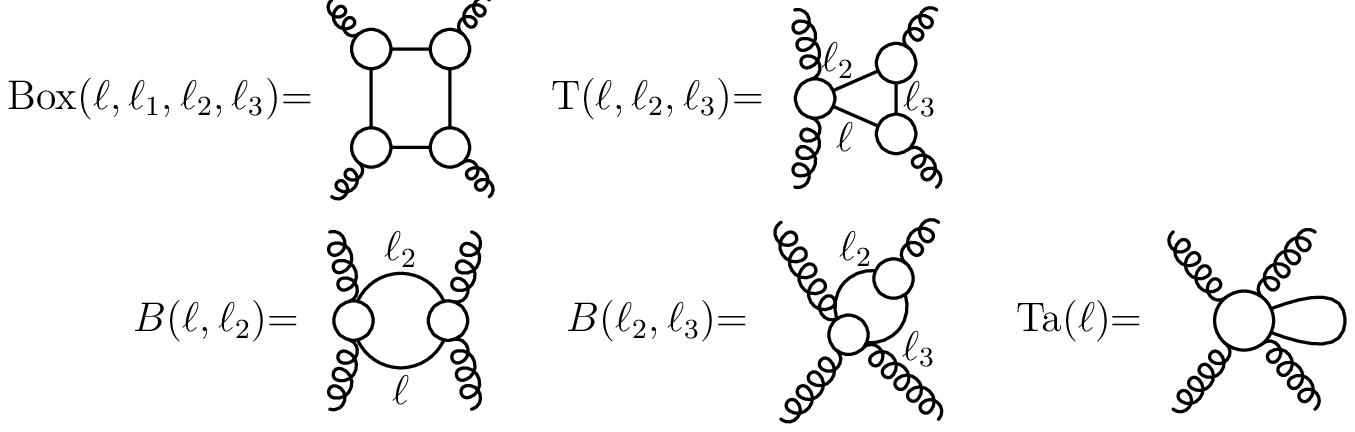}
  \caption{Types of diagrams contributing to the four-point all-plus amplitude}
  \label{fig:all-plus}
}
The calculation simplifies if the spinor $\eta$ used to define the off-shell continuation is fixed in terms of one of the momentum of one of the gluons since this eliminates all diagrams where this gluon enters a three point vertex, cf. the
discussion below~\eqref{eq:csw-cubic}. In particular this eliminates the box diagram.
For the choice $\ket{\eta-}=\ket{1-}$ one remains with two triangle diagrams (T$(\ell,\ell_2,\ell_3)$, T$(\ell_1,\ell_2,\ell_3)$), two bubble diagrams with four-point vertices (B$(\ell,\ell_2)$, B$(\ell_1,\ell_3)$) and three bubble diagrams with a five-point vertex (B$(\ell_2,\ell_3)$, B$(\ell_1,\ell_2)$ and  B$(\ell,\ell_3)$) as shown in figure~\ref{fig:all-plus-nonzero}.
\FIGURE[t]{
\parbox{0.9\textwidth}{
\begin{center}
  \includegraphics[width=0.5\textwidth]{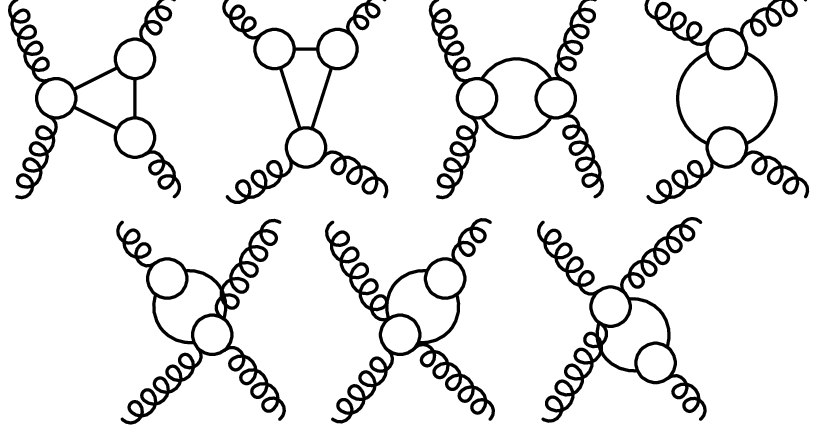}
  \caption{Non-vanishing diagrams contributing to the four-point all-plus amplitude for $\ket{\eta+}=\ket{1+}$}
  \label{fig:all-plus-nonzero}
\end{center}
}}

Focus first on the two triangle diagrams. Applying the CSW prescription with $\ket{\eta-}=\ket{1-}$ to the 
 loop-momenta we find for the triangle T$(\ell,\ell_2,\ell_3)$
\begin{equation}
\label{eq:t-023}
\begin{aligned}
  \text{T}(\ell,\ell_2,\ell_3)
  &=\int \frac{d^D\ell}{(2\pi)^D}
\frac{(i)^3}{d(\ell)d(\ell_2)d(\ell_2)}
\frac{-2i \mu^2\braket{\ell\ell_2}}{
\braket{\ell 1}\braket{12}\braket{2\ell_2}}
\frac{-\sqrt 2i \mu^2\braket{\ell_2\ell_3}}{
\braket{\ell_2 3}\braket{3\ell_3}}
 \frac{-\sqrt 2i \mu^2\braket{\ell_3\ell}}{
\braket{\ell_34}\braket{4\ell}}\\ 
&=\frac{4\sbraket{12}}{\braket{12}\braket{34}}
\int \frac{d^D\ell}{(2\pi)^D}
\frac{\mu^4}{d(\ell)d(\ell_2)d(\ell_3) 2(\ell\cdot k_1)}
\frac{- \mu^2\braket{1-|\fmslash k_3\fmslash k_4|1-}}{
\braket{3-|\fmslash \ell_3|1-}
\braket{4-|\fmslash \ell|1-}} 
\end{aligned}
\end{equation}

To eliminate the additional factor of $\mu^2$ and the spinor products
containing the loop-momentum in the denominator one can use the
formula~\cite{Brandhuber:2006bf}
\begin{multline}
\label{eq:cancel-propagators}
   \mu^2\braket{\eta+|\fmslash k_{i}\fmslash k_{i+1}|\eta-}=
\sbraket{i(i+1)}\braket{\eta+|\fmslash \ell_i|i+}
\braket{(i+1)-|\fmslash \ell_{i+1}|\eta-}\\
+d(\ell_{i-1}) \braket{\eta+|\fmslash \ell_{i}\fmslash k_{i+1}|\eta-}
-d(\ell_i) \braket{\eta+|\fmslash \ell_{i-1}\fmslash k_{i,i+1}|\eta-}
+d(\ell_{i+1}) \braket{\eta+|\fmslash \ell_{i-1}\fmslash k_i|\eta-}
\end{multline}
valid for an arbitrary spinor $\eta$ and momentum $k_i$. 
The labels of the momenta are considered as cyclic, e.g.$\ell_{0}\equiv \ell$.

Applying~\eqref{eq:cancel-propagators} to the triangle
diagram~\eqref{eq:t-023} results in one term without spinor products
with loop momenta in the denominator and three terms with cancelled
propagators:
\begin{multline}
\label{eq:t-023-simple}
  \text{T}(\ell,\ell_2,\ell_3)=
\frac{-4\sbraket{12}}{\braket{12}\braket{34}}
  \int \frac{d^D\ell}{(2\pi)^D}
\frac{\mu^4}{d(\ell)d(\ell_2)d(\ell_3) 2(\ell\cdot k_1)}
\left[ \sbraket{34}
+ d(\ell_2)\frac{\sbraket{41}}{\braket{3-|\fmslash \ell_3|1-}} \right.\\
\left. -d(\ell_3)\frac{\braket{1-|\fmslash \ell_2\fmslash k_{3,4}|1-}}{
\braket{3-|\fmslash \ell_3|1-}
\braket{4-|\fmslash \ell|1-}}
+d(\ell)  \frac{\sbraket{31}}{\braket{4-|\fmslash \ell|1-}} 
\right]
\end{multline}
Since $2(\ell\cdot k_1)=\ell^2 -\ell_1^2$  the denominator
is not yet in the right shape so the first term does
not  give the correct scalar integral at this stage.
The second triangle diagram can be treated in a similar fashion and
is given by
\begin{multline}
\label{eq:t-123}
  \text{T}(\ell_1,\ell_2,\ell_3)=
\frac{4\sbraket{41}}{\braket{23}\braket{41}}\int 
 \frac{d^D\ell}{(2\pi)^D}
\frac{\mu^4}{d(\ell_1)d(\ell_2)d(\ell_3) 2(\ell\cdot k_1)}
\left[ \sbraket{23}
+ d(\ell_1)\frac{\sbraket{31}}{\braket{2-|\fmslash \ell_2|1-}} \right.\\
\left. -d(\ell_2)\frac{\braket{1-|\fmslash \ell_1\fmslash k_{2,3}|1-}}{
\braket{2-|\fmslash \ell_2|1-}
\braket{3-|\fmslash \ell_3|1-}}
+d(\ell_3)  \frac{\sbraket{21}}{\braket{3-|\fmslash \ell_3|1-}} 
\right]
\end{multline}
Using momentum conservation to write $\sbraket{23}\sbraket{41}/\braket{23}\braket{41}=\sbraket{12}\sbraket{34}/\braket{12}\braket{34}$
 and applying a partial fraction identity 
\begin{equation}
  \frac{1}{\ell^2-\ell_1^2}
\left[\frac{1}{\ell_1^2-\mu^2}-\frac{1}{\ell^2-\mu^2}\right]
=\frac{1}{(\ell^2-\mu^2)(\ell_1^2-\mu^2)}
\end{equation}
one finds that the sum of the two triangle diagrams produces the
correct result~\eqref{eq:4-plus} (up to a factor two since we consider
only one scalar degree of freedom) and a sum of terms with 
cancelled propagators.

The left-over terms are all of the right form to be cancelled by
bubble-diagrams and this is indeed what happens.
For example the bubble with propagators $d(\ell)$ and $d(\ell_3)$
exactly cancels the corresponding term in~\eqref{eq:t-023-simple}
\begin{equation}
\begin{aligned}
\text{B}(\ell,\ell_3)
&=\int  \frac{d^D\ell}{(2\pi)^D}
\frac{-4i \mu^2\braket{\ell\ell_3}}{
\braket{\ell 1}\braket{12}\braket{23}\braket{3\ell_3}}
\frac{(i)^2}{d(\ell)d(\ell_3)}
\frac{-i \mu^2\braket{\ell_3\ell}}{
\braket{\ell_3 4}\braket{4\ell}} \\
&=\frac{-4\sbraket{12} }{
\braket{12}\braket{34}}
\int \frac{d^D\ell}{(2\pi)^D}
 \frac{1}{2(\ell\cdot k_1)d(\ell) d({\ell_3})}
 \frac{\sbraket{14}}{\braket{3-|\fmslash \ell_3|3-}}
\end{aligned}
\end{equation}
Similarly
the diagram B$(\ell_1,\ell_2)$ cancels the term with missing $\ell_3$
 propagator in~\eqref{eq:t-123}.
The term with missing  
$\ell_3$ propagator in~\eqref{eq:t-023-simple} is cancelled 
by a bubble with two four point vertices:
\begin{equation}
\begin{aligned}
\text{B}(\ell,\ell_2)
  &=\int \frac{d^D\ell}{(2\pi)^D}
 \frac{-2i \mu^2\braket{\ell\ell_2}}{
\braket{\ell 1}\braket{12}\braket{2\ell_2}}
\frac{(i)^2}{d(\ell) d(\ell_2)}
\frac{-2i \mu^2\braket{\ell_2\ell}}{
\braket{\ell_2 3}\braket{34}\braket{4\ell}} \\
&=\frac{4\sbraket{12}}{\braket{12}\braket{34}}
\int \frac{d^D\ell}{(2\pi)^D}
\frac{\braket{1+|\fmslash \ell_2\fmslash\ell|1-}}{2(\ell\cdot k_1)
d(\ell) d(\ell_2)\braket{3-|\fmslash \ell_3|1-}\braket{4-|\fmslash \ell|1-}}
\end{aligned}
\end{equation}
In the same way, the diagram B$(\ell_1,\ell_3)$ cancels the term with 
missing $\ell_2$ propagator in~\eqref{eq:t-123}.
The last remaining bubble diagram is given by
\begin{equation}
  \begin{aligned}
 \text{B}(\ell_2,\ell_3)
   &= \int\frac{d^D\ell}{(2\pi)^D} \frac{-4i \mu^2\braket{\ell_3\ell_2}}{
\braket{\ell_3 4}\braket{41}\braket{12}\braket{2\ell_2}}
\frac{(i)^2}{d(\ell_2)d(\ell_3)}
\frac{-i \mu^2\braket{\ell_2\ell_3}}{
\braket{\ell_2 3}\braket{3\ell_3}} \\
&=\frac{-4\sbraket{12}}{\braket{12}\braket{34}}
\int \frac{d^D\ell}{(2\pi)^D}
 \frac{1}{d(\ell_2) d(\ell_3)}
 \frac{ \sbraket{31}^2}{\sbraket{32}\braket{1+|\fmslash \ell_3|4+}
\braket{2-|\fmslash \ell_2|1-}}
  \end{aligned}
\end{equation}
This time it requires slightly more work to see the cancellation since both
triangle diagrams~\eqref{eq:t-023-simple} and~\eqref{eq:t-123} contain
a term with $\ell_2$ and $\ell_1$ propagators. The sum is found to cancel
against the the bubble B$(\ell_2,\ell_3)$, as expected:
\begin{multline}
\left[ 
\text{T}(\ell,\ell_2,\ell_3)+\text{T}(\ell_1,\ell_2,\ell_3)
\right]_{d(\ell_2),d(\ell_3)}\\
=
-\frac{4\sbraket{12}\sbraket{34}}{\braket{12}\braket{34}}
\int \frac{d^D\ell}{(2\pi)^D}
\frac{\mu^4\sbraket{31}}{d(\ell_2)d(\ell_3) 2(\ell\cdot k_1)}
\left[\frac{1}{\braket{3+|\fmslash k_2\fmslash \ell|1-}}+
\frac{1}{\braket{3+|\fmslash k_4\fmslash \ell|1-}}
\right]\\
=
\frac{4\sbraket{12}\sbraket{34}}{\braket{12}\braket{34}}
\int  \frac{d^D\ell}{(2\pi)^D}
\frac{\mu^4}{d(\ell_2)d(\ell_3)}
\frac{\sbraket{31}^2}{\braket{3+|\fmslash k_2\fmslash \ell|1-}
\braket{3+|\fmslash k_4\fmslash \ell|1-}}
\end{multline}
This completes the demonstration that the result~\eqref{eq:4-plus} is
correctly reproduced by the application of the CSW rules for a massive
scalar to one-loop diagrams.

Although the number of contributing diagrams is not particularly small, the final result is obtained in a
rather transparent way since cancellations among diagrams take
place before any loop integral is performed.  This is in contrast
to the calculation of the all-minus amplitude using light-cone gauge
Feynman rules~\cite{Brandhuber:2006bf} where the final result was
obtained after applying extensive Passarino-Veltman reductions to box,
triangle and bubble diagrams separately.  The essential
identity~\eqref{eq:cancel-propagators} always can be applied to
products of three-point mass-vertices so the pattern of cancellations
is expected to persist in the calculation of more general amplitudes,
although a more systematical approach to the combinatorics might be
needed. Note however that collinear limits are easily seen to be correctly reproduced in this calculation as these can be read off directly from the vertices. Hence the only ambiguity in the all-multiplicity calculation through massive CSW vertices is the usual ambiguity for the five point amplitude. 

\subsection{Remarks on other amplitudes}
\label{sec:loop-remarks}

We briefly comment on some features of the calculation of the four-point all
plus amplitude give above and on the structure of other amplitudes in the CSW formalism for massive scalars.

\paragraph{The all-plus amplitudes as an anomaly}
The finite all-plus amplitude  arises from the cancellation of a numerator factor
$\epsilon$ with an UV $1/\epsilon$ pole. This feature has been used to suggest that this
 amplitude arises as an anomaly~\cite{Bardeen:1995gk}. In our formalism these cancellations are implied by the fact that
the mass-vertices~\eqref{eq:csw-mass} are proportional to $\mu^2$ and
 one-loop integrals with an insertion of a power of $\mu^2$ are related
to integrals in higher dimensions times explicit pre-factors of $\epsilon$.
The situation is somewhat subtle since some  factors of $\mu^2$ might be eliminated trough Dirac algebra,
 c.f.~\eqref{eq:cancel-propagators} in order to cancel some of the unphysical $\frac{1}{\braket{(\eta p) i}}$ singularities in the integrand so that the power-counting in $\mu^2$ is
 not unique. Nevertheless the calculation in section~\ref{sec:four-point}
 clearly shows how the finite all-plus amplitude arises from the interplay of
 explicit $\mu^2$ factors supplied by the vertices and the $D$-dimensional
 loop-integral measure.
Our formalism could also be useful to clarify the anomalous symmetry responsible for the anomaly.  A symmetry dimensional regularisation breaks explicitly in our formulation is scaling invariance, and this suggests that this is the anomalously broken symmetry. Note that it cannot be the twistor gauge symmetry as our massive scalar regulator is explicitly invariant under this symmetry. In the canonical formalism this can be interpreted as the absence of Jacobian factors.

\paragraph{The all-minus amplitudes}
Using the massive scalar regulator, the all-minus amplitudes are generated only by three-point MHV vertices as in the usual application of the one-loop MHV rules. Note that the propagator is now properly dimensionally regulated. Since in this particular case this calculation is equivalent diagram by diagram to the light-cone calculation it follows that the correct answer will be obtained.

\paragraph{The one-minus amplitudes}
The diagrams for amplitudes with one negative helicity gluon contain precisely one ordinary MHV
vertex and at least one vertex proportional to $\mu^2$. This suggests that these amplitudes arise, as in the all-plus case, manifestly as a $\epsilon\times 1/\epsilon$ 'anomaly' in our formalism. For this conclusion it is important that an ordinary MHV vertex with two adjacent legs tied together vanishes. Since this vertex involves a $\frac{1}{\braket{1 n}}$ term, at face value this diagram does not vanish, but is actually infinite. However, one should regulate the poles of this particular type by a well-defined prescription \cite{Leibbrandt:1987qv}. Using the regulator it follows that this particular diagram is a finite tad-pole and hence vanishes in dimensional regularisation.

\paragraph{General structure of the integrands}
Note that combining two of our massive scalar vertices into a CSW loop gives almost the same integrand as in the MHV $\mathcal{N}=4$ case \cite{Brandhuber:2004yw}. The difference between the calculations is completely in line with \cite{Bern:1996ja}. However, since for massive scalars one is bound to keep the propagator in $D$ dimensions, one cannot simply apply the same reasoning as in \cite{Brandhuber:2004yw} which  applies dimensional regularisation only after performing the spinor algebra (which is perfectly fine for $\mathcal{N}=4$). It would be interesting to explore these similarities further since this could lead to a more streamlined evaluation of the integrals encountered here. Combining the rewriting  of the loop integrals as dispersion integrals and phase-space integrals of  \cite{Brandhuber:2004yw} with the formulation  of the massive phase-space integral used in~\cite{Anastasiou:2006jv} could be helpful in this respect.

\subsection{Direct construction of pure Yang-Mills amplitudes}
\label{sec:pure-ym} 
We would like to observe that the methods used in the derivation of the
 massive CSW rules suggest also a more direct way for the quantum completion
 of the CSW rules independent of the SUSY decomposition. 
This could be e.g. useful for the application of MHV-methods to two-loop
diagrams where a SUSY decomposition is less straightforward.
We use the
 four-dimensional helicity scheme variant of dimensional regularisation since
 we want to preserve as much of the four dimensional structure, including
 supersymmetry, as possible.  Therefore we keep the vector fields
 four-dimensional so that the light-cone vertices can be expressed in terms of
 the usual spinor products.  This is in contrast to \cite{Ettle:2007qc} for
 instance, who treat everything in $D$ dimensions. On the level of the action,
 this is implemented for lightcone Yang-Mills by adding a term,
\begin{equation}                       
\mathcal{L}_{\mu}= -2\mu^2 \tr A_z A_{\bar z}
\end{equation}
to the Lagrangian. Hence the theory which we'd like to canonically transform
contains an extra term which also needs to be transformed. This is of
course analogous to the massive scalar case. 
However, in the present case we also continue the momenta of external gluons to $D$-dimensions which is in contrast to the computation of the rational piece using a massive scalar loop and massless gluons.
Inserting the expansions \eqref{eq:em-coeff} into the above expression will yield a series of terms
which are all proportional to $\mu^2$. 
Exchanging some of the summations the resulting expression can be 
brought into the form
\begin{equation}\label{eq:dimregpureYM}
2\mu^2\tr[ A_{\bar z,p}A_{z,-p}]=
\sum_{n=2}^{\infty} \prod_{i=1}^n
\int\widetilde{d k_i}
2(-i)g^n  V_{\bar B_1, B_2,\dots B_n}
\tr[\bar B_1,B_2,\ldots B_n]
\end{equation}
where the vertex is given in terms of the expansion coefficients as
\begin{multline}
(-i)g^n V_{\bar B_1, B_2,\dots B_n}=  \mu^2 \sum_{j=1}^{n-1}\sum_{s=1}^j
  \frac{k_{1+}}{p_+}\mathcal{X}^s(p,k_{n+2-s},\dots,k_n,k_1,\dots,k_{j+1-s})\\
\times  \mathcal{Y}(-p,k_{j+2-s},\dots k_{n+1-s})
\end{multline}
Here the indices have to be interpreted in a cyclic way, i.e $k_{n+1}=k_1$.
The quadratic contribution gets re-summed into the $B \bar{B}$ propagator,
similar to the massive scalar case.

Inserting the explicit expressions for the expansion coefficients one finds for the
three-point vertex
\begin{equation}
\label{eq:gluon-mass-3}
\begin{aligned}
 V_{\bar B_1, B_2,B_3}= (i g^{-3}\mu^2) \left[
\mathcal{Y}(k_1,k_2,k_3)
+\frac{k_{1,+}}{k_{3,+}}\mathcal{X}^1(k_3,k_1,k_2)+
\frac{k_{1,+}}{k_{2,+}}\mathcal{X}^2(k_2,k_3,k_1)\right]\\
=\frac{\sqrt 2i \mu^2 \braket{\eta 1}^3}{\braket{12}\braket{23}\braket{31}}
\left(
\frac{\braket{12}\braket{31}}{\braket{\eta 1}\braket{\eta 2}\braket{3\eta}}
+\frac{\braket{23}\braket{31}}{\braket{\eta 3}^2\braket{\eta 2}}
+\frac{\braket{12}\braket{23}}{\braket{\eta 2}^2\braket{\eta 3}}
 \right)
\end{aligned}
\end{equation}
We will now argue that this vertex gives the correct
three-point tree-level vertex in the 'four-dimensional' 
limit $\mu\to 0$ and consequently the
correct quadruple-cut contribution to the four-point all-plus amplitude.
Naively, the tree amplitudes generated by the above vertices vanish in the $\mu\to 0$ limit  but one has  to be careful about the order of the four-dimensional and the on-shell limits.
It is easy to see that in the four dimensional limit there are poles in this expression since in that limit one obtains 
\begin{equation}
\braket{1 2} \sbraket{1 2} \rightarrow 0
\end{equation}
Hence one should be more careful about the limit. It is not hard to show using the on-shell condition $p_{D,i}^2 = 0 \Rightarrow p_i^2=\mu^2$ and \eqref{eq:braket-eta} that one obtains exactly the same type of expression as derived in \cite{Ettle:2007qc} (in their notation),
\begin{equation}
A = \frac{1^+}{\left(1 2\right)} \left(\frac{p_1^2}{1^+} + \frac{p_2^2}{2^+} + \frac{p_3^2}{3^+}  \right)
\end{equation}
Just as in the `massive' scalar case, we arrive at a complete picture of where the missing tree three particle amplitudes are. If one takes the four dimensional limit first and the LSZ reduction next, then these amplitudes arise from equivalence theorem violations. If one takes LSZ first and the four dimensional limit next as in the above example, these amplitudes arise from the dimensional regularisation vertices \eqref{eq:dimregpureYM}. Note that the same restriction to three particle tree amplitudes as in the 'violations' case is expected to apply to the tree level contributions generated by \eqref{eq:dimregpureYM}. At the loop level however there is no choice as one needs to integrate over the off-dimensional momenta.  

For the all-plus four point amplitude it is easy to see that a quadruple cut of the terms generated by the terms in equation \eqref{eq:dimregpureYM} will give the same answer as in Yang-Mills theory, since the contributing cut-box diagram only features the on-shell Yang-Mills three point function. Hence it is fully expected that the total expression integrates to the Yang-Mills answer. To proof this one would need to show in addition that triple and double cuts do not give
rise to extra contributions.  

Let us now briefly comment on amplitudes with more than four legs. 
For the general case one can use the
eikonal identity to write the vertex~\eqref{eq:dimregpureYM} in the form
\begin{equation}
\begin{aligned}
V_{\bar B_1,B_2,\dots\dots B_n}
&=\frac{-i  2^{n/2-1}\mu^2\braket{\eta 1}^3}{
\braket{12}\dots\braket{(n-1)n}\braket{n1}}
 \sum_{i=1}^{n-1}
\frac{\braket{(i+1)1}\braket{i(i+1)}}{\braket{\eta (1+i)}^2\braket{\eta i}}
\end{aligned}
\end{equation}
There is an analogous twistor expression. This vertex is not yet of the same simplicity as the scalar mass-vertex but recall that here all the gluons are treated as 'massive'. We conjecture but are as yet unable to proof that these vertices provide a consistent completion of the CSW formalism at the loop level.

As favourable circumstantial evidence this is the case let us note that collinear singularities are manifestly generated by terms in the vertices. Therefore given that the four-point all plus amplitude comes out correctly, the n-particle all plus amplitude will also be contained in the formalism up to the usual ambiguity for the 5 particle amplitude. Furthermore the observations made in section~\ref{sec:loop-remarks} in the context of the massive scalar regulator continue to apply, for instance the all-minus amplitude is still given by three-point MHV vertices alone and the all-plus and one-minus amplitudes are expected to arise as a $\epsilon\times 1/\epsilon$ cancellation. Hence our picture of loop level completion is a direct mixture of observations in \cite{Ettle:2007qc} and \cite{Brandhuber:2007vm}, with the added bonus of using standard dimensional regularisation and standard vertices. One further difference to \cite{Brandhuber:2007vm} is that our regulator is parity asymmetric, while there it is parity symmetric. The major difference to \cite{Ettle:2007qc} is basically the variant of dimensional regularisation employed: we employ a version of the four dimensional helicity scheme, they treat all their fields in D dimensions.

Finally we would like to remark that the regularisation using the vertex~\eqref{eq:dimregpureYM} should in principle apply also at higher loops. This suggests an interesting picture of higher loop-amplitudes, where contributions containing the leading $1/\epsilon$ divergences \emph{always} follow from MHV vertices combined into loops, with contributions to sub-leading divergences, finite pieces and higher orders in $\epsilon$ arising from the vertices~\eqref{eq:dimregpureYM}. This certainly deserves further study, including a more careful treatment of the regularisation then employed here since it is known that dimensional reduction in non-supersymmetric gauge theories requires some additional care at higher loop levels, see e.g.~\cite{Jack:1994bn}.

\section{Higgs-gluon couplings}
\label{sec:higgs}
One of the reasons scalar particles are interesting and by extension MHV methods for these is of course the fact that there is a scalar in the standard model: the Higg's particle. Apart from the direct application of our methods to the weak sector, one can study an effective coupling of the Higg's particle to gluons mediated by a top quark loop which will be done in this section. 

In the approximation of a large top quark mass, this interaction can be modelled nicely by an effective interaction vertex between Higg's field $H$ and the gluons of the form~\cite{Wilczek:1977zn, Shifman:1979eb}
\begin{equation}\label{eq:effhiggscoupl}
\frac{1}{2} \frac{\alpha_s}{6 \pi v}\int dx^4 \tilde{H} F^2 
\equiv \frac{1}{2}\int dx^4 H F^2.
\end{equation}
with $v\sim 246 GeV$\cite{Dixon:2004za}. The above vertex contains of course the simplest possible gauge invariant dimension 5 operator we can construct out of the fields at hand.  We have reabsorbed the effective coupling constant into the (now dimensionless) Higg's field $H$.  As advocated in \cite{Dixon:2004za} to derive the CSW rules from the space-time action one should split $H = \phi+\phi^{\dagger}$ and write
\begin{equation}
\label{eq:higgs-action}
\int dx^4 H F^2 \sim \int dx^4 \phi F_+^2 + \phi^{\dagger} F_{-}^2 
\end{equation}
The spinor-components of the (anti) self-dual Yang-Mills field strength $F_{\mp}$ appear in the decomposition
\begin{equation}
F_{\alpha\dot\alpha \beta\dot\beta}=\epsilon_{\alpha\beta}
F_{+\dot\alpha \dot\beta}+\epsilon_{\dot\alpha\dot\beta}
F_{-\alpha \beta}
\end{equation}
Using the explicit expression of the symmetrical self-dual field-strength 
\begin{equation}
  F_{+\dot\alpha \dot\beta}=
\frac{1}{2}({F_{\dot\alpha \alpha\dot\beta}}^\alpha)
=\frac{1}{2}
\left(\partial_{\dot\alpha \alpha}A^{\alpha}_{\dot\beta}
 +\partial_{\dot\beta \alpha}A^{\alpha}_{\dot\alpha}-i g (A_{\dot\alpha\alpha}A_{\dot\beta}^\alpha
+A_{\dot\beta\alpha} A_{\dot\alpha}^\alpha)\right)
\end{equation}
it is easy to read off the effective two-gluon $\phi$ coupling from
the expression
\begin{equation}
\label{eq:spinor-higgs-glue}
 \frac{1}{4} \phi F_+^2=\frac{1}{2}\phi\; \partial_{\dot\alpha \alpha}A_{\dot\beta}^\alpha
  (\partial^{\dot\alpha}_{\beta}A^{\dot\beta\beta} 
+\partial^{\dot\beta}_{\beta}A^{\dot\alpha\beta} )+\dots
\end{equation}
Note that there are only two terms contributing, whereas in the corresponding expression in the four-vector formalism~\cite{Dixon:2004za} there are three terms including one with an $\epsilon$-tensor.
For the amplitude of one $\phi$ and two on-shell gluons one obtains
\begin{equation}
  A_3(\phi, A_1, A_2)
= -\frac{i}{2} 
(p_{1\dalpha\alpha}\epsilon^\alpha_{1\dbeta})
  (p^{\dot\alpha}_{2,\beta}\epsilon_2^{\dot\beta\beta} 
+p^{\dbeta}_{2,\beta}\epsilon_2^{\dot\alpha\beta} )
= i \braket{1+|\fmslash \epsilon_1|2+}
  \braket{2+|\fmslash \epsilon_2|1+}
\end{equation}
where the Schouten-identity was used. Inserting the expressions for the
polarisation vectors one reproduces in a simple way the known fact that
the vertex vanishes as soon as one gluon has positive helicity while
for two negative helicity gluon one finds
\begin{equation}
  A_3(\phi, A^-_1, A^-_2)= -2 i \braket{12}^2
\end{equation}
Similarly the MHV-amplitudes with an extra $\phi$-particle 
are just the same as the purely gluonic ones
\begin{equation}
    A_{n+1}(\phi,A^+_1,\dots, A_i^-,\dots, A_{j}^-,\dots, A^+_n)=
i 2^{n/2-1}  \frac{ \braket{ij}^4}{
  \braket{12}\dots\braket{(n-1)n}\braket{n1}}
\end{equation}
as can be shown using the BCFW relations~\cite{Berger:2006sh}. In~\cite{Dixon:2004za} it was proposed to use off-shell continuation of these amplitude as vertices in a CSW-like construction of $\phi$+gluon amplitudes. This approach has been extended to amplitudes including quarks~\cite{Badger:2004ty} and to the one-loop level~\cite{Berger:2006sh,Badger:2007si}.
In the following we will derive these rules from the action~\eqref{eq:higgs-action} using the twistor Yang-Mills approach and will comment on the relation to the canonical approach at the end of this section. In the course of the derivation we obtain additional vertices with an arbitrary number of phi-fields 
\begin{equation}\label{eq:higgsgluonstower}
    V_{n}(\phi_1,\dots\phi_{l-1},B_{l},\dots, \bar{B}_i,\dots, \bar{B}_{j},\dots, B_n)=
i 2^{n/2-1}  \frac{ \braket{ij}^4}{
  \braket{l(l+1)}\dots\braket{(n-1)n}\braket{nl}}
\end{equation}
While puzzling at first, one can convince oneself that these vertices have
to be present in order to calculate amplitudes with more than one Higgs boson. 
Calculating the four point amplitude with two $\phi$ fields in ordinary Feynman
gauge using the interaction~\eqref{eq:spinor-higgs-glue} one finds that
the gluon propagator is cancelled and the same expression as for the three
point amplitude is obtained:
\begin{equation}\label{eq:contractselfdual}
\begin{aligned}
  A_4(\phi_{1},\phi_2,A_3,A_4)&=-\frac{i}{4}
(p_{3,\dalpha\alpha}\epsilon_{3\dbeta}^\alpha+p_{3\dbeta\alpha}\epsilon_{3\dalpha}^\alpha)K^{\dalpha}_\beta  \;
\frac{\varepsilon^{\beta\gamma}\varepsilon^{\dbeta\dgamma}}{K^2}\;
K^{\ddelta}_\gamma  
(p_{4,\ddelta\delta}\epsilon_{4\dgamma}^\delta+p_{4\dgamma\delta}\epsilon_{4\ddelta}^\delta)\\
&=i \braket{3+|\fmslash \epsilon_3|4+}\braket{4+|\fmslash \epsilon_4|3+}
\end{aligned}
\end{equation}
This type of cancellation can be shown to hold even off-shell. It therefore applies directly to the full tower of amplitudes with two gluons and the rest $\phi$ particles. A BCFW type argument as in ~\cite{Berger:2006sh} then yields the scattering amplitudes of MHV type easily. By extension, it is then also easy to conjecture that these extra MHV amplitudes can be promoted to MHV vertices. 

\subsection{Twistor Yang-Mills derivation}
The twistor lifting formulae of section \ref{sec:twistor} (and a sneak peek at the known result) indicate that we'd rather consider terms like
\begin{equation}
\sim \phi C^2
\end{equation}
added to the Chalmers and Siegel action \eqref{eq:chalsiegaction} as these lift easily to twistor space. As we have a field equation equating $C = F^+$ in the Chalmers and Siegel action this is also a quite plausible term. Integrating out $C$ from 
\begin{equation}
 \int d^4\!x  CF_+ - \frac{1}{2} C^2  + \frac{1}{2} \phi C^2
\end{equation}
however does not yield the action we are interested in. Rather, this yields
\begin{equation}\label{eq:wrongresulthiggs}
\frac{1}{2}\int d^4\!x \frac{1}{1 - \phi} (F_+)^2
\end{equation}
where the inverse is defined by it's Taylor series. Instead, one has to integrate out $C$ from 
\begin{equation}\label{eq:higgstolift}
 \int d^4\!x  C F_+ - \frac{1}{2} C^2  + \frac{1}{2} \frac{\phi}{1 + \phi} C^2
\end{equation}
to obtain the action of interest in this section. In these computations a field dependent determinant arises which however does not contribute to scattering amplitudes when calculated in dimensional regularisation since it involves a trivial kinetic term. 

The action thus obtained can be lifted easily to twistor space by the same formulae as in \ref{sec:twistor}. Since the Higgs is not charged under the strong gauge group, it can either be left on space-time or simply be lifted by
\begin{equation}
\phi(x) = \int dk \xi_0
\end{equation}
without any of the strong gauge group frames. In particular, the coupling of the Higgs fields $\xi$ to the twistor field $\bar{B}_0$ is given by
\begin{equation}\label{eq:higgscouplingtower}
- \tr \int d^4x dk_1 dk_2  \frac{\int \xi_0}{1 + \int \xi_0} \, <\!\pi_1 \pi_2\!>^2 \left(H \bar{B}_0 H^{-1}\right)_{(1)} \wedge \left(H \bar{B}_0 H^{-1}\right)_{(2)}.
\end{equation}
Expanding out the inverse yields a sum over terms with increasing numbers of $\xi$ particles.

In the obtained twistor action two different gauges are again available as the entire formalism is invariant under the appropriate gauge group by construction. Picking space-time gauge will reduce the action down to the space-time action we started of from. However, the real fun begins when one selects CSW gauge. It is not hard to see that the structure of the vertices in \ref{eq:higgscouplingtower} are all the same as the '$C^2$' term in Yang-Mills theory. Hence it is also not hard to see that in CSW gauge the entire tower of vertices generates scattering amplitudes of the same MHV form as in pure gauge theory, the only difference being the momentum conservation delta function which involves all participating particles, including the Higgs particles. Furthermore, the MHV rules are also easy to derive by the same analogy and are exactly those given in \cite{Dixon:2004za} in the case of one Higg's particle. As a special case one obtains the vanishing of the equal helicity and one-unequal helicity Higgs-gluon amplitudes. 

What is new in the above derivation is that there is actually a tower of '$\phi^i$ glue' amplitudes, all of the same MHV form. It is not known yet how practical this is as for instance the $H^2$ amplitudes would also involve calculating a $\phi \phi^{\dagger}$ amplitude. In addition, other effects than the effective interaction considered here might start dominating when one considers 2-Higgs processes. One such effect is the contribution of a top quark loop with two Higgs particles attached that leads to an $F_{\mu\nu}F^{\mu\nu}H^2$ term in the effective action. 
 Of course, what is easy to calculate is the case of gluon amplitudes with one  pseudo-scalar Higgs ($\sim \phi - \phi^{\dagger}$) and one Higgs ($\sim \phi + \phi^{\dagger}$) since this requires only  $\phi^2+g$ vertices and the complex conjugate . In fact, many MHV-type calculations for this case will probably only require minimal effort once the single Higg's calculation has been done. 

\subsection{Extension to more general vertices}
The arguments above generalise easily to one particular class of space-time vertices. Consider instead of the coupling studied above a scalar field coupled through a higher dimension operator,
\begin{equation}
\mathcal{L}_{\textrm{int}} = \kappa \tr \phi(x)  (F_+)^i
\end{equation}
for some natural number $i$ with some particular contraction amongst the indices of $F$. The field $\phi$ can be charged or uncharged under the gauge group. This type of operator arises for instance in studying sub-leading effects in the effective action \eqref{eq:effhiggscoupl} suppressed by powers of the top quark mass. As above, in order to lift this easily to twistor space, $F_+$ needs to be replaced by the Chalmers and Siegel $C$ field. It is not hard to guess that one obtains an infinite tower of couplings with increasing numbers of $\phi$ fields. Lifting these to twistor space and choosing CSW gauge then easily generates towers of MHV vertices and scattering amplitudes. In particular, the interaction written above will generate amplitudes which involve at least $i$ negative helicity gluons, while for precisely $i$ the amplitudes will be holomorphic. The contractions between indices in the higher dimension operator translates into a particular contraction of the spinor momenta of the negative helicity gluons in the numerator. A Noether argument will then also yield couplings of the $\phi$ field to quarks or scalars in any representation if so desired. 

As an example of the above general argument, consider the term
\begin{equation}\label{eq:interactiontermex}
\mathcal{L}_{\textrm{int}} = \kappa  \phi \tr \left(F_{\dalpha}^{\phantom{\dalpha}\dbeta} F_{\dbeta}^{\phantom{\dalpha}\dgamma} F_{\dgamma}^{\phantom{\dalpha}\dalpha}\right)
\end{equation}
A complication with respect to the quadratic analysis is the fact that the cubic interaction cannot be integrated out exactly. However, as we are only interested here in perturbation theory, this obstacle can be overcome. For this it is easier to consider the inverse problem,
\begin{align}
Z[J] &= \int dCdA e^{\int C_{\dalpha \dbeta} F^{\dalpha \dbeta } - \frac{1}{2} C_{\dalpha \dbeta} C^{\dalpha \dbeta} + \kappa  \phi(x) \tr \left(C_{\dalpha}^{\phantom{\dalpha}\dbeta} C_{\dbeta}^{\phantom{\dalpha}\dgamma} C_{\dgamma}^{\phantom{\dalpha}\dalpha}\right) + J_{\dalpha \dbeta} C^{\dalpha \dbeta}}\nonumber \\
& = \int dCdA \sum_{i=0}^{\infty} \frac{1}{i!}\left( \int \kappa  \phi  \frac{\partial}{\partial J} \frac{\partial}{\partial J}\frac{\partial}{\partial J} \right)^i e^{\int C_{\dalpha \dbeta} F^{\dalpha \dbeta } - \frac{1}{2} C_{\dalpha \dbeta}C^{\dalpha \dbeta} + J_{\dalpha \dbeta} C^{\dalpha \dbeta}}\nonumber \\
& = \int dA \sum_{i=0}^{\infty} \frac{1}{i!} \left( \int \kappa  \phi \frac{\partial}{\partial J} \frac{\partial}{\partial J}\frac{\partial}{\partial J} \right)^i e^{\int (F_++J)^2}
\end{align}
At this point some thought is needed. First of all, if the $J^2$ term was absent, then the result would be obvious. It is is also clear that the case considered in the first part of this section one obtains increasing numbers of $\phi$ fields as in \eqref{eq:wrongresulthiggs}: the $J^2$ term is a 'unity' propagator. Actually, this simply echoes the result in \eqref{eq:contractselfdual}. This propagator cannot connect a vertex to itself as this would yield a tadpole loop contribution of the form
\begin{equation}
\sim \int d^d p
\end{equation}
which vanishes in dimensional regularisation. Restricting to tree level one finds that `$J^2$' always must connect two different vertices for a non-trivial result and therefore increases the number of external $\phi$ and $C$ fields. 

With this analysis the Lagrangian in terms of mostly $C$ fields which reproduces \eqref{eq:interactiontermex} can be constructed perturbatively,
\begin{equation}\label{eq:lifttotwspacexcoupl}
S = \int \int C_{\dalpha \dbeta} F^{\dalpha \dbeta } - \frac{1}{2} C_{\dalpha \dbeta}C^{\dalpha \dbeta} + \kappa  \phi(x) \tr \left(C_{\dalpha}^{\phantom{\dalpha}\dbeta} C_{\dbeta}^{\phantom{\dalpha}\dgamma} C_{\dgamma}^{\phantom{\dalpha}\dalpha}\right) + \textrm{terms with more {$\phi,C$}'s}
\end{equation}
This can be lifted to twistor space. It is furthermore easy to see that in CSW gauge the scattering amplitude with 3 negative and an arbitrary number of positive helicity gluons generated by \eqref{eq:interactiontermex} is
\begin{equation}
<\phi  g^+_{1} \ldots g^-_{r} \ldots g^-_{s} \ldots g^-_{t} \ldots  g_n> = i \sqrt{2}^n \kappa  \frac{\braket{r s}^2\braket{s t}^2\braket{t r}^2}{\braket{1 2} \ldots \braket{n 1}}
\end{equation}
confirming an expression in \cite{Dixon:2004za}. 

More general cases than the example considered here follow easily. For combinations of quadratic vertices with any number of higher higher dimensional operators of the form \eqref{eq:interactiontermex} it is convenient to perform the path integral over the quadratic ones exactly. This situation occurs for instance when one wants to study sub-leading terms in the effective action for Higg's particles coupled to glue through a massive top quark loop.

\subsection{Coupling in other matters}
In~\cite{Badger:2004ty} the extension of the above analysis in the quadratic case to include Higg's particle-quark interactions was discussed. In the context of the twistor action approach to gauge theory it is now very simple to derive these results directly from the action as well. The main observation needed is that within the lifting approach to the gauge theory one should keep at all times the extra twistor space gauge symmetry manifest. In the glue coupled to matter case, this requirement was shown in~\cite{Boels:2007qn} to lead to towers of vertices via the Noether procedure. These towers translate to CSW-style scattering amplitudes and vertices. For $\mathcal{N}=1$ for instance, this will reproduce exactly the gluon-gluino and gluino-gluino couplings needed for the appropriate CSW rules in this case. For $\mathcal{N}=4$, application of the same argument reproduces Nair's super-vertex \cite{Nair:1988bq} in the action. It is then a simple matter to extend the same type Noether calculation to the vertex leading to \eqref{eq:higgscouplingtower}. It is again important here that the Higgs is uncharged under the strong gauge group. Hence in the case of the vertex \eqref{eq:effhiggscoupl} the CSW formalism as described in~\cite{Badger:2004ty} is easily derived with the added bonus of having an arbitrary number of $\phi$ particles. Furthermore, this derivation can be made to work directly for matter transforming in the fundamental. A worked out example of applying the Noether procedure to the twistor action can be found in appendix \ref{app:noetherproc}.

One point to bear in mind in all this is that the generated MHV vertices are derived for a particular space-time action which can be re-obtained by imposing space-time gauge. As a consequence, just as noted before, there is an ambiguity for deriving the couplings of colored scalars since there is more than one gauge-invariant term on twistor space for four colored scalars\footnote{One can show there is no such term for the fermions.}. On space-time this corresponds to the effective dimension $5$ vertex,
\begin{equation}
\mathcal{L}_{\textrm{ambiguous}} \sim \int \phi(x) \tr \left(s \bar{s}\right)^2
\end{equation} 
where $s$ and $\bar{s}$ denote the space-time scalar fields. In principle the value of this coupling would have to be calculated separately. However, due to the lack of scalar fields charged under the strong gauge group in the standard model this ambiguity is at present of mostly theoretical value.

The analysis for the general class of higher dimensional operators follows along similar lines. In terms of the example studied above, starting from the twistor lift of \eqref{eq:lifttotwspacexcoupl} one obtains twistor vertices of the form,
\begin{align}
  &V_{\text{CSW}} (\bar{\lambda}_1 B_2,\ldots \bar B_{i} \ldots \bar B_{j} \ldots B_{n-1} \lambda_n)= \\&
 3 \kappa  \phi(x)  \nonumber \int_{\left(\CP^1\right)^4}  
\frac{\braket{\pi_1 \pi_2} \braket{\pi_2 \pi_3} 
\braket{\pi_3 \pi_1}}{\braket{\pi_1 \pi_4}} \left(\bar{\lambda}_0 H^{-1} \right)_{(1)} \left(H \bar{B} H^{-1}\right)_{(2)}  \left(H \bar{B} H^{-1}\right)_{(3)}   \left( H \lambda_0 \right)_{(4)}\nonumber    \\
 & V_{\text{CSW}}(\bar{\lambda}_1 B_2,\ldots \lambda_i \bar{\lambda}_{i+1} B_{i+2}\ldots \bar B_{j} \ldots B_{n-1} \lambda_n) 
= 3 \sqrt{2} \kappa  \phi(x)  \\
&\int_{\left(\CP^1\right)^5} \!\!\!\!
 \frac{\braket{\pi_1 \pi_2} \braket{\pi_2 \pi_4} \braket{\pi_4 \pi_1}}{\braket{\pi_3 \pi_4} \braket{\pi_5 \pi_1}} \left(\bar{\lambda}_0 H^{-1} \right)_{(1)} \left(H \bar{B} H^{-1}\right)_{(2)} \left( H \lambda_0 \right)_{(3)} 
\left(\bar{\lambda}_0 H^{-1} \right)_{(4)}   \left( H \lambda_0 \right)_{(5)}
 \nonumber\\
&V_{\text{CSW}}(\bar{\lambda}_1 B_2,\ldots \lambda_i \bar{\lambda}_{i+1} B_{i+2}\ldots \lambda_j \bar{\lambda}_{j+1} B_{j+2},\ldots \lambda_n)= 3 \kappa  \phi(x)  \int_{\left(\CP^1\right)^6}   \\
& \frac{\braket{\pi_1 \pi_3} \braket{\pi_3 \pi_5} \braket{\pi_5 \pi_1}}{\braket{\pi_2 \pi_3} \braket{\pi_4 \pi_5} \braket{\pi_6 \pi_1}} \left(\bar{\lambda}_0 H^{-1} \right)_{(1)}   \left( H \lambda_0 \right)_{(2)}\left(\bar{\lambda}_0 H^{-1} \right)_{(3)}   \left( H \lambda_0 \right)_{(4)}  \left(\bar{\lambda}_0 H^{-1} \right)_{(5)}   \left( H \lambda_0 \right)_{(6)} \nonumber
\end{align}
with $\lambda$ and $\bar\lambda$ a positive and negative helicity quark transforming in the fundamental respectively. These vertices have to be integrated over space-time. It is straightforward to derive scattering amplitudes from these vertices in CSW gauge. In addition, more general vertices and more general forms of matter are a straightforward extension. 

\subsection{Charged scalar vertices}
Although of less phenomenological relevance presently, it is easy to see that the methods in this section extend also to non-minimally coupled colored scalars transforming in the adjoint or in the 'bi-fundamental, bi-anti-fundamental' representation. The difference is that the twistor lift of 
\begin{equation}
 \tr \phi F_{+}^2
\end{equation}
in the case of an adjoint scalar after an analogous analysis now reads
\begin{equation}
\tr \int \dk_1 \dk_2 \dk_3 \left(H \xi_0 H^{-1} \right)_1 \left(H \bar{B}_0 H^{-1}  \right)_2 \left(H \bar{B}_0 H^{-1}  \right)_3 \braket{\pi_2 \pi_3}^2 
\end{equation}
In addition, there are analogous terms with more $\phi$ fields inserted. The above term will lead to holomorphic scattering amplitudes of the form
\begin{equation}
A(\phi_1 B_2 \ldots \bar{B}_i \ldots \bar{B}_j \ldots) \sim \frac{ \braket{\pi_1 \pi_i }  \braket{\pi_i \pi_j}^3 \braket{\pi_j \pi_1}} {\braket{1 2} \ldots \braket{n 1}}
\end{equation}
and, more generally, to MHV vertices of the indicated shape. Coupling in additional matters is more involved in this particular case since the scalar is charged under the gauge group. 

\subsection{Comment on canonical transformation method}
In principle the above twistor derivation of the Higgs-glue couplings can be derived using the canonical transformation. This however turns out to be fairly tedious and, in keeping with our general observations on the applicability of CSW-rules- derivation methods, \emph{if} the twistor approach works it will generate complete vertices at once. For comparison reasons however, let us derive here the E-M transformation coefficients for the case of the twistor lift of the action in equation \eqref{eq:higgstolift}. 

The twistor calculation uses the same gauge transformation as before to trivialise the self-dual Yang-Mills vertex. Therefore the transformation for $A_z$ is also given by the same coefficients. In the twistor derivation of the E-M coefficients however, the field equation for $\bar{B}_0$ was used. As there is in the case at hand an extra tower of terms which depends on this field, the E-M coefficients will be changed. However, it is also easy to see that the change will only be by a multiplicative function,
\begin{equation}
A_{\bar z} = A_{\bar z}^{\textrm{pure glue}} \left(\frac{1}{1+\phi} \right).
\end{equation}
We expect, but do not attempt to prove here, that the above canonical transformations would arise in a attempt to derive the CSW rules from the lightcone Yang-Mills theory with an effective interaction \eqref{eq:effhiggscoupl}. One immediate problem one has to face there is that integrating out the $A_-$ component is complicated by this interaction, for instance due to derivatives of the $\phi$ field generated by integrating by parts.
Coupling in matter fields seems to be even a step beyond that. 
On the other hand, if one would be able to apply a canonical transformation to 
the Chalmers-Siegel action, clearly the same results as from the twistor approach would be obtained.

\hfill

\section{Conclusions}
In this article several aspects of deriving CSW rules for massive
matter have been discussed. As an example, we focused on scattering of
massive scalars, but we expect that our methods can be applied to much
wider classes of interesting gauge theories and to a wider class of
questions within these. 
As another application of deriving CSW rules we have shown how all CSW rules which involve an effective Higgs-gluon coupling studied in the literature can be easily reproduced from the action. In addition, we have found some CSW vertices with more than one Higgs boson that have not been studied so far in the literature. It would be interesting to see if these extra vertices can be applied to phenomenologically interesting amplitudes such as multi-gluon amplitudes with one pseudo-scalar Higgs and one Higgs.

From the phenomenological point of view, an immediate next step would be the extension to amplitudes with top quarks. 
 Another interesting avenue to pursue  would be to study questions on
 electroweak amplitudes that are especially relevant for applications at the LHC (see e.g.~\cite{Bern:2008ef}). 
 In particular, it would be interesting to see if
the enlarged  gauge freedom of the twistor action allows 
to choose a gauge which simplifies the
electroweak action beyond what is currently used. 
Since the CSW rules can be used to simplify the proof of the
BCFW on-shell recursions, extensions of our approach to massive particles 
with spin could also provide further insight into their treatment in the BCFW approach~\cite{Badger:2005zh,Forde:2005ue, Ferrario:2006np, Schwinn:2007ee}.

On the methodical side, our analysis has shown how the canonical
transformation method and the twistor action method are related by
deriving the transformation formulae of one from the other. We found
it interesting to see how two completely differently motivated
derivations can lead to the same result. Both methods have their own
advantages and disadvantages, which we elaborated upon already
earlier.  It would be desirable to obtain a more systematical
understanding of the canonical transformation method in order to allow
a simpler derivation of the expansion coefficients without the need to
solve recursion relations.
 The most pressing issue from the twistor point of view is to find a twistor
action for Einstein gravity. Since there is a
close relationship between twistor methods and harmonic superspace
methods (see also \cite{Boels:2007gv}) and actions have been
constructed for harmonic superspace supergravity, it is not hard to
imagine a similar reasoning will lead to a twistor supergravity
action. This would be very interesting to obtain, as the amplitudes
generated by this action should give some form of the CSW rules for
gravity.

Yet another set of questions raised by this article concerns loop effects in pure glue Yang-Mills theory. As shown, the massive scalar CSW rules can be used to calculate one loop amplitudes at least in principle, and it would be interesting to see if the computation can be streamlined more, both for more external particles as for addition of more 'minus' gluons. 
 In addition, the rules are compatible with the proposal that some amplitudes arise as an anomaly in dimensional regularisation and suggest this is an anomaly in the conformal symmetry.
Motivated by the massive CSW rules we initiated the study of a direct regulator of the pure glue Yang-Mills Lagrangian that,  at least in principle, could extend to higher loop computations. 
However, there are still open questions that have to be clarified, such as finding a more compact expression for the  regulating terms and verifying the consistency of the proposed regularisation scheme.

\section*{Acknowledgements}
It is a pleasure to thank Kasper Risager, German Rodrigo, Stefan Weinzierl and Costas Zoubos for discussions. We also would like to thank the Arnold Sommerfeld Center for Theoretical Physics in Munich for organizing the workshop on ``Twistors, perturbative gauge theories, supergravity and superstrings'' where this work was initiated. The work of CS is supported by the DFG Sonder\-forschungs\-bereich/Trans\-regio~9 ``Computergest\"utzte Theoretische Teilchenphysik''  and the BMBF grant 05HT6PAA. The work of RB was partly supported by the European Community through the FP6 Marie Curie RTN {\it ENIGMA} (contract number MRTN-CT-2004-5652).

\appendix

\section{Yang-Mills conventions}
\label{app:yang-mills}

For the generators
$T^a$ of the fundamental representation of $SU(N)$ we use the normalisation
\begin{equation}
\label{eq:trace-t}
\tr\;T^a T^b  =  \frac{1}{2} \delta^{a b}
\end{equation}
and the Lie algebra $ \left[ T^a, T^b \right] = i f^{abc} T^c$.
One can translate to these conventions by replacing $T^a\to (-i)T^a$
in~\cite{Ettle:2006bw} 
while in comparison to~\cite{Brandhuber:2006bf,Brandhuber:2007vm} one
needs to replace $T\to \sqrt 2 T$. 
The  Yang-Mills Lagrangian is
\begin{equation}
\mathcal{L}=-\frac{1}{2}\tr [F^{\mu\nu}F_{\mu\nu}]
\end{equation}
with $F_{\mu\nu}=\partial_\mu A_\nu-\partial_\nu A_\mu -i g[A_\mu,A_\nu]$ 
and $A_\mu = T^a A^a_\mu$. 
Expressions from~\cite{Ettle:2006bw} can be converted to these conventions by
replacing $A_\mu\to -i g A_\mu$.

For  amplitudes with scalars transforming under
the fundamental representation of $SU(N)$ 
we employ the usual decomposition
of the full amplitude ${\cal A}_n$ into gauge invariant partial 
amplitudes $A_n$ defined 
by~\cite{Berends:1987me,Dixon:1996wi}: 
\begin{multline}
\label{eq:color-ordered}
 {\cal A}_{n}(\phi_1,g_2,g_3,...,g_{n-1},\phi_{n}) 
= \\ 
 g^{n-2} \sum_{\sigma\in S_{n-2}(2,...,n-1)}  \; \left(
 T^{a_{\sigma(2)}} ... T^{a_{\sigma(n-1)}} \right)_{ij}
 A_{n}\left(\phi_1, g_{\sigma(2)}, ..., g_{\sigma(n-1)}, \phi_n \right). 
\end{multline}
where the sum is over all  permutations of the external gluon legs.

\section{Spinor Identities}
A useful consequence of the Schouten identity $
0=\braket{ab}\braket{cd}+\braket{ac}\braket{db}+\braket{ad}\braket{bc}$
is the so called eikonal identity (see e.g.~\cite{Dixon:1996wi})
\begin{equation}
\label{eq:eikonal}
  \sum_{i=j}^{k-1}\frac{\braket{i (i+1)}}{\braket{i\eta}\braket{\eta(i+1)}}
  =\frac{\braket{jk}}{\braket{j\eta}\braket{\eta k}}
\end{equation}
As an application of the eikonal identity one can derive the formula
\begin{equation}
\label{eq:eikonal-b}
  \sum_{i=j}^{n-1}\frac{(k_{j,i})_+
\braket{i (i+1)}}{\braket{i\eta}\braket{\eta(i+1)}}
= \frac{\braket{\eta +|\fmslash k_{j,n-1}|n+}}{2\braket{\eta n}} 
\end{equation}
where we have used the translation between light-cone components and
spinor-products to write $2(k_k)_+\braket{kn}=\braket{k
  \eta}\braket{\eta+|\fmslash k_k|n+}$.
As a generalisation one has (for $l<j$)
\begin{equation}
\label{eq:eikonal-c}
  \sum_{i=j}^{n-1}\frac{(k_{l,i})_+
\braket{i (i+1)}}{\braket{i\nu}\braket{\nu(i+1)}}
= \frac{ \braket{\nu +|\fmslash k_{j,n-1}|n+}}{2\braket{\nu n}}
+ \frac{(k_{l,j-1})_+\braket{jn}}{\braket{j\nu}\braket{\nu n}}
\end{equation}

\section{Derivation of the transformation of $A_{\bar z}$ in the canonical method}
\label{app:derive-a-bar}
In this appendix we show that the transformations of the scalars
~\eqref{eq:phi-trafo} and~\eqref{eq:phibar-trafo} together with the additional
piece in the transformation of the gluon momentum~\eqref{eq:bbar-trafo-scalar}
leave the sum of scalar and gluon kinetic terms invariant, c.f.~\eqref{eq:transform-kinetic}.
 
The kinetic term of the scalars alone is not transformed just into
a quadratic term but into a sum of terms
\begin{equation}
\label{eq:kinetic-phi}
  (p_+\bar\phi_p)(p_-\phi_{-p})= \sum_{n=2}^\infty\int \prod_{i=1}^n 
\widetilde{d k_i} (g\sqrt 2)^{n-2}
\frac{\mathcal{K}^{\phi}_{k_1,\dots k_n}\braket{\eta 1}\braket{\eta n}}
{\braket{12}\dots\braket{(n-1)n}} 
\left(\bar\xi_{k_1}B_{k_2}\dots B_{k_{n-1}}\xi_{k_n}\right)
\end{equation}
where the coefficients simplify using the eikonal identity and~\eqref{eq:eikonal-b}
\begin{equation}
\label{eq:kinetic-phi-coeff}
\begin{aligned}
 \mathcal{K}^{\phi}_{k_1,\dots k_n}&=
\sum_{j=1}^{n-1}\frac{(k_{1,j})_+(k_{j+1,n})_-\braket{j(j+1)}}{
\braket{\eta j}\braket{\eta (j+1)}}
=\sum_{l=2}^{n} \sum_{k=1}^{l-1} \frac{(k_{k})_+(k_l)_-\braket{kl}}{
\braket{\eta l}\braket{\eta k}}\\
&= \sum_{l=2}^{n-1}(k_l)_-  \frac{ \braket{\eta+|\fmslash k_{1,l-1}|l+}}{2
\braket{\eta l}}
\end{aligned}
\end{equation}
 The term $l=n$ has
been dropped since it vanishes due to momentum conservation.
While the $n=2$ term in~\eqref{eq:kinetic-phi} gives the correct kinetic
Lagrangian for the $\xi$ scalars, the terms with
$n\geq 3$ have to be cancelled by the  scalar
contribution to $\partial_+A_{\bar z} $
in   the transformation of the kinetic term of
the gluons. For this we make an Ansatz of the form
\begin{multline}
\label{eq:bbar-em-scalar}
\left.  p_+ (A_{p,\bar z})_{ij}\right|_{\bar\xi\xi}
=\frac{1}{2}\sum_{n=2}^\infty
\sum_{s=1}^{n-1} \int \prod_{i=1}^n \;\widetilde {dk_i}
\mathcal{W}^s(p,k_1,\dots k_n)\times\\
\left(k_{1+}\bar\xi_{-k_1}\dots B_{-k_s}\right)_k
\left(B_{-k_{s+1}}\dots \xi_{-k_n}\right)_l
\left(\delta_{il}\delta_{jk}-\frac{1}{N}\delta_{ij}\delta_{kl}\right)
\end{multline}
Here the group-index structure has been made explicit.
 Inserting the expansions of $A_z$ and $A_{\bar z}$
 from~\eqref{eq:b-trafo} 
and~\eqref{eq:bbar-em-scalar} 
into the kinetic
term of the gluons leads to the expression 
\begin{multline}
\label{eq:a-kinetic-phi}
2\tr[( p_+ A_{p,\bar z})(p_-A_{-p,z})]=
2\tr[( p_+ \bar B_p) (p_-B_{-p})]+\sum_{n=3}^\infty
 \sum_{s=1}^{n-2} \sum_{j=s+1}^{n-1}
\;\left(\bar\xi_{k_1}B_{k_2}\dots B_{k_{n-1}}\xi_{k_n}\right)\\
(k_1)_+(k_{s+1,j})_-\;\mathcal{W}^s(k_{s+1,j},k_1,\dots k_s,k_{j+1},\dots k_n)
\mathcal{Y}(-k_{s+1,j},k_{s+1},\dots k_j)
\end{multline}

The coefficients of a given $(k_l)_-$ must cancel between~\eqref{eq:a-kinetic-phi} and~\eqref{eq:kinetic-phi}. 
For instance,   the cancellation of the coefficient of $(k_2)_-$ against 
 the $n=3$ term in~\eqref{eq:kinetic-phi}
 determines the first term in the Ansatz~\eqref{eq:bbar-em-scalar}: 
\begin{equation}
  \mathcal{W}^1(k_2,k_1,k_3)=
-(g\sqrt 2)
\frac{\braket{\eta 1}\braket{\eta 3}\braket{\eta+|\fmslash k_1|2+}}{
\braket{\eta 2}\braket{1 2}\braket{23}2(p_1)_+} 
=(g\sqrt 2)\frac{\braket{\eta 3}}{\braket{\eta 2}\braket{23}} 
=(g\sqrt 2)\frac{\braket{\eta 3}}{\braket{\eta 1}\braket{31}}
\end{equation}
where momentum conservation and the identity~\eqref{eq:braket-eta} were
used in the last step.
With some more work one determines the next coefficients in the Ansatz
 from the cancellation
of the terms proportional to $(k_2)_-$ and $(k_3)_-$ in the
 $n=4$ terms as
\begin{equation}
   \mathcal{W}^1(p,k_1,k_2,k_3)=2 g^2\frac{\braket{\eta 3}^2 }{
\braket{\eta 1}\braket{\eta 2} \braket{23}\braket{31} }\; ;\qquad
  \mathcal{W}^2(p,k_1,k_2,k_3)=2 g^2
\frac{\braket{\eta 3}}{\braket{\eta 2}\braket{12}\braket{31}}
\end{equation}
These explicit results motivate an all-multiplicity 
Ansatz for
the coefficients $\mathcal{W}$:
\begin{equation}
\label{eq:def-w}
\mathcal{W}^s(p,k_1,\dots k_n)=
 (g\sqrt 2)^{n-1} \frac{\braket{\eta n}^2 \braket{s(s+1)}}
{\braket{\eta s}\braket{\eta (s+1)}
\braket{12}\dots\braket{(n-1)n}\braket{n1}}
\end{equation}
After some relabelling  one obtains the form 
quoted in~\eqref{eq:bbar-trafo-scalar}.
Inserting the result for $\mathcal{Y}$ and the Ansatz for $\mathcal{W}$
into~\eqref{eq:a-kinetic-phi} 
results in  an expression of the same form
as~\eqref{eq:kinetic-phi} with $\mathcal{K}^\phi$ replaced by
\begin{equation}
\label{eq:kinetic-a}
\mathcal{K}^{A}_{k_1,\dots k_n}
= \frac{\braket{\eta 1}\braket{\eta n}}{\braket{n1}}
\sum_{l=2}^{n-1} (k_l)_-
\sum_{s=1}^{l-1} \sum_{j=l}^{n-1} \frac{(k_{s+1,j})_+
    \braket{s(s+1)}\braket{j(j+1)}}{\braket{\eta s}\braket{(s+1)\eta}
    \braket{\eta j}\braket{(j+1)\eta}}
\end{equation}
Here the order of the summations was exchanged in order to
isolate the coefficient of a given $(k_l)_-$.
 Performing the summation over $j$ 
using~\eqref{eq:eikonal-c} leads to the identity required for
the cancellation of the $n\geq 3$ terms in~\eqref{eq:kinetic-phi-coeff}:
\begin{equation}
\begin{aligned}
\mathcal{K}^{A}_{k_1,\dots k_n}
&= \frac{\braket{\eta 1}}{\braket{n1}}
\sum_{l=2}^{n-1} (k_l)_-\sum_{s=1}^{l-1}\frac{
    \braket{s(s+1)}}{\braket{\eta s}\braket{(s+1)\eta}}
\left(\frac{1}{2}\braket{\eta +|\fmslash k_{l,n} |n+}
 +(k_{s+1,l-1})_+\frac{ \braket{ln}}{\braket{l \eta }}\right)\\
&=
\sum_{l=2}^{n-1} \frac{(k_l)_-}{2\braket{l\eta}\braket{n1} }\left(
\braket{\eta +|\fmslash k_{1,l-1} |n+}\braket{l1} 
+\braket{\eta+|\fmslash k_{2,l-1}|1+}\braket{nl}
\right)
=- \mathcal{K}^{\phi}_{k_1,\dots k_n}
\end{aligned}
\end{equation}
Here  $k_{i,j}$ is understood to be zero for $j<i$.
Momentum conservation and the eikonal identity have be used in the second step 
while the last step uses the Schouten identity.
This concludes the demonstration that the sum of the kinetic terms~\eqref{eq:transform-kinetic}
stays invariant and the
 transformations~\eqref{eq:phi-trafo} and~\eqref{eq:bbar-trafo-scalar} are indeed canonical.

\section{Harmonic expansions and Noether procedure for actions on twistor space}
\label{app:noetherproc}

In this appendix we show that given the twistor action for pure Yang-Mills theory, 
\begin{align}\nonumber S_{\text{YM}} = 2 \tr & \int d^4x k \,  \bar{B} \wedge \left(\dbar B -i \sqrt{2} g B \wedge B \right)  \\ & - \tr \int d^4x k_1 k_2 \wedge \, \braket{\pi_1 \pi_2}^2 \left(H \bar{B} H^{-1}\right)_{(1)} \wedge \left(H \bar{B} H^{-1}\right)_{(2)} 
\end{align}
which was constructed in section \ref{sec:twistor} the action for a massive scalar coupled to Yang-Mills follows by the Noether procedure. In the course of the derivation we will derive the twistor lifting formula for scalars used in the main text. The Noether procedure will of course need a free action as an input. It is easy to verify that one should consider
\begin{equation}\label{eq:uncoupledscalar}
S_{\textrm{scalar}} = \int \Omega \wedge \bar{\xi} \wedge \dbar \xi
\end{equation}
for two as yet uncharged weight $-2$ twistor $(0,1)$ form fields $\bar{\xi}, \xi$. There are two related reasons for this. The more abstract one is that the Penrose transform will relate cohomology classes on twistor space with solutions to the field equations on space-time. The field equations of the above action give exactly those classes, since it has a symmetry
\begin{equation}\label{eq:gaugesymscal}
\bar{\xi} \rightarrow \bar{\xi} + \dbar \bar{f}^{-2} \quad \xi \rightarrow \xi + \dbar f^{-2} 
\end{equation}
with $f^{-2}$ and $\bar{f}^{-2}$ scalar functions of the indicated weight so solutions to the field equations 
\begin{equation}
\dbar \bar{\xi} = \dbar \xi = 0
\end{equation}
up to gauge equivalence are the definition of the Dolbeault cohomology class. The more concrete reason \eqref{eq:uncoupledscalar} is the correct action to consider is that the gauge symmetry \eqref{eq:gaugesymscal} can be fixed to 'space-time gauge',
\begin{equation}
\partial_0^{\dagger} \xi_0 = \partial_0^{\dagger} \bar{\xi}_0 = 0
\end{equation}
Solutions to these conditions are harmonic functions on the sphere, so 
\begin{equation}
\xi_0 = \phi(x) \quad \bar{\xi}_0 = \bar{\phi}(x)
\end{equation}
by a straightforward cohomology argument on the Riemann sphere for fields of this specific weight. The fields $\xi_\alpha$ and $\bar{\xi}_{\alpha}$ can be integrated out to yield
\begin{align}
\xi_{\alpha} &= \frac{\hat{\pi}^{\dalpha} }{\braket{\pi \hat{\pi}}} \partial_{\alpha \dalpha} \phi(x)\\
\bar{\xi}_{\alpha} &= \frac{\hat{\pi}^{\dalpha} }{\braket{\pi \hat{\pi}}} \partial_{\alpha \dalpha} \bar{\phi}(x)
\end{align}
Plugging back into the action this gives 
\begin{equation}
\int d^4x \partial_{\alpha \dalpha} \bar{\phi}(x) \partial^{\alpha \dalpha} \phi(x)
\end{equation}
which is the free field Lagrangian. This is basically the argument in \cite{Boels:2006ir}. 

\subsection*{Harmonic expansions and the twistor transform}
Another way to understand the gauge fixing argument is through a harmonic expansion on the Riemann sphere of the weight $0$ form,
\begin{equation}\label{eq:harmphiexp}
\xi_0 = \phi(x) + \frac{\pi^{\dalpha} \hat{\pi}^{\dbeta}}{\braket{\pi \hat{\pi}}} \phi_{\dalpha \dbeta}(x) + \frac{\pi^{\dalpha} \pi^{\dbeta}\hat{\pi}^{\dgamma} \hat{\pi}^{\ddelta}}{\braket{\pi \hat{\pi}}^2} \phi_{\dalpha \dbeta \dgamma \ddelta}(x) + \ldots
\end{equation}
the weight $-1$ form
\begin{equation}\label{eq:harmphiexp2}
\xi_\alpha = \frac{\hat{\pi}^{\dalpha}}{\braket{\pi \hat{\pi}} } \phi_{\alpha \dalpha}(x) + \frac{\pi^{\dalpha} \hat{\pi}^{\dbeta} \hat{\pi}^{\dgamma}}{\braket{\pi \hat{\pi}}^2} \phi_{\alpha \dalpha \dbeta \dgamma}(x)  + \ldots
\end{equation}
and the weight $-2$ gauge function 
\begin{equation}
f = \frac{\hat{\pi}^{\dalpha} \hat{\pi}^{\dbeta}}{\braket{\pi \hat{\pi}}^2} f_{\dalpha \dbeta}(x) + \ldots
\end{equation}
Note that an harmonic expansion for fields on the Riemann sphere is equal to the perhaps more familiar expansion in spherical harmonics since $SU(2) \sim  SO(3)$. All coefficients in the expansions above are completely symmetric in the primed indices, since they would otherwise not be independent. In principle one has to expand the given action to all orders in the harmonic expansion. However, the gauge symmetry here comes to the rescue. Since 
\begin{equation}
\dbar_0 = \braket{\pi \hat{\pi}} \pi^{\dalpha} \frac{\partial}{\partial \hat{\pi}^{\dalpha}}
\end{equation}
one easily sees by explicit calculation that the gauge symmetry implies that the transformation
\begin{equation}
\xi_0 \rightarrow \xi_0 + \dbar_0 f^{-2}
\end{equation}
can be used to set to zero all higher modes of the harmonic expansion~\eqref{eq:harmphiexp}. The field equations for $\xi_\alpha$ and $\bar{\xi}_\alpha$ in this gauge then set to zero all higher modes of these fields. The action collapses therefore into just the space-time action written above. The harmonic expansion argument is of course equivalent to the gauge fixing argument given before. However, in addition it makes clear that one can invert the expansion by integrating both sides over the Riemann sphere. This leads to a lifting formula for the scalar fields,
\begin{equation}\label{eq:liftforphiwithouta}
\phi(x) = \int_{\CP^1} \xi_0 \quad  \bar{\phi}(x) = \int_{\CP^1} \bar{\xi}_0
\end{equation}
By construction, this is invariant under the gauge transformations of the scalars which can be verified explicitly. 

\subsection*{Penrose transform}
Besides illustrating the action based derivation of the free field equations for a free massless scalar on twistor space, harmonic expansions also offer a convenient way to illustrate Penrose's original observations about the map between cohomology classes and solutions to the wave equation in four dimensions directly. The field equation derived from \eqref{eq:uncoupledscalar} reads,
\begin{equation}
\dbar \xi =0
\end{equation}
As noted before the solution to this equation up to gauge equivalence is precisely the definition of the Dolbeault cohomology class. In terms of the basis utilised throughout this article the equation can be written as (see \eqref{eq:fieldeqtwi})
\begin{align}
\dbar_0 \xi_{\alpha} - \dbar_{\alpha} \xi_0 & = 0 \nonumber \\
\dbar_{\alpha} \xi^{\alpha}& =0 \nonumber 
\end{align}
Inserting the harmonic expansions of \eqref{eq:harmphiexp} and \eqref{eq:harmphiexp2} into these equations yields a system of equations which do not depend on $\pi$ any more. The first few equations read,
\begin{align}\label{eq:remainfieldeq}
\phi_{\alpha \dalpha}(p) - i p_{\alpha \dalpha} \phi(p) = 0 \nonumber\\ 
p_{\alpha \dalpha} \phi^{\alpha}_{\dbeta} = 0
\end{align}
for the first non-trivial one, and 
\begin{align}
\phi_{\alpha \dalpha \dbeta \dgamma}(p) - i p_{\alpha \dalpha} \phi_{\dbeta \dgamma}(p) = 0 \nonumber \\ 
p_{\alpha \dalpha} \phi^{\alpha}_{\dbeta \dgamma \ddelta} = 0
\end{align}
for the second where the top line is symmetrised over the dotted indices. The higher order equations simply contain more symmetrised dotted indices. Above we have used a Fourier transform for only the space-time coordinate $x$. Therefore,
\begin{equation}
\dbar_\alpha = \pi^{\dalpha} \frac{\partial}{\partial x^{\alpha \dalpha}} \rightarrow i \pi^{\dalpha} p_{\alpha \dalpha}.
\end{equation}

By the same argument as above, the gauge symmetry shows that all the higher modes of the harmonic expansion of $\xi_0$~\eqref{eq:harmphiexp} are `pure gauge'. This starts with the $\phi_{\dbeta \dgamma}$ in the equation directly above. Through the field equation, this translates immediately to almost all the harmonic modes of $\xi_\alpha$. Above we used this gauge symmetry to set all these modes to zero. In fact, the only non-trivial equation is \eqref{eq:remainfieldeq} which is unaffected by the gauge transformations. Contracting the upper equation with $p^{\alpha \dgamma}$ gives
\begin{equation}
p^2 \phi(p) = 0
\end{equation}
which is of course the free wave equation of a scalar. So in the special case of a massless scalar field in four dimensions we have shown how the Penrose transform works. 

\subsection{Coupling the scalar twistor fields into Yang-Mills}
We would now like to couple the scalars to the gauge field. The starting point is of course changing the transformation law for the scalars by adding a transformation term under the gauge symmetry
\begin{equation}
\xi^i \rightarrow \xi^i +   f^{0,i}_j \xi^j \quad \bar{\xi}_j \rightarrow \bar{\xi}_j -  \bar{\xi}_i f^{0,i}_j 
\end{equation}
Note that the only term allowed by a simple scaling argument is actually the weight zero gauge transformation parametrised by $f^{0}$, not the weight $-4$ one. Of course, the action is now not invariant under this transformation. 
\begin{equation}
\delta S_{\text{scalar}} = 2  \tr \int d\Omega \wedge \bar{\xi}_i \wedge \left(\dbar f^{0,i}_j  \right)\wedge \xi^j
\end{equation}
It is an easy guess that one has to change the derivative in the action into a covariant one,
\begin{equation}
\delta^{i}_j \dbar \rightarrow \delta^{i}_j \dbar - i \sqrt{2} g B^{i}_j
\end{equation}
This leads however to another problem: the action is now not invariant any more under the scalar gauge formations \eqref{eq:gaugesymscal}. This symmetry is crucial to keep the connection to space-time fields. A halfway solution to the problem is to turn the derivatives in that transformation into covariant ones by adding,
\begin{equation}
\delta \xi^i \,+\!=  - i \sqrt{2} g B^{i}_j f^{-2,j} \quad \delta \bar{\xi}_j \,+\!=  - i \sqrt{2} g \bar{f}^{-2}_i B^{i}_j
\end{equation}
Although it is almost there now, the action is still not invariant under the gauge symmetry. It transforms however into a nice field strength,
\begin{equation}
\delta S_{\text{scalar}} = - 2 i \sqrt{2} g \tr \int    \xi_i  \bar{f}^{-2}_j \wedge F^{ij}  +  \bar{\xi}_j f_i^{-2} \wedge F^{ij}
\end{equation}
This suggests to modify the transformation law of the field $\bar{B}$
\begin{equation}
\delta \bar{B}_{ij} \,+\!=  - i \sqrt{2} g \xi_i \bar{f}^{-2}_j +  - i \sqrt{2} g \bar{\xi}_j f^{-2}_i 
\end{equation}
so that the first term in the Yang-Mills action and the scalar action are invariant under all the gauge symmetries. The second `$C^2$' term of the Yang-Mills action  is not any more because of the just added term to the transformation law of $\bar{B}$. Instead, this term transforms to 
\begin{equation}
\delta S_{\text{YM}}
 = g \int_{\CP^1 \times \CP^1} \left(H \bar{B}_0 H^{-1}\right)^{ij}_1 \left(H (\xi_{0,i} \bar{f}^{-2}_j + \bar{\xi}_{0,j} f^{-2}_i) H^{-1} \right)_2 \braket{\pi_1 \pi_2}^2
\end{equation} 
This term can be cancelled by adding an interaction term to the action,
\begin{equation}
S \,+\!= g \int_{\left(\CP^1\right)^3 } \left(\bar{\xi}_{0} H^{-1} \right)_1 \left(H \bar{B}_0 H^{-1} \right)_2 \left(H \xi_0\right)_3  \frac{\braket{\pi_1 \pi_2} \braket{\pi_2 \pi_3}}{\braket{\pi_1 \pi_3}}
\end{equation}
The reason this works is that a partial integration of the variation will yield a delta function through
\begin{equation}
\dbar_0 \frac{1}{\braket{\pi \kappa}} = \delta(\braket{\pi \kappa})
\end{equation}
for some spinor kappa. The added term again yields a problem through the transformation of $\bar{B}$, but applying the same remedy as above again yields,
\begin{equation}
S \,+\!= \int_{\left(\CP^1\right)^4 } \left(\bar{\xi}_0 H^{-1} \right)_1 \left(H \xi_0\right)_2 \left(\bar{\xi}_0 H^{-1} \right)_3 \left(H \xi_0\right)_4  \frac{\braket{\pi_1 \pi_2} \braket{\pi_3 \pi_4}}{\braket{\pi_1 \pi_4} \braket{\pi_2 \pi_3}}
\end{equation}
Now the action is invariant under the following transformations,
\begin{equation}
\xi \rightarrow \xi + \dbar_B f^{-2} \quad \bar{B} \rightarrow \bar{B}  - i \sqrt{2} g [\bar{\xi}, f^{-2}]
\end{equation}
and
\begin{equation}
\bar{\xi} \rightarrow \bar{\xi} + \dbar_B \bar{f}^{-2} \quad \bar{B} \rightarrow \bar{B}  - i \sqrt{2} g [\xi, \bar{f}^{-2}]
\end{equation}
in addition to the gauge symmetry already present in the pure glue twistor action. However, since we started from a non-minimally coupled action \eqref{eq:uncoupledscalar} it has to be verified what the coupled action has the right space-time interpretation. It is easy to check that gauge-fixing space-time gauge in the action will yield a $\phi^4$ interaction term in the space-time Lagrangian. The reason the Noether procedure is ambiguous in this case is that there exists a separate term invariant under the requested symmetries,
\begin{equation}\label{eq:phi^4}
S_{\phi^4} = g^2 \int_{(\CP^1)^4} (\bar{\xi}_0 H^{-1})_1(H \xi_0)_2(\bar{\xi}_0 H^{-1})_3(H \xi_0 )_4.
\end{equation}
Hence this term can be added to the action with a constant designed to eliminate the $\phi^4$ term in space-time gauge. Of course, this is the procedure employed in the main text. 

\subsection{Lifting formula}
Note that \eqref{eq:liftforphiwithouta} is now not invariant any more under the gauge transformation. This can easily be remedied by adding frames,
\begin{equation}
\phi(x) = \int_{\CP^1} H^{-1} \xi_0 \quad  \bar{\phi}(x) = \int_{\CP^1} \bar{\xi}_0 H
\end{equation}
In effect, the frames are needed to gauge-transform back to the case where $B_0$ is zero. These are the formulae used in the main text \eqref{eq:liftforphi}. Lifting a $\phi^4$ term directly using these gives \eqref{eq:phi^4}. 

From this appendix it should be clear that in principle scalars in any representation of the gauge group can be lifted to twistor space. In particular, it can be verified (or easily guessed!) that in the adjoint representation the lifting formulae are 
\begin{equation}\label{eq:liftforphiwithaadj}
\phi(x) = \int_{\CP^1} H^{-1} \xi_0 H \quad  \bar{\phi}(x) = \int_{\CP^1} H^{-1} \bar{\xi}_0 H
\end{equation}
In more general cases, one needs the Wilson link operator in the required representation.

From the starting point \eqref{eq:uncoupledscalar} one can see that massless fields of in principle \emph{any} spin can be coupled to Yang-Mills theory through a similar Noether procedure. In practice this needs a bit of care when one needs to construct interaction terms to balance the transformation of $\bar{B}$ for higher spin fields. In principle this procedure yields consistent Lagrangians for higher spin fields coupled to Yang-Mills theory. Following the same gauge fixing steps as before, it also easily yields MHV amplitudes for these. A more pedestrian application of these observations would be for spin $\frac{1}{2}$ fermions. We hope to return to this issue in future work.

\providecommand{\href}[2]{#2}\begingroup\raggedright\endgroup


\end{document}